\documentclass[english,aps,manuscript]{revtex4}
\usepackage[T1]{fontenc}
\usepackage[latin9]{inputenc}
\usepackage[a4paper]{geometry}
\usepackage{amsmath}
\usepackage{amssymb}
\usepackage{graphicx}
\usepackage{esint}

\makeatletter

\providecommand{\tabularnewline}{\\}

\@ifundefined{textcolor}{}
{%
 \definecolor{BLACK}{gray}{0}
 \definecolor{WHITE}{gray}{1}
 \definecolor{RED}{rgb}{1,0,0}
 \definecolor{GREEN}{rgb}{0,1,0}
 \definecolor{BLUE}{rgb}{0,0,1}
 \definecolor{CYAN}{cmyk}{1,0,0,0}
 \definecolor{MAGENTA}{cmyk}{0,1,0,0}
 \definecolor{YELLOW}{cmyk}{0,0,1,0}
}

\@ifundefined{showcaptionsetup}{}{%
 \PassOptionsToPackage{caption=false}{subfig}}
\usepackage{subfig}
\makeatother

\usepackage{babel}
\begin{document}

\title{Stability change of a multi-charged vortex due to coupling with quadrupole
mode}

\author{Rafael Poliseli Teles$^{1}$, F. E. A. dos Santos$^{2}$ and V. S.
Bagnato$^{1}$}

\affiliation{$^{1}$Instituto de F�sica de S�o Carlos, USP, Caixa Postal 369,
13560-970 S�o Carlos, S�o Paulo, Brazil\\
$^{2}$Departamento de F�sica, UFSCar, Caixa Postal 676, CEP 13565-905
S�o Carlos, S�o Paulo, Brazil}
\begin{abstract}
We have studied collective modes of quasi-2D Bose-Einstein condensates
with multiply-charged vortices using a variational approach. Two of
the four collective modes considered exhibit coupling between the
vortex dynamics and the large-scale motion of the cloud. The vortex
presence causes a shift in all frequencies of collective modes even
for the ones that do not couple dynamically with the vortex-core.
The coupling between vortex and large-scale collective excitations
can induce the multi-charged vortex to decay into singly-charged vortices
with the quadrupole mode being one possible channel for such a decay.
Therefore a thorough study was done about the possibility to prevent
the vortex decay by applying a Gaussian potential with its width proportional
to the vortex-core radius and varying its height. In such way, we
created a stability diagram of height versus interaction strength
which has stable regions due the static Gaussian potential. Furthermore,
by using a sinusoidal time-modulation around the average height of
the Gaussian potential, we have obtained a diagram for the parametric
resonance which can prevent the vortex decay in regions where static
potential can not.
\end{abstract}
\maketitle

\section{Introduction}

\indent

The dynamics of a trapped Bose-Einstein condensate (BEC) containing
a vortex line at its center has been the object of our studies. We
have studied the effects of a multi-charged vortex in free expansion
dynamics. These central vortices contribute with the quantum pressure
(kinetic energy) which increases the expansion velocity of the condensate
\cite{rpteles1}. Consequently, our work culminates in describing
the collective excitations of a vortex state as well. Here the vortex-core
dynamics couples with the well known collective modes \cite{rpteles2}.
Furthermore, we shows that it is possible to excite these modes using
modulation of the s-wave scattering length. Such a technique has been
already applied to excite the lowest-lying quadrupole mode in a lithium
experiment \cite{cm1}. The motivation for these works is the possibility
of experimental realization. Now our focus is in the anisotropic oscillations
of the vortex-core. In other words, oscillations that lead the vortex
shape from circular to elliptical. Such deformation is a symmetry
breaking of vortex state, and can result in changes of dynamical stability. 

The presence of vortices in condensates can also shift the frequency
of collective excitations. The frequency shift of quadrupole oscillations
have been analytically explored for positive scattering lengths by
using the sum-role approach \cite{smv2}, as well as the effects of
lower-dimensional geometry in the frequency splitting of quadrupole
oscillations \cite{smv3}.

First of all, multi-charged vortices in trapped ultracold Bose gases
are thermodynamically unstable, which means that a single $\ell$-charged
vortex tends to decay into $\ell$ singly quantized vortices. Thus
the configuration of separated singly-charged vortices has lower energy
instead a single vortex with the same angular momentum. Although such
a state with multiple singly-charged vortices is also thermodynamical
unstable when compared with a vortex-free condensate. These multiple
vortices spiral outward from the condensate until remain only the
ground state.

The vortex dynamic instability has so far been studied in the context
of Bogoliubov excitations \cite{mcv06,mcv07,stab03}. Indeed, the
vortex state possess certain Bogoliubov eigenmodes which grow exponentially
and become unstable against infinitesimal perturbations \cite{mcv03}.
These vortices present several unstable modes being a quadrupole mode
the most unstable. For instance, let us consider the work in Ref.\cite{mcv03}.
There the authors studied the modes of quadruple-charged vortex. Among
of them, only three modes are unstable. These unstable modes have
complex eigenfrequencies (CE) and are associated with $l$-fold symmetries.
These symmetries are:
\begin{itemize}
\item Two-fold symmetry; the quadruply-charged vortex splits into four single
vortices arranged in a straight line configuration.
\item Three-fold symmetry; the quadruply-charged vortex splits into four
single vortices arranged in a triangular configuration, i.e., there
are three vortices forming a triangle with each vortex representing
a vertex. The fourth one is at center.
\item Four-fold symmetry; the quadruply-charged vortex splits into four
single vortices arranged in a square configuration with each vortex
placed in a vertex.
\end{itemize}
\indent

Our target is to describe them as a result of the coupling between
the vortex-core dynamics with the collective modes of the condensate.
In order to achieve this goal we have used variational calculations
focusing on the description of only one of the unstable mode (specially
two-fold symmetry). The variational description becomes very complicated
as we increase the number of parameters. Fortunately the most relevant
unstable mode is also the easiest one to calculate within the variational
approximation.

Furthermore, there are some works which add a static Gaussian potential
centered in the core of a vortex with a large circulation which results
into a stable configuration for the multi-charged vortex \cite{mcv05,mcv09}.
Based on these works we checked the dynamical stability for a static
as well as dynamic potential due to a Gaussian laser beam placed in
the vortex-core, when compared with the multiple vortices state. 

This paper is organized as follows: In section \ref{sec:quasi2D},
the quasi-2D approach is introduced. We discussed the wave-function
used with the variational method in section \ref{sec:breakingS} and
detailed the calculation of the Lagrangian in section \ref{sec:lagrangian}.
Section \ref{sec:collectiveM} contains equations of the motion and
their solutions, i.e. the stationary solution, collective modes, and
the fully numerical calculation of Gross-Pitaevskii equation (GPE).
In section \ref{sec:staticLG}, we made a dynamical stability diagram
considering a static Gaussian potential while in section \ref{sec:dynamicLG}
we made a parametric resonance diagram due to a dynamical Gaussian
potential where its height is sinusoidally time dependent.

\section{Quasi-2d condensate}

\indent

\label{sec:quasi2D}

The presence of a large number of atoms in the ground state allows
us to use a classical field description \cite{pit-str}. Where the
non-uniform Bose gas of atomic mass $m$ and s-wave scattering length
$a_{s}$. The scattering length is smaller than the average inter-particle
distance at absolute zero temperature. Its dynamics is given by the
Gross-Pitaevskii equation \cite{pethick}:
\begin{equation}
i\hbar\frac{\partial\Psi\left(\mathbf{r},t\right)}{\partial t}=\left[-\frac{\hbar^{2}}{2m}\nabla^{2}+V\left(\mathbf{r}\right)+U_{0}\left|\Psi\left(\mathbf{r},t\right)\right|^{2}\right]\Psi\left(\mathbf{r},t\right),\label{eq:GPE}
\end{equation}
where the interaction strength between two atoms is 
\begin{equation}
U_{0}=\frac{4\pi\hbar^{2}a_{s}}{m}.
\end{equation}

In order to suppress possible effects due to motions along the axial
direction, we consider a highly anisotropic harmonic confinement of
the form
\begin{eqnarray}
V\left(\mathbf{r}\right) & = & V_{\bot}\left(\mathbf{r}_{\bot}\right)+V_{z}\left(z\right)\nonumber \\
 & = & \frac{1}{2}m\omega_{\rho}^{2}\rho^{2}+\frac{1}{2}m\omega_{z}^{2}z^{2},\label{eq:2d-1}
\end{eqnarray}
with $\omega_{z}\gg\omega_{\rho}$. With this condition the condensate
wave-function can be separated as a product of radial and axial functions,
which are entirely independent. This yields a quasi-2D Bose-Einstein
condensate \cite{quasi-2D.collisions,2Dtempvortex,smv3}, and leads
to
\begin{equation}
\Psi\left(\mathbf{r},t\right)=N\Phi\left(\mathbf{r}_{\bot},t\right)W\left(z,t\right),\label{eq:2d-2}
\end{equation}
where
\begin{equation}
W\left(z,t\right)=\frac{1}{d_{z}\sqrt{\pi}}\exp\left(-\frac{z^{2}}{2d_{z}^{2}}-\frac{i\omega_{z}t}{2}\right).\label{eq:2d-3}
\end{equation}
By replacing (\ref{eq:2d-1}) and (\ref{eq:2d-2}) with (\ref{eq:2d-3})
into the Gross-Pitaevskii equation (\ref{eq:GPE}), we obtain 
\begin{eqnarray}
i\hbar W\frac{\partial\Phi}{\partial t}+\frac{\hbar\omega_{z}}{2}\Phi W & = & \left[-\frac{\hbar^{2}}{2m}\nabla_{\bot}^{2}+\frac{\hbar^{2}}{2md_{z}^{2}}-\frac{\hbar^{2}z^{2}}{2md_{z}^{4}}+V\left(\mathbf{r}\right)+NU_{0}\left|\Phi W\right|^{2}\right]\Phi W.\label{eq:2d-4}
\end{eqnarray}
The product $\Phi\left(\mathbf{r}_{\bot},t\right)W\left(z,t\right)$
is normalized to unity, thus the number of atoms appears multiplying
the coupling constant $U_{0}$. Now we can multiply Eq. (\ref{eq:2d-4})
by $W^{*}\left(z,t\right)$ and integrate this equation over the entire
z domain. Since $\hbar^{2}/2md_{z}^{2}=\hbar\omega_{z}/2$, and $\hbar^{2}/2md_{z}^{4}=m\omega_{z}^{2}/2$
we obtain the following simplified equation
\begin{equation}
i\hbar\frac{\partial\Phi\left(\mathbf{r}_{\bot},t\right)}{\partial t}=\left[-\frac{\hbar^{2}}{2m}\nabla_{\bot}^{2}+V_{\bot}\left(\mathbf{r}_{\bot}\right)+NU_{2D}\left|\Phi\left(\mathbf{r}_{\bot},t\right)\right|^{2}\right]\Phi\left(\mathbf{r}_{\bot},t\right),\label{eq:2D-GPE}
\end{equation}
where
\begin{equation}
U_{2D}=\frac{U_{0}}{d_{z}\sqrt{2\pi}}=2\sqrt{2\pi}\frac{\hbar^{2}a_{s}}{md_{z}}.
\end{equation}
Let us then write the Lagrangian density which leads to quasi-2D Gross-Pitaevskii
equation (\ref{eq:2D-GPE}) for a complex field $\Phi\left(\mathbf{r}_{\bot},t\right)$
normalized to unity. So the Lagrangian is given by
\begin{eqnarray}
\mathcal{L}_{2D} & = & -\frac{i\hbar}{2}\left[\Phi^{*}\left(\mathbf{r}_{\bot},t\right)\frac{\partial\Phi\left(\mathbf{r}_{\bot},t\right)}{\partial t}-\Phi\left(\mathbf{r}_{\bot},t\right)\frac{\partial\Phi^{*}\left(\mathbf{r}_{\bot},t\right)}{\partial t}\right]\nonumber \\
 &  & +\frac{\hbar^{2}}{2m}\left|\nabla_{\bot}^{2}\Phi\left(\mathbf{r}_{\bot},t\right)\right|^{2}+V_{\bot}\left(\mathbf{r}_{\bot}\right)\left|\Phi\left(\mathbf{r}_{\bot},t\right)\right|^{2}+\frac{NU_{2D}}{2}\left|\Phi\left(\mathbf{r}_{\bot},t\right)\right|^{4}.\label{eq:dL2D}
\end{eqnarray}

\section{Breaking wave-function symmetry}

\label{sec:breakingS}

In order to examine the coupling between the vortex-core dynamics
and the collective modes as well as their stability, we choose the
situation where a multi-charged vortex is created at the center of
a condensate. Its wave-function can be written in cartesian coordinates
as
\begin{equation}
\Phi_{\ell}\left(\mathbf{r}_{\bot},t\right)\propto\left\{ 1-\frac{1}{\left[x/\xi_{x}\left(t\right)\right]^{2}+2xy/\xi_{xy}\left(t\right)+\left[y/\xi_{y}\left(t\right)\right]^{2}+1}\right\} ^{\ell/2}\sqrt{1-\left[\frac{x}{R_{x}\left(t\right)}\right]^{2}-\left[\frac{y}{R_{y}\left(t\right)}\right]^{2}}e^{iS\left(\mathbf{r}_{\bot},t\right)}.\label{eq:wfA}
\end{equation}
The sizes in each direction are given by $R_{i}\left(t\right)$. They
are known as Thomas-Fermi radii, since the wave-function vanishes
for $x>R_{x}$ and $y>R_{y}$. The vortex-core sizes are given by
$\xi_{i}\left(t\right)$. They are of the order of the healing length
for a singly charged vortex. The parameter $\xi_{xy}\left(t\right)$
is responsible for a complete description of the quadrupole symmetries
between vortex-core and condensate. The wave-function phase $S\left(\mathbf{r},t\right)$
must be carefully chosen within the context of the variational method.
Because the phase must contain the same number of degrees of freedom
as the wave-function amplitude. Since we have one pair of variational
parameters for each direction in the wave-function amplitude ($\xi_{i}$
and $R_{i}$), we also need a pair of variational parameter in the
wave-function phase ($B_{i}$ and $C_{i}$):
\begin{equation}
S\left(\mathbf{r}_{\bot},t\right)=\ell\arctan\left(\frac{y}{x}\right)+B_{x}\left(t\right)\frac{x^{2}}{2}+B_{xy}\left(t\right)xy+B_{y}\left(t\right)\frac{y^{2}}{2}+C_{x}\left(t\right)\frac{x^{4}}{4}+C_{y}\left(t\right)\frac{y^{4}}{4}.\label{eq:wfP}
\end{equation}
Thus $B_{i}\left(t\right)$ and $C_{i}\left(t\right)$ compose the
variations of the condensate velocity field allowing the components
$\xi_{i}\left(t\right)$ and $R_{i}\left(t\right)$ to oscillate with
opposite directions. While $B_{xy}\left(t\right)$ gives us the contribution
of the distortion $\xi_{xy}\left(t\right)$ for velocity field which
changes the axis of the quadrupole oscillation. Note that we are not
using a parameter which yields a scissor motion to the external components
of the condensate, since it has already been shown that such a motion
is not coupled with neither breathing nor quadrupole modes \cite{scissors1,scissor2}.

This choice for our wave-function implies that our vortex-core might
have an elliptical shape. It is enough to destabilize a multi-charged
vortex and allow it to decay splitting itself into several vortices,
each one with unitary charge.

Following the variational method used in Ref. \cite{perez1,perez2,rpteles1,rpteles2},
we substituted (\ref{eq:wfA}) into (\ref{eq:dL2D}), and performed
the integration over the spacial coordinates, $L_{2D}=\int\mathcal{L}_{2D}d\mathbf{r}_{\bot}$.
Although the Lagrangian density (\ref{eq:dL2D}) cannot be analytically
integrated since it does not keep the polar symmetry. One way to proceed
is to introduce small fluctuations around the polar-symmetry solutions
into the wave-function, and to then to make a Taylor expansion. Thus
we can take advantage of the approximate polar symmetry of the vortex-core
while the fluctuations act breaking the vortex-core symmetry. These
calculations are discussed in detail in the next section.

\section{Expanding the Lagrangian around the polar-symmetry solution}

\indent

\label{sec:lagrangian}

Within the Thomas-Fermi approximation the trapping potential shape
determines the condensate dimensions. The wavefunction (\ref{eq:wfA})
is approximated by an inverted parabola except for the central vortex.
So that its integration domain is defined by $1-x^{2}/R_{x}^{2}-y^{2}/R_{y}^{2}\geq0$.
Some care should be taken when calculating the kinetic energy $\left|\nabla_{\bot}\Phi\left(\mathbf{r}_{\bot},t\right)\right|^{2}$
before integrating. The vortex presence inserts an important term
in the gradient, while the rest of the gradient is neglected in the
Thomas-Fermi approximation. That means that the density varies smoothly
along the condensate except in the vortex. 

By introducing deviations from the equilibrium position in our parameters
\begin{eqnarray}
\xi_{j}\left(t\right) & \approx & \xi_{0}+\delta\xi_{j}\left(t\right),\\
R_{j}\left(t\right) & \approx & R_{0}+\delta R_{j}\left(t\right),
\end{eqnarray}
we can expand in Taylor series the deviations of the Lagrangian. In
this way we have
\begin{equation}
L=L^{\left(0\right)}+L^{\left(1\right)}+L^{\left(2\right)}+...\label{eq:L-1}
\end{equation}
The linear approximation is obtained by truncating the series in second
order terms, this leads to:
\begin{itemize}
\item Terms of zeroth order in $L^{\left(0\right)}$ being responsible for
the equilibrium energy per number of atoms.
\item Terms of first order $L^{\left(1\right)}$ that vanish due to the
stationary solution of Euler-Lagrange equations. The equilibrium configuration
has polar symmetry.
\item Terms of second order $L^{\left(2\right)}$ carries the collective
excitations. Their Euler-Lagrange equations result in a eigensystem
whose eigenvectors are composed of deviations ($\delta R_{j}$ and
$\delta\xi_{j}$).
\end{itemize}
Notice that $B_{i}$ and $C_{i}$ from phase (\ref{eq:wfP}) also
must be considered as first order terms since they lead to deviations
in the velocity field. In order to evaluate all the necessary integrals
in Eq.(\ref{eq:L-1}), it is convenient to use $\xi_{i}\left(t\right)/R_{i}\left(t\right)=\alpha_{i}\left(t\right)$
instead of $\xi_{i}\left(t\right)$. This change is explained due
to all these integrals result in functions of $\alpha_{0}=\xi_{0}/R_{0}$.
Thus, a we use $\alpha_{i}\left(t\right)\approx\alpha_{0}+\delta\alpha_{i}\left(t\right)$
instead of $\xi_{i}\left(t\right)\approx\xi_{0}+\delta\xi_{i}\left(t\right)$.
The same happens for $\alpha_{xy}\left(t\right)=R_{x}\left(t\right)R_{y}\left(t\right)/\xi_{xy}\left(t\right)$
where $\alpha_{xy}\left(t\right)\approx\delta\alpha_{xy}\left(t\right)$.
Hereafter we omit the time dependences for simplicity, and we named
zeroth order functions as $A_{i}\equiv A_{i}\left(\ell,\alpha_{0}\right)$
as well as the other integrated results as $I_{i}\equiv I_{i}\left(\ell,\alpha_{0}\right)$.
Such functions are described in Appendix \ref{sec:A}.

The proportionality constant in wave-function (\ref{eq:wfA}) is found
through normalization, being
\begin{eqnarray}
N_{0} & = & R_{x}^{-1}R_{y}^{-1}\left[A_{0}+I_{1}\left(\delta\alpha_{x}+\delta\alpha_{y}\right)+I_{2}\left(\delta\alpha_{x}^{2}+\delta\alpha_{y}^{2}\right)+I_{3}\delta\alpha_{x}\delta\alpha_{y}+I_{4}\delta\alpha_{xy}^{2}\right]^{-1}\nonumber \\
 & \approx & R_{x}^{-1}R_{y}^{-1}A_{0}^{-1}\left[1-\frac{I_{1}}{A_{0}}\left(\delta\alpha_{x}+\delta\alpha_{y}\right)+\left(\frac{I_{1}^{2}}{A_{0}^{2}}-\frac{I_{2}}{A_{0}}\right)\left(\delta\alpha_{x}^{2}+\delta\alpha_{y}^{2}\right)\right.\nonumber \\
 &  & \left.+\left(\frac{2I_{1}^{2}}{A_{0}^{2}}-\frac{I_{3}}{A_{0}}\right)\delta\alpha_{x}\delta\alpha_{y}-\frac{I_{4}}{A_{0}}\delta\alpha_{xy}^{2}\right].
\end{eqnarray}

By calculating the Lagrangian integrals we obtain
\begin{eqnarray}
\int\rho^{2}\left|\Phi\right|^{2}d\mathbf{r}_{\bot} & = & N_{0}R_{x}R_{y}\left\{ R_{x}^{2}\left[A_{1}+I_{5}\left(\delta\alpha_{x}+\frac{1}{3}\delta\alpha_{y}\right)+I_{6}\delta\alpha_{x}^{2}+I_{7}\delta\alpha_{y}^{2}+I_{8}\delta\alpha_{x}\delta\alpha_{y}+I_{9}\delta\alpha_{xy}^{2}\right]\right.\nonumber \\
 &  & \left.R_{y}^{2}\left[A_{1}+I_{5}\left(\frac{1}{3}\delta\alpha_{x}+\delta\alpha_{y}\right)+I_{7}\delta\alpha_{x}^{2}+I_{6}\delta\alpha_{y}^{2}+I_{8}\delta\alpha_{x}\delta\alpha_{y}+I_{9}\delta\alpha_{xy}^{2}\right]\right\} ,\label{eq:Etrap}
\end{eqnarray}
\begin{eqnarray}
-i\int\left[\Phi^{*}\frac{\partial\Phi}{\partial t}-\Phi\frac{\partial\Phi^{*}}{\partial t}\right]d\mathbf{r}_{\bot} & = & N_{0}R_{x}R_{y}\left\{ R_{x}^{2}\dot{B_{x}}\left[A_{1}+I_{5}\left(\delta\alpha_{x}+\frac{1}{3}\delta\alpha_{y}\right)\right]\right.\nonumber \\
 &  & +2\dot{B_{xy}}I_{10}\delta\alpha_{xy}+R_{y}^{2}\dot{B_{y}}\left[A_{1}+I_{5}\left(\frac{1}{3}\delta\alpha_{x}+\delta\alpha_{y}\right)\right]\nonumber \\
 &  & +\frac{1}{2}R_{x}^{4}\dot{C_{x}}\left[A_{2}+I_{11}\left(\delta\alpha_{x}+\frac{1}{5}\delta\alpha_{y}\right)\right]\nonumber \\
 &  & \left.+\frac{1}{2}R_{y}^{4}\dot{C_{y}}\left[A_{2}+I_{11}\left(\frac{1}{5}\delta\alpha_{x}+\delta\alpha_{y}\right)\right]\right\} ,
\end{eqnarray}
\begin{eqnarray}
\int\left|\nabla_{\bot}\Phi\right|^{2}d\mathbf{r}_{\bot} & = & N_{0}R_{x}R_{y}\left\{ A_{1}R_{0}^{2}\left(B_{x}^{2}+2B_{xy}^{2}+B_{y}^{2}\right)+2A_{2}R_{0}^{4}\left(B_{x}C_{x}+B_{y}C_{y}\right)+A_{3}R_{0}^{6}\left(C_{x}^{2}+C_{y}^{2}\right)\right.\nonumber \\
 &  & +\frac{\ell^{2}}{R_{x}^{2}}\left[A_{4}+I_{12}\delta\alpha_{x}+I_{13}\delta\alpha_{y}+I_{14}\delta\alpha_{x}^{2}+I_{15}\delta\alpha_{y}^{2}+I_{16}\delta\alpha_{x}\delta\alpha_{y}+I_{17}\delta\alpha_{xy}^{2}\right]\nonumber \\
 &  & +\frac{\ell^{2}}{R_{y}^{2}}\left[A_{4}+I_{13}\delta\alpha_{x}+I_{12}\delta\alpha_{y}+I_{15}\delta\alpha_{x}^{2}+I_{14}\delta\alpha_{y}^{2}+I_{16}\delta\alpha_{x}\delta\alpha_{y}+I_{17}\delta\alpha_{xy}^{2}\right]\nonumber \\
 &  & +\frac{\ell^{2}}{R_{0}^{2}}\left[A_{5}-\frac{A_{5}}{R_{0}}\left(\delta R_{x}+\delta R_{y}\right)+\frac{A_{5}}{R_{0}^{2}}\left(\delta R_{x}^{2}+\delta R_{y}^{2}+\delta R_{x}\delta R_{y}\right)\right.\nonumber \\
 &  & +\frac{I_{18}}{R_{0}}\left(\delta R_{x}\delta\alpha_{x}+\frac{1}{3}\delta R_{x}\delta\alpha_{y}+\frac{1}{3}\delta R_{y}\delta\alpha_{x}+\delta R_{y}\delta\alpha_{y}\right)\nonumber \\
 &  & \left.\left.-\frac{2}{3}I_{18}\left(\delta\alpha_{x}+\delta\alpha_{y}\right)+I_{19}\left(\delta\alpha_{x}^{2}+\delta\alpha_{y}^{2}\right)+I_{20}\delta\alpha_{x}\delta\alpha_{y}+I_{21}\delta\alpha_{xy}^{2}\right]\right\} ,\label{eq:Ekin}
\end{eqnarray}
and
\begin{equation}
\int\left|\Phi\right|^{4}d\mathbf{r}_{\bot}=N_{0}^{2}R_{x}R_{y}\left[A_{6}+I_{22}\left(\delta\alpha_{x}+\delta\alpha_{y}\right)+I_{23}\left(\delta\alpha_{x}^{2}+\delta\alpha_{y}^{2}\right)+I_{24}\delta\alpha_{x}\delta\alpha_{y}+I_{25}\delta\alpha_{xy}^{2}\right].\label{eq:Eint}
\end{equation}

By scaling according to Table \ref{tab:scale2D}, each of the three
first terms from (\ref{eq:L-1}) are given by
\begin{equation}
L^{\left(0\right)}=A_{0}^{-1}\left[A_{1}r_{\rho0}^{2}+\frac{\ell^{2}}{r_{\rho0}^{2}}\left(A_{4}+\frac{1}{2}A_{5}\right)\frac{\sqrt{2\pi}\gamma A_{6}}{r_{\rho0}^{2}A_{0}}\right],
\end{equation}
\begin{eqnarray}
L^{\left(1\right)} & = & \frac{1}{2}A_{0}^{-1}\left[A_{1}r_{\rho0}^{2}\left(\dot{B_{x}}+\dot{B_{y}}\right)+\frac{1}{2}A_{2}r_{\rho0}^{4}\left(\dot{C_{x}}+\dot{C_{y}}\right)\right.\nonumber \\
 &  & \left.+S_{\rho}\left(\delta R_{x}+\delta R_{y}\right)+S_{\alpha}\left(\delta\alpha_{x}+\delta\alpha_{y}\right)\right],
\end{eqnarray}
\begin{eqnarray}
L^{\left(2\right)} & = & \frac{1}{2}A_{0}^{-1}\left[A_{1}r_{\rho0}^{2}\dot{\beta}_{x}\left(2\frac{\delta r_{x}}{r_{\rho0}}+F_{1}\delta\alpha_{x}+F_{2}\delta\alpha_{y}\right)+A_{1}r_{\rho0}^{2}\dot{\beta}_{y}\left(2\frac{\delta r_{y}}{r_{\rho0}}+F_{2}\delta\alpha_{x}+F_{1}\delta\alpha_{y}\right)\right.\nonumber \\
 &  & +\frac{1}{2}A_{2}r_{\rho0}^{4}\dot{\zeta}_{x}\left(4\frac{\delta r_{x}}{r_{\rho0}}+F_{3}\delta\alpha_{x}+F_{4}\delta\alpha_{y}\right)+\frac{1}{2}A_{2}r_{\rho0}^{4}\dot{\zeta}_{y}\left(4\frac{\delta r_{y}}{r_{\rho0}}+F_{4}\delta\alpha_{x}+F_{3}\delta\alpha_{y}\right)\nonumber \\
 &  & +A_{1}r_{\rho0}^{2}\left(\beta_{x}^{2}+\beta_{y}^{2}\right)+2A_{2}r_{\rho0}^{2}\left(\beta_{x}\zeta_{x}+\beta_{y}\zeta_{y}\right)+A_{3}r_{\rho0}^{6}\left(\zeta_{x}^{2}+\zeta_{y}^{2}\right)+V_{\rho}\left(\delta r_{x}^{2}+\delta r_{y}^{2}\right)\nonumber \\
 &  & +V_{\rho\rho}\delta r_{x}\delta r_{y}+V_{\alpha}\left(\delta\alpha_{x}^{2}+\delta\alpha_{y}^{2}\right)+V_{\alpha\alpha}\delta\alpha_{x}\delta\alpha_{y}+V_{\rho\alpha}\left(\delta r_{x}\delta\alpha_{x}+\delta r_{y}\delta\alpha_{y}\right)\nonumber \\
 &  & \left.+V_{\alpha\rho}\left(\delta r_{x}\delta\alpha_{y}+\delta r_{y}\delta\alpha_{x}\right)+2r_{\rho0}^{2}\left(A_{1}\beta_{xy}^{2}+I_{10}\dot{\beta}_{xy}\delta\alpha_{xy}\right)+V_{xy}\delta\alpha_{xy}^{2}\right],
\end{eqnarray}
where
\begin{equation}
S_{\rho}=2A_{1}r_{\rho0}-\frac{\ell^{2}}{r_{\rho0}^{3}}\left(2A_{4}+A_{5}\right)-\frac{2\sqrt{2\pi}\gamma A_{6}}{r_{\rho0}^{3}A_{0}},
\end{equation}
\begin{equation}
S_{\alpha}=A_{1}r_{\rho0}^{2}\left(F_{1}+F_{2}\right)+\frac{\ell^{2}}{r_{\rho0}^{2}}\left[A_{4}\left(F_{5}+F_{6}\right)-A_{5}\left(F_{7}+F_{8}\right)\right]+\frac{2\sqrt{2\pi}\gamma A_{6}}{r_{\rho0}^{2}A_{0}}F_{9},
\end{equation}
\begin{equation}
V_{\rho}=A_{1}+\frac{\ell^{2}}{r_{\rho0}^{4}}\left(3A_{4}+A_{5}\right)+\frac{2\sqrt{2\pi}\gamma A_{6}}{r_{\rho0}^{4}A_{0}},
\end{equation}
\begin{equation}
V_{\rho\rho}=\frac{\ell^{2}}{r_{\rho0}^{4}}A_{5}+\frac{2\sqrt{2\pi}\gamma A_{6}}{r_{\rho0}^{4}A_{0}},
\end{equation}
\begin{eqnarray}
V_{\alpha} & = & A_{1}r_{\rho0}^{2}\left[\frac{I_{6}+I_{7}}{A_{1}}-2\frac{I_{2}}{A_{0}}-\frac{I_{1}}{A_{0}}\left(F_{1}+F_{2}\right)\right]\nonumber \\
 &  & +\frac{A_{4}\ell}{r_{\rho0}^{2}}^{2}\left[\frac{I_{14}+I_{15}}{A_{4}}-2\frac{I_{2}}{A_{0}}-\frac{I_{1}}{A_{0}}\left(F_{5}+F_{6}\right)\right]\nonumber \\
 &  & +\frac{A_{5}\ell^{2}}{r_{\rho0}^{2}}\left[\frac{I_{19}}{A_{5}}-\frac{I_{2}}{A_{0}}-\frac{I_{1}}{A_{0}}\left(F_{7}+F_{8}\right)\right]\nonumber \\
 &  & +\frac{2\sqrt{2\pi}\gamma A_{6}}{r_{\rho0}^{2}A_{0}}\left[\frac{I_{23}}{A_{6}}-2\frac{I_{2}}{A_{0}}-\frac{I_{1}}{A_{0}}\left(2F_{9}+\frac{I_{1}}{A_{0}}\right)\right],
\end{eqnarray}
\begin{eqnarray}
V_{\alpha} & = & A_{1}r_{\rho0}^{2}\left[\frac{I_{6}+I_{7}}{A_{1}}-2\frac{I_{2}}{A_{0}}-\frac{I_{1}}{A_{0}}\left(F_{1}+F_{2}\right)\right]\nonumber \\
 &  & +\frac{A_{4}\ell}{r_{\rho0}^{2}}^{2}\left[\frac{I_{14}+I_{15}}{A_{4}}-2\frac{I_{2}}{A_{0}}-\frac{I_{1}}{A_{0}}\left(F_{5}+F_{6}\right)\right]\nonumber \\
 &  & +\frac{A_{5}\ell^{2}}{r_{\rho0}^{2}}\left[\frac{I_{19}}{A_{5}}-\frac{I_{2}}{A_{0}}-\frac{I_{1}}{A_{0}}\left(F_{7}+F_{8}\right)\right]\nonumber \\
 &  & +\frac{2\sqrt{2\pi}\gamma A_{6}}{r_{\rho0}^{2}A_{0}}\left[\frac{I_{23}}{A_{6}}-2\frac{I_{2}}{A_{0}}-\frac{I_{1}}{A_{0}}\left(2F_{9}+\frac{I_{1}}{A_{0}}\right)\right],
\end{eqnarray}
\begin{eqnarray}
V_{\alpha\alpha} & = & 2A_{1}r_{\rho0}^{2}\left[\frac{I_{8}}{A_{1}}-\frac{I_{3}}{A_{0}}-\frac{I_{1}}{A_{0}}\left(F_{1}+F_{2}\right)\right]\nonumber \\
 &  & +2\frac{A_{4}\ell}{r_{\rho0}^{2}}^{2}\left[\frac{I_{16}}{A_{4}}-\frac{I_{3}}{A_{0}}-\frac{I_{1}}{A_{0}}\left(F_{5}+F_{6}\right)\right]\nonumber \\
 &  & +\frac{A_{5}\ell^{2}}{r_{\rho0}^{2}}\left[\frac{I_{20}}{A_{5}}-\frac{I_{3}}{A_{0}}+\frac{I_{1}}{A_{0}}\left(F_{7}+F_{8}\right)\right]\nonumber \\
 &  & +\frac{2\sqrt{2\pi}\gamma A_{6}}{r_{\rho0}^{2}A_{0}}\left[\frac{I_{24}}{A_{6}}-2\frac{I_{3}}{A_{0}}-2\frac{I_{1}}{A_{0}}\left(2F_{9}+\frac{I_{1}}{A_{0}}\right)\right],
\end{eqnarray}
\begin{equation}
V_{\rho\alpha}=2A_{1}r_{\rho0}F_{1}-\frac{\ell^{2}}{r_{\rho0}^{3}}\left(2A_{4}F_{5}-A_{5}F_{7}\right)-\frac{2\sqrt{2\pi}\gamma A_{6}}{r_{\rho0}^{3}A_{0}}F_{9},
\end{equation}
\begin{equation}
V_{\alpha\rho}=2A_{1}r_{\rho0}F_{2}-\frac{\ell^{2}}{r_{\rho0}^{3}}\left(2A_{4}F_{6}-A_{5}F_{8}\right)-\frac{2\sqrt{2\pi}\gamma A_{6}}{r_{\rho0}^{3}A_{0}}F_{9},
\end{equation}
\begin{eqnarray}
V_{xy} & = & 2A_{1}r_{\rho0}^{2}\left(\frac{I_{9}}{A_{1}}-\frac{I_{4}}{A_{0}}\right)+2\frac{A_{4}\ell^{2}}{r_{\rho0}^{2}}\left(\frac{I_{17}}{A_{4}}-\frac{I_{4}}{A_{0}}\right)\\
 &  & +\frac{A_{5}\ell^{2}}{r_{\rho0}^{2}}\left(\frac{I_{21}}{A_{5}}-\frac{I_{4}}{A_{0}}\right)+\frac{2\sqrt{2\pi}\gamma A_{6}}{r_{\rho0}^{2}A_{0}}\left(\frac{I_{25}}{A_{6}}-2\frac{I_{4}}{A_{0}}\right),
\end{eqnarray}
 with
\begin{eqnarray}
F_{1} & = & \frac{I_{5}}{A_{1}}-\frac{I_{1}}{A_{0}},\\
F_{2} & = & \frac{I_{5}}{3A_{1}}-\frac{I_{1}}{A_{0}},\\
F_{3} & = & \frac{I_{11}}{A_{2}}-\frac{I_{1}}{A_{0}},\\
F_{4} & = & \frac{I_{11}}{5A_{2}}-\frac{I_{1}}{A_{0}},\\
F_{5} & = & \frac{I_{12}}{A_{4}}-\frac{I_{1}}{A_{0}},\\
F_{6} & = & \frac{I_{13}}{A_{4}}-\frac{I_{1}}{A_{0}},\\
F_{7} & = & \frac{I_{18}}{A_{5}}-\frac{I_{1}}{A_{0}},\\
F_{8} & = & \frac{I_{18}}{3A_{5}}-\frac{I_{1}}{A_{0}},\\
F_{9} & = & \frac{I_{22}}{A_{6}}-2\frac{I_{1}}{A_{0}}.
\end{eqnarray}

The terms proportional to $\ell^{2}/r_{\rho0}^{2}$ and $\ell^{2}/r_{\rho0}^{3}$
are due to the centrifugal energy added by the multi-charged vortex,
and the interaction parameter in dimensionless units is given by $\gamma=Na_{s}/d_{z}$.
In the next section, we discuss the Euler-Lagrange equations for the
deviations that lead to the four collective modes. Where one of them
is dynamically unstable.

\begin{table}
\centering
\begin{tabular}{|c|c|}
\hline 
 & Dimensionless scale\tabularnewline
\hline 
\hline 
$t$ & $\omega_{\rho}^{-1}\tilde{t}$\tabularnewline
\hline 
$\mu$ & $\hbar\omega_{\rho}\tilde{\mu}$\tabularnewline
\hline 
$\Omega$ & $\omega_{\rho}\tilde{\Omega}$\tabularnewline
\hline 
$R_{0}$ & $d_{\rho}r_{\rho0}$\tabularnewline
\hline 
$\delta R_{j}$ & $d_{\rho}\delta r_{j}$\tabularnewline
\hline 
$\xi_{0}$ & $d_{\rho}r_{\xi0}$\tabularnewline
\hline 
$\delta\xi_{j}$ & $d_{\rho}\delta r_{\xi j}$\tabularnewline
\hline 
$B_{j}$ & $d_{\rho}^{-2}\beta_{j}$\tabularnewline
\hline 
$C_{j}$ & $d_{\rho}^{-4}\zeta_{j}$\tabularnewline
\hline 
$\varpi$ & $\omega_{\rho}\tilde{\varpi}$\tabularnewline
\hline 
\end{tabular}\qquad{}%
\begin{tabular}{|c|}
\hline 
Dimensionless parameters\tabularnewline
\hline 
\hline 
$\gamma=Na_{s}/d_{z}$\tabularnewline
\hline 
$\alpha=\xi/R$\tabularnewline
\hline 
\end{tabular}

\protect\caption{Scale table.}

\label{tab:scale2D}
\end{table}

\section{Energy per atoms, collective modes, and instability of a quadrupole
mode}

\indent

\label{sec:collectiveM}

\begin{figure}
\centering
\subfloat[$\ell=2$]{\includegraphics[scale=0.75]{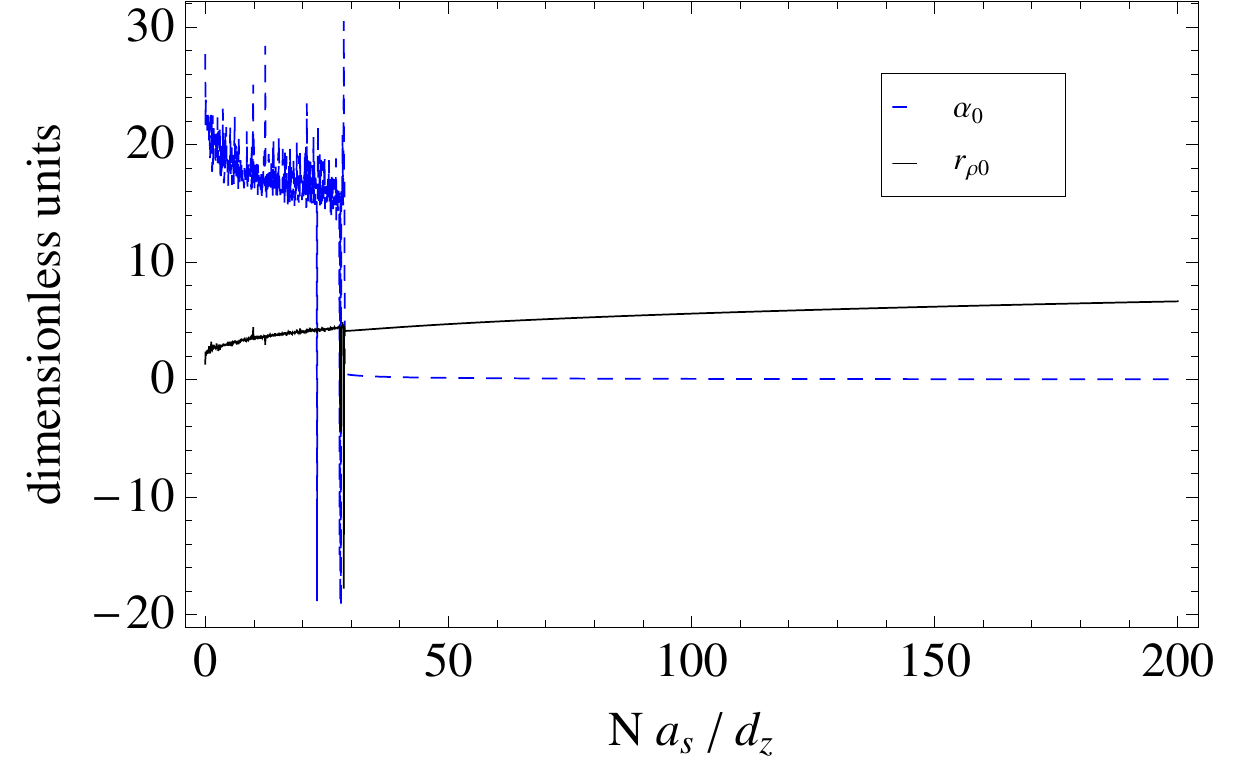}

}

\subfloat[$\ell=4$]{\includegraphics[scale=0.75]{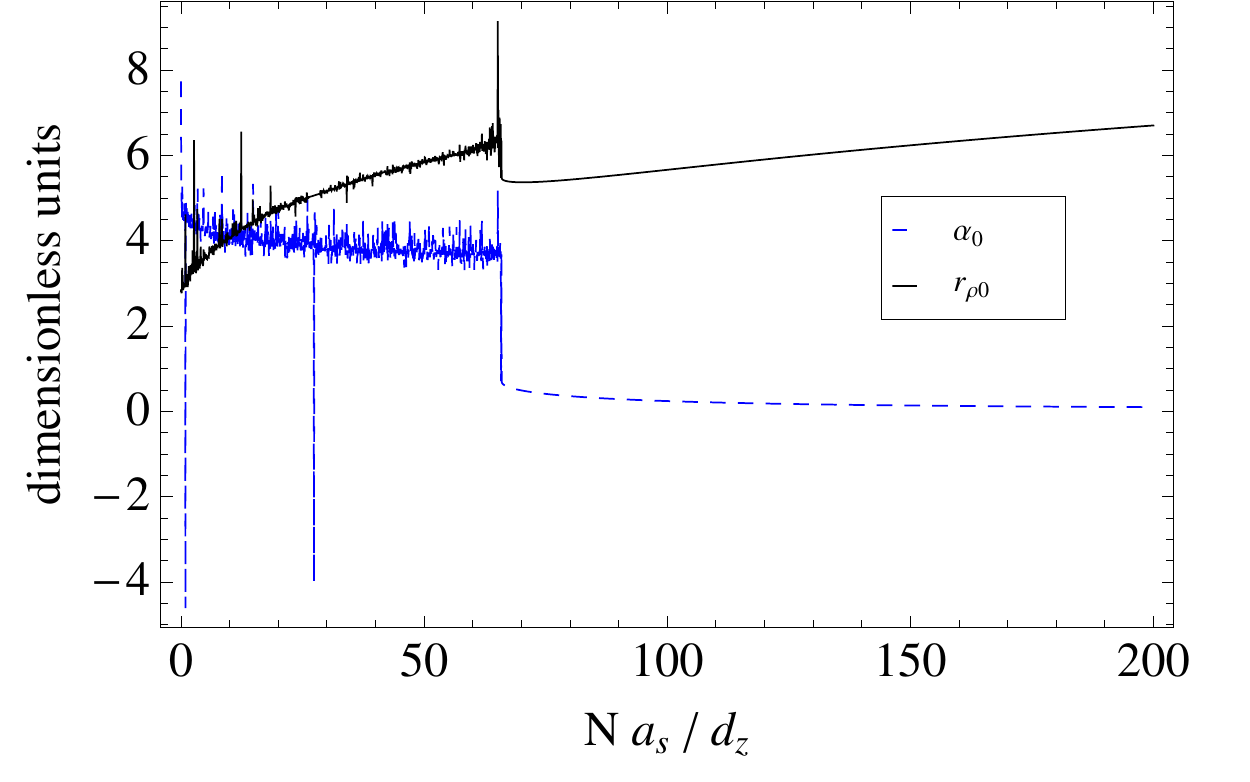}

}

\subfloat[$\ell=7$]{\includegraphics[scale=0.75]{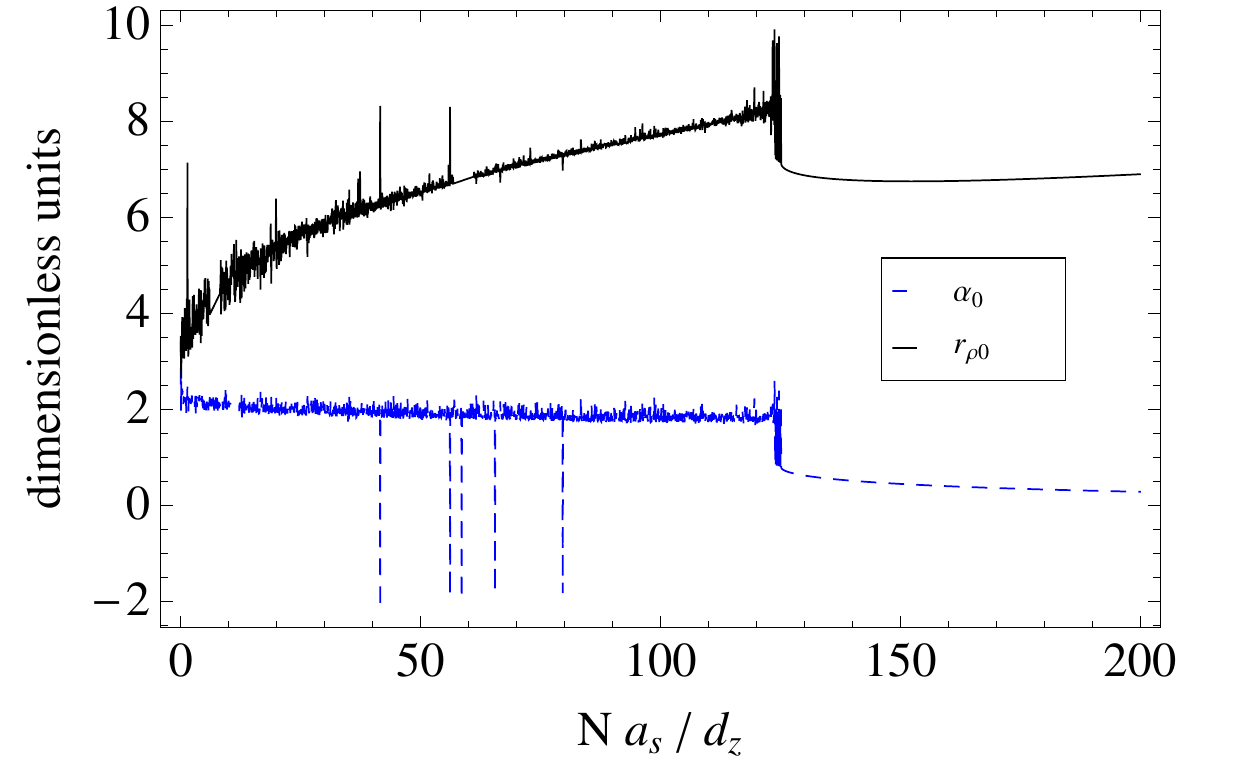}

}

\protect\caption{(Color online) Equilibrium point of parameters ($r_{\rho0}$, and
$\alpha_{0}$) by atomic interaction. Solid (black) line represents
$r_{\rho0}$, and dashed (blue) line represents $\alpha_{0}$. Both
are calculated from (\ref{eq:se}) where $a_{0}$ must be smaller
than $r_{\rho0}$ and near to zero value. This approach shows itself
valid for $Na_{s}/d_{z}>29$ ($Na_{s}/d_{z}>66$, and $Na_{s}/d_{z}>125$)
when we have $\ell=2$ ($\ell=4$ and $\ell=7$).}

\label{fig:equilibrium points}
\end{figure}

First, in order to calculate both energy per atoms and collective
modes, we need to know the equilibrium points $r_{\rho0}$ and $\alpha_{0}$.
They are obtained from Euler-Lagrange equations for $\delta r_{i}$
and $\delta\alpha_{i}$, resulting in 
\begin{equation}
S_{\rho}=0,\text{ and }S_{\alpha}=0.\label{eq:se}
\end{equation}
Thus we have different pairs of $r_{\rho0}$ and $\alpha_{0}$ for
each value of $\ell$ and $\gamma$, which are obtained by applying
Newton's method to solve these coupled stationary equations (\ref{eq:se}).
Note that for $\ell=0$ its solution is trivial, given by 
\begin{equation}
r_{\rho0}=2\left(2/\pi\right)^{1/8}\gamma^{1/4}.
\end{equation}
 These equations (\ref{eq:se}) do not have physically consistent
solutions for low values of $\gamma$ depending on the value of $\ell$,
as can be seen in fig.\ref{fig:equilibrium points}. We have evaluated
the values of the pair $r_{\rho0}$ and $\alpha_{0}$ for the vortex-states
with $\ell=2,4,7$, where the lowest values of interaction are around
$\gamma\equiv Na_{s}/d_{z}=29,76,125$, respectively.

\begin{figure}
\centering

\includegraphics[scale=1.25]{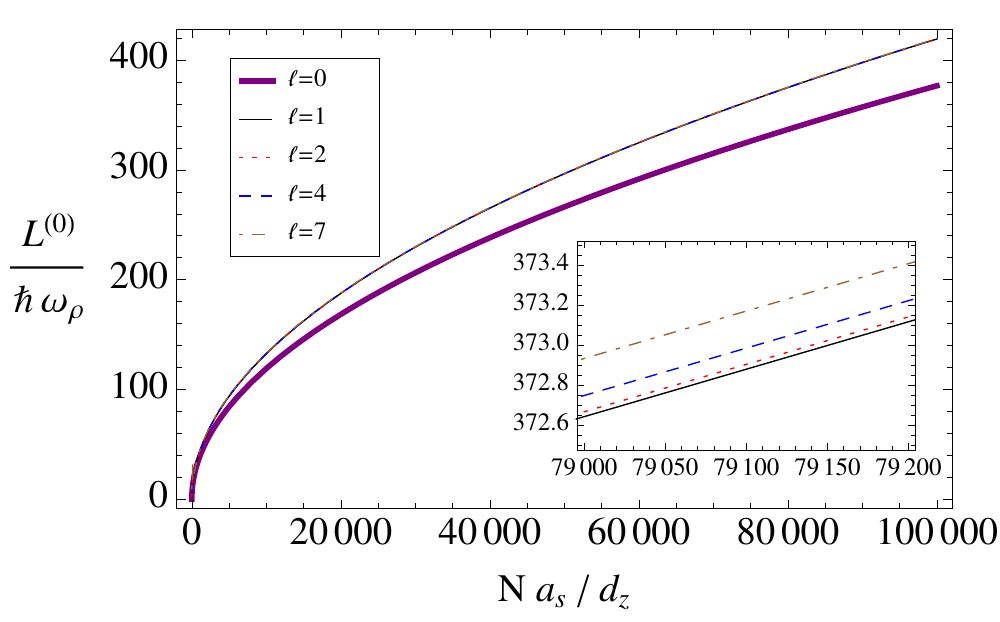}

\protect\caption{(Color online) Energy per atom as function of interaction parameter.}

\label{fig:energia}
\end{figure}

The energy per atom $L^{\left(0\right)}$ increases proportionally
to $\gamma^{1/2}$ being more evident for the vortex-free state ($\ell=0$),
where 
\begin{equation}
L^{\left(0\right)}=4\left(2/\pi\right)^{1/4}\gamma^{1/2}/3.
\end{equation}
 We show this behavior for others values of $\ell$ in fig.\ref{fig:energia}.
The energy gap between the vortex-free state and the remaining states
corresponds to the amount of energy needed to create the $\ell$-charged
vortex states. For instance, if a focused laser beam is used to stir
a Bose-Einstein condensate in order to nucleate vortices, the stirring
frequency must exceeds a critical value \cite{vf1}, which is defined
by difference of energy between the vortex-free state and the singly
vortex state.

\begin{figure}
\centering

\subfloat[Breathing mode ($B_{c}$)]{\includegraphics[scale=0.75]{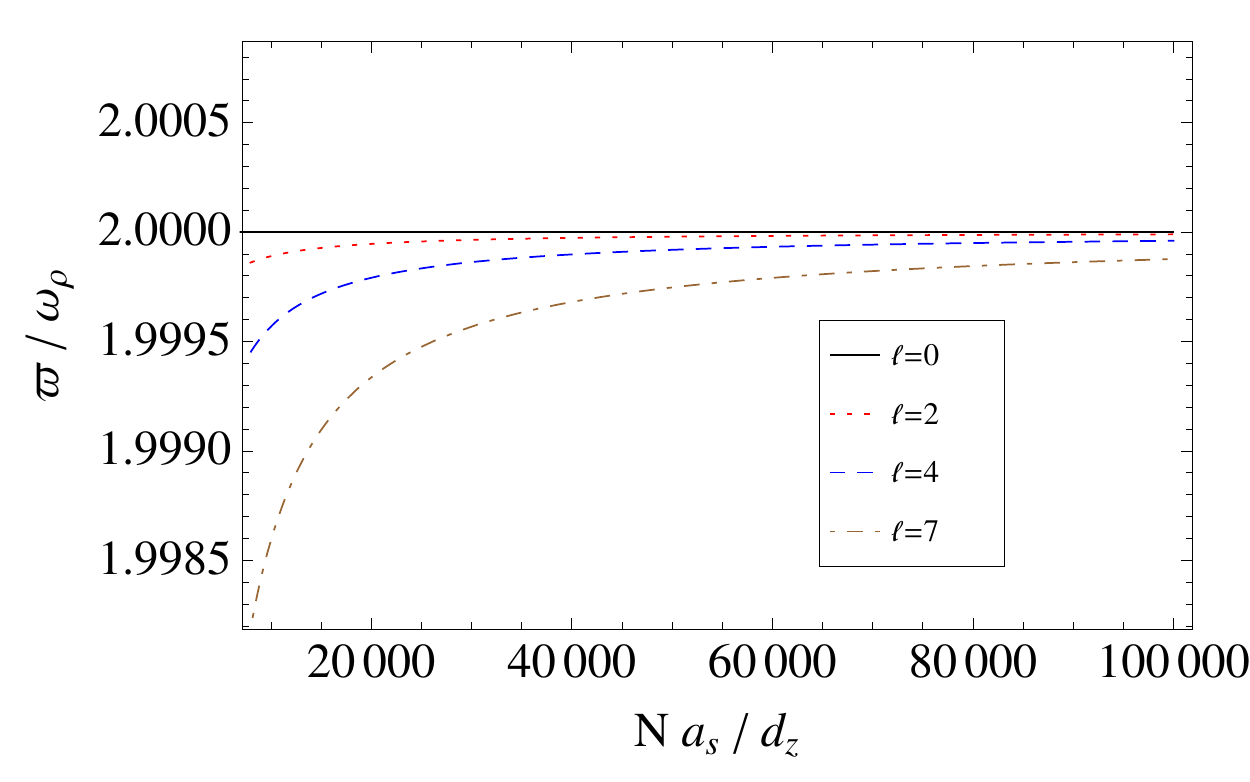}

}\subfloat[Breathing mode ($B_{c}$)]{\includegraphics[scale=0.21]{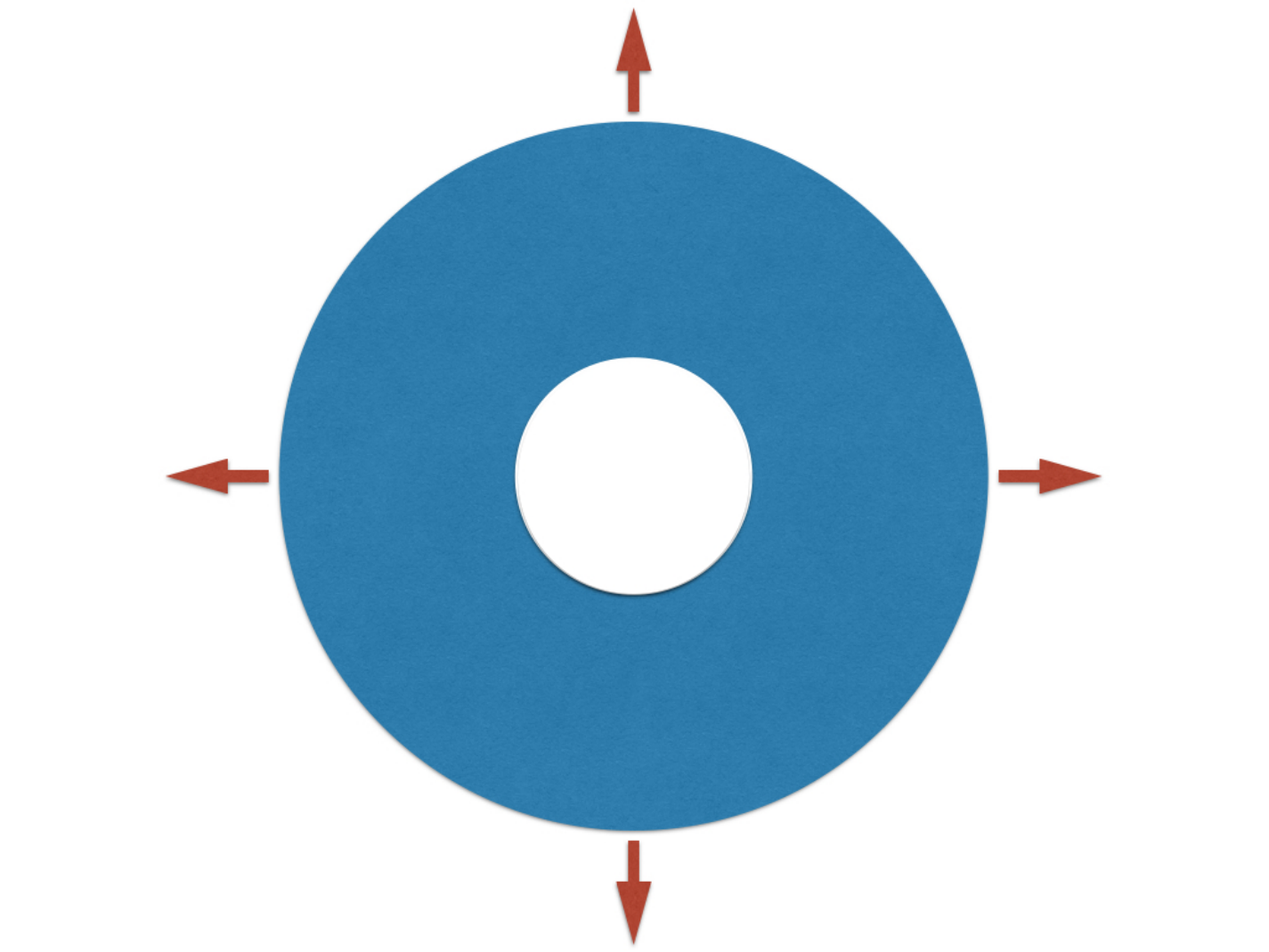}

\label{fig.2DcmBCs}}

\subfloat[Quadrupole mode ($Q_{c}$)]{\includegraphics[scale=0.75]{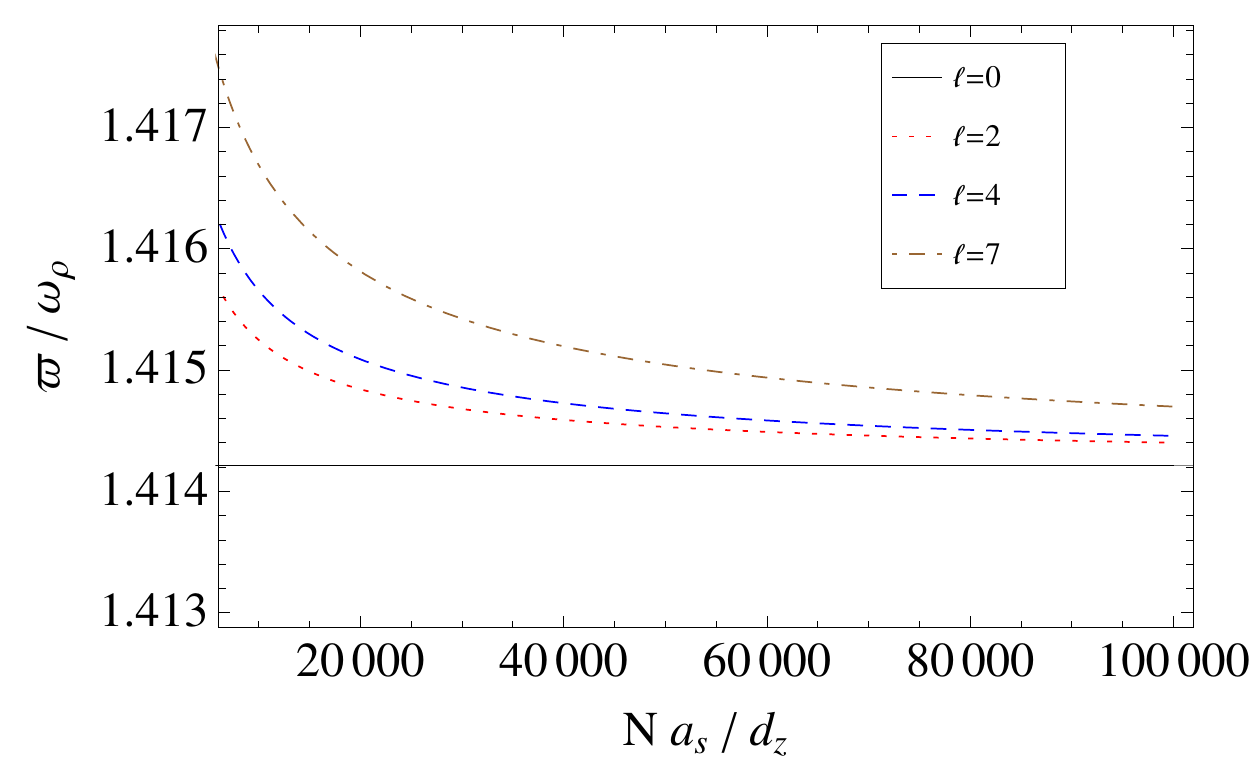}

}\subfloat[Quadrupole mode ($Q_{c}$)]{\includegraphics[scale=0.21]{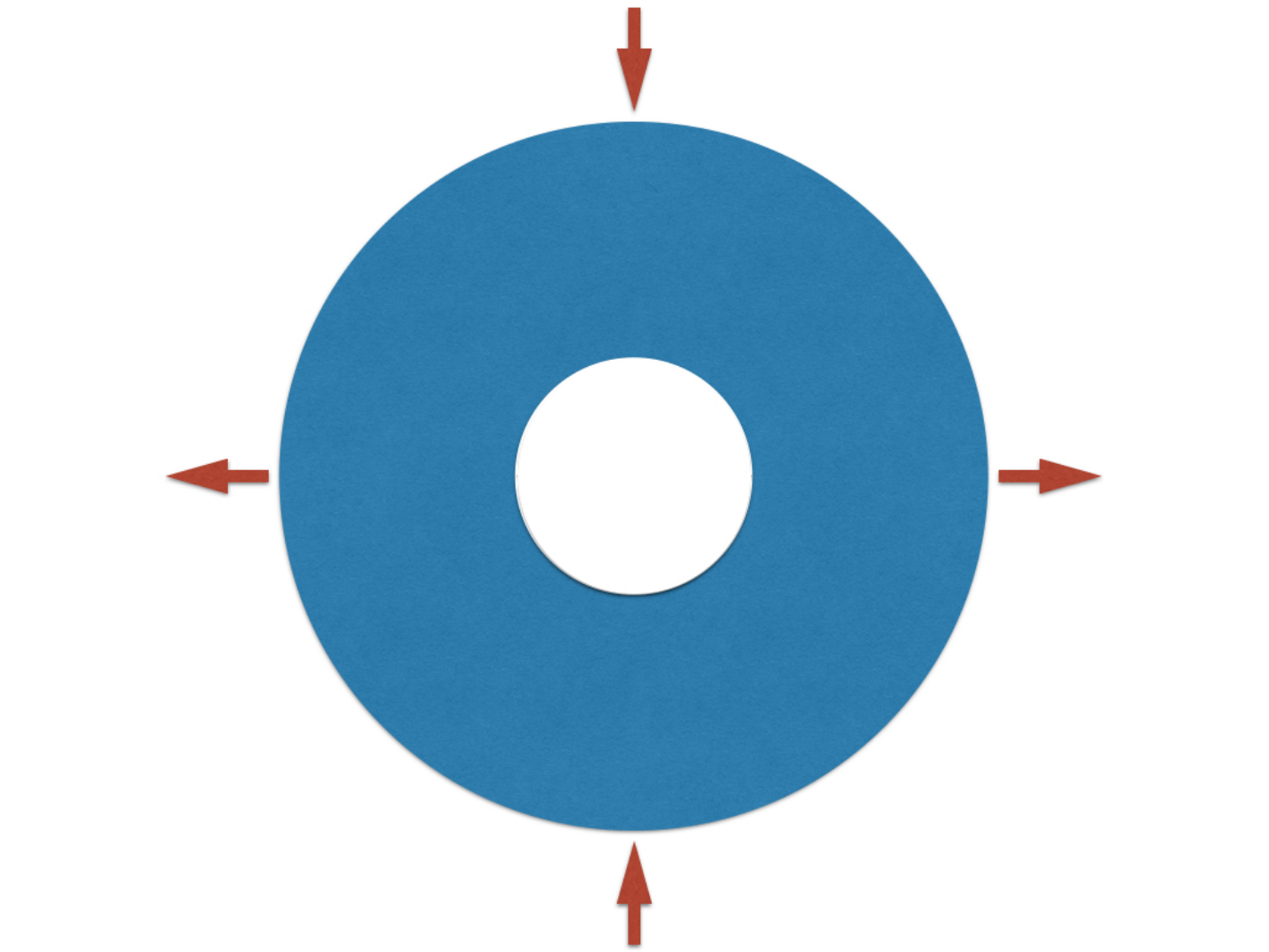}

\label{fig.2DcmQCs}}

\protect\caption{(Color online) Frequency as functions of interaction parameter with
respect to cloud's collective modes in (a) and (c). These two modes
have real frequencies in domain of positive interaction ($Na_{s}/d_{z}>0$).
Schematic representation of each collective mode is in (b) and (d).}

\label{fig:MC}
\end{figure}

\begin{figure}
\centering

\subfloat[Breathing mode ($B_{v}$)]{\includegraphics[scale=0.75]{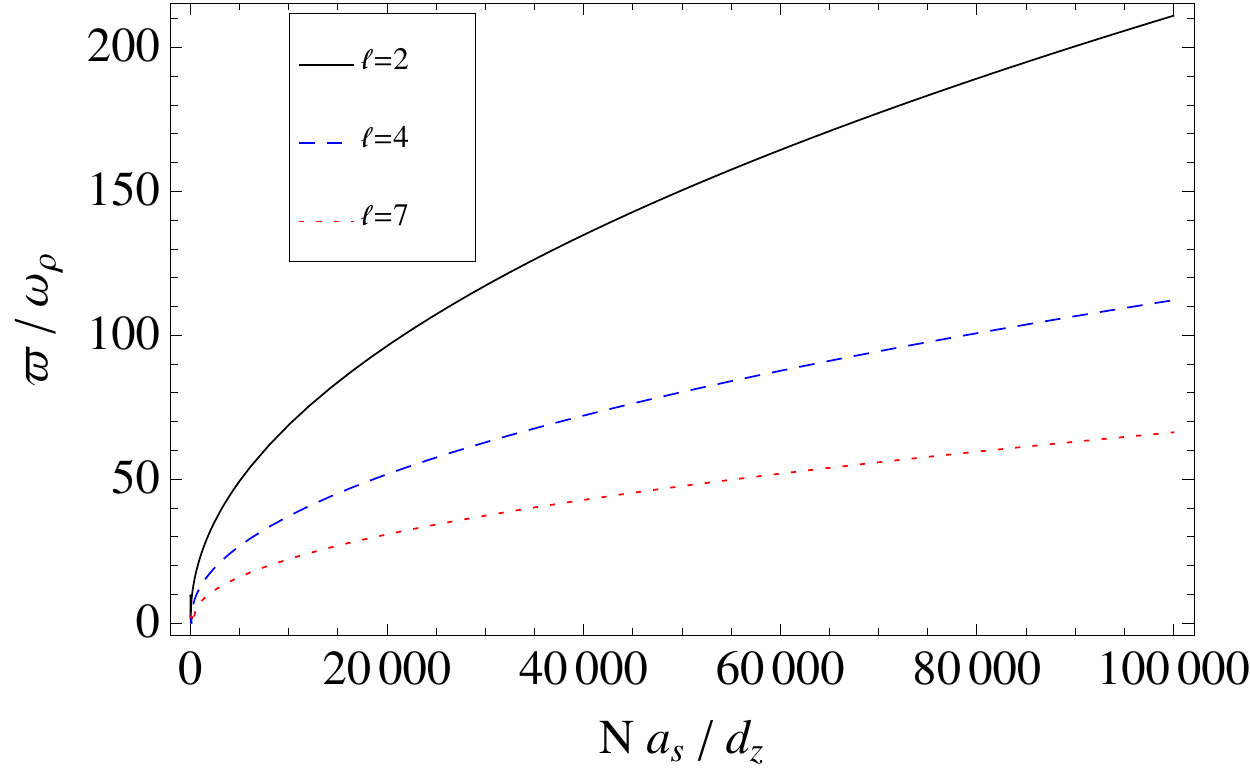}

}\subfloat[Breathing mode ($B_{v}$)]{\includegraphics[scale=0.21]{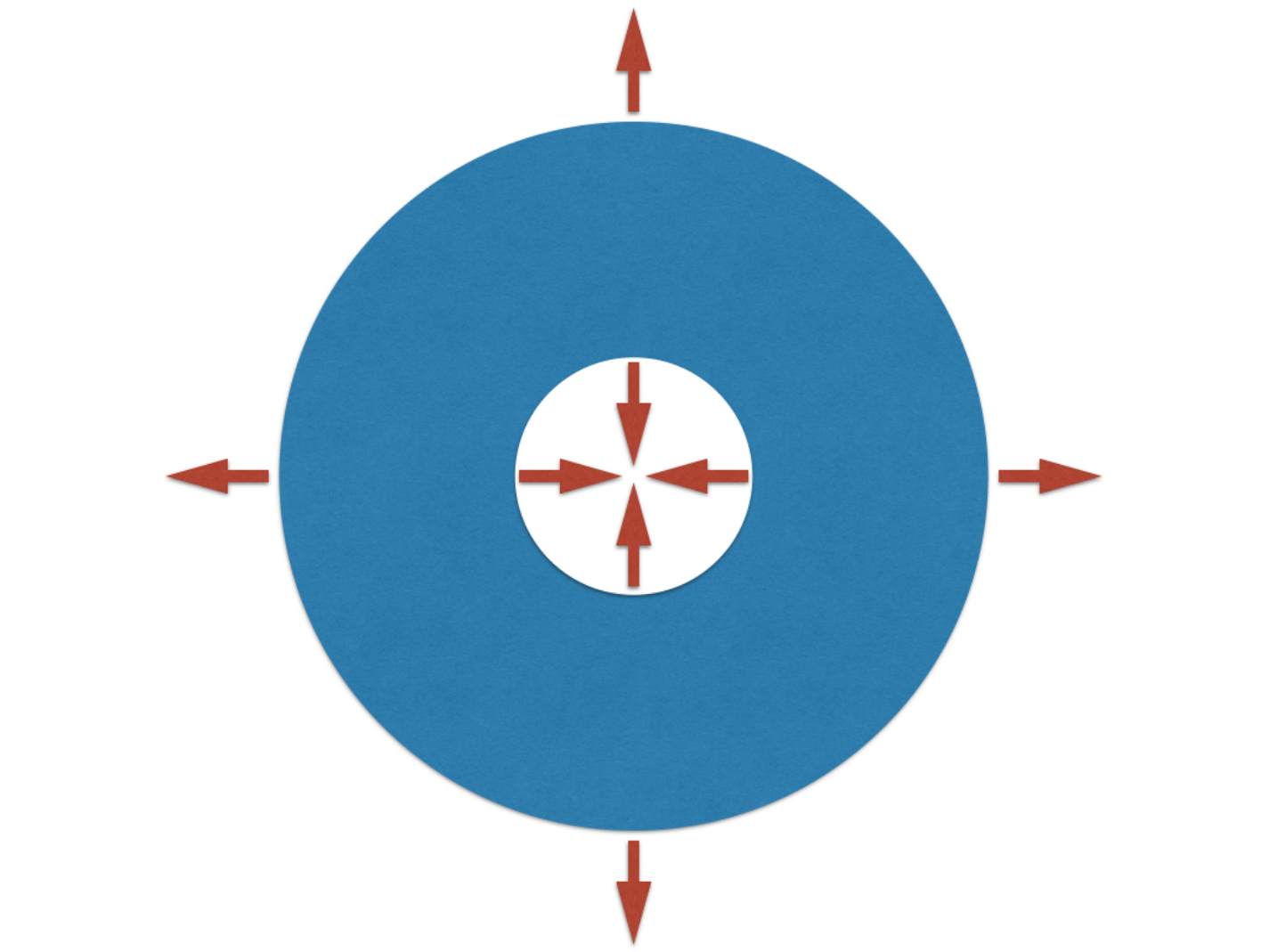}

\label{fig.2DcmBVs}}

\subfloat[Quadrupole mode ($Q_{v}$)]{\includegraphics[scale=0.75]{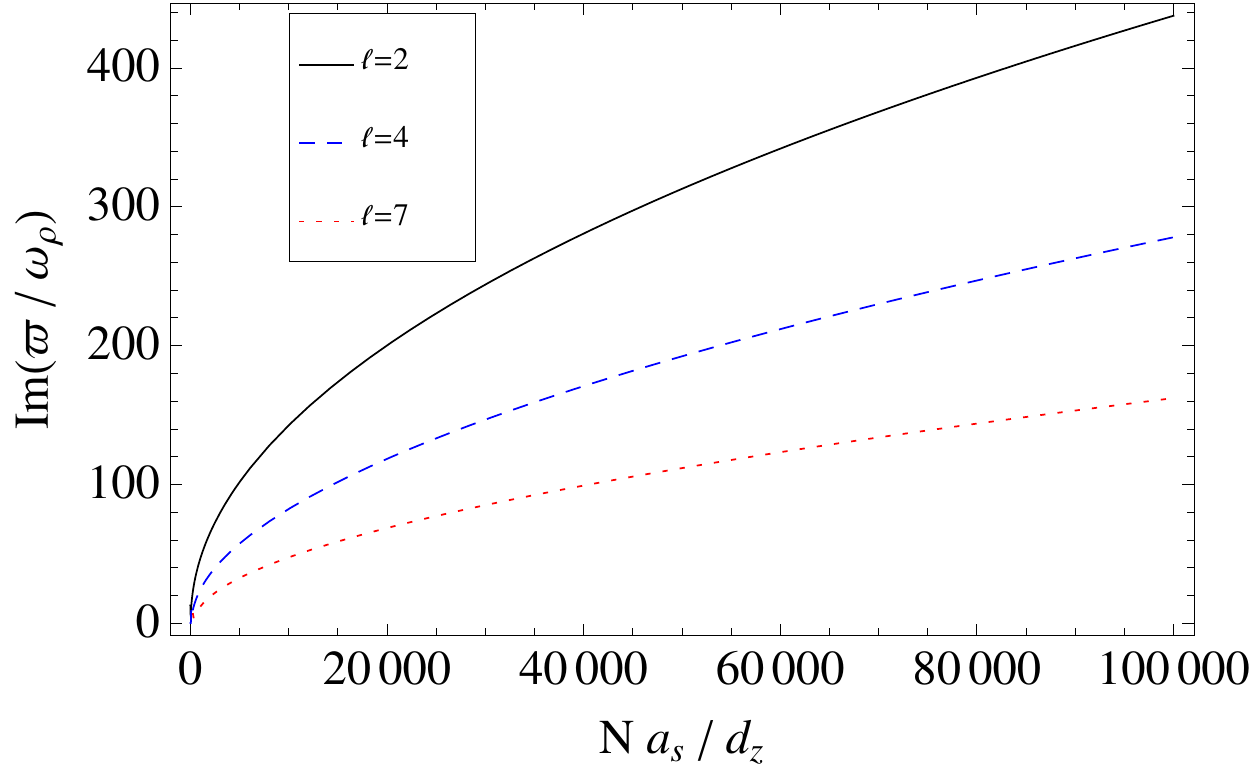}

\label{fig.2DcmQV}}\subfloat[Quadrupole mode ($Q_{v}$)]{\includegraphics[scale=0.21]{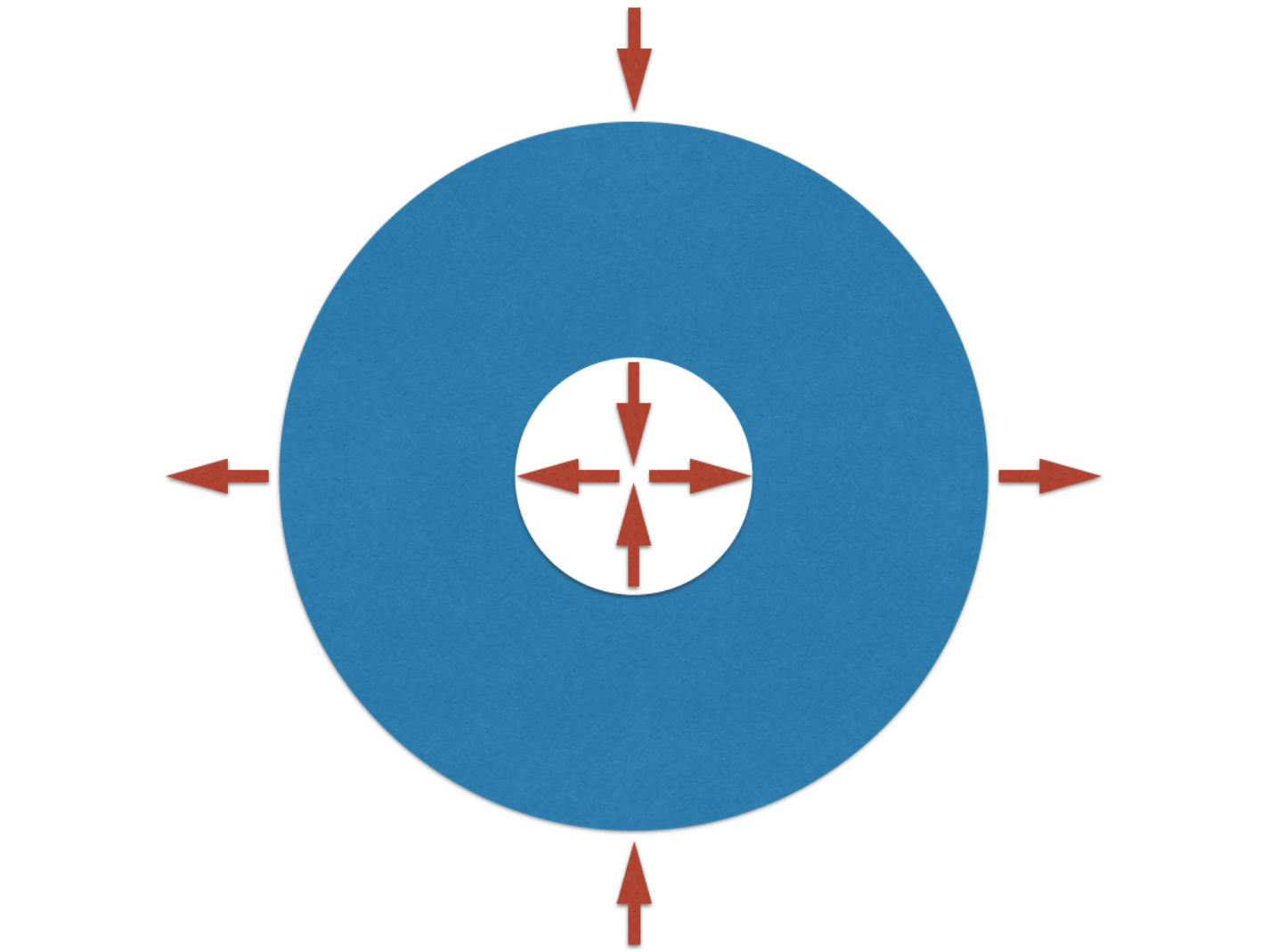}

\label{=00005Bfig.2DcmQVs}}

\protect\caption{(Color online) Frequency as function of interaction parameter for
collective modes coupling the dynamics of the vortex-core with the
oscillation of atomic cloud radii. Only the quadrupole mode (c) is
unstable with imaginary frequency. Schematic representation of collective
modes are shown in (b) and (d). $B_{v}$ mode has the vortex core
oscillating out of phase with cloud radii. $Q_{v}$ mode is a quadrupole
oscillation where vortex core is in phase with cloud radii.}

\label{fig:MV}
\end{figure}

Calculating the Euler-Lagrange equations from $L^{\left(2\right)}$
we obtain ten coupled equations, being five equations for phase
\begin{eqnarray}
\frac{\dot{\delta r_{x}}}{r_{\rho0}}+\frac{F_{1}}{2}\dot{\delta\alpha_{x}}+\frac{F_{2}}{2}\dot{\delta\alpha_{y}} & = & \beta_{x}+\frac{A_{2}}{A_{1}}r_{\rho0}^{2}\zeta_{x},\\
\frac{\dot{\delta r_{y}}}{r_{\rho0}}+\frac{F_{2}}{2}\dot{\delta\alpha_{x}}+\frac{F_{1}}{2}\dot{\delta\alpha_{y}} & = & \beta_{y}+\frac{A_{2}}{A_{1}}r_{\rho0}^{2}\zeta_{y},\\
\frac{\dot{\delta r_{x}}}{r_{\rho0}}+\frac{F_{3}}{4}\dot{\delta\alpha_{x}}+\frac{F_{4}}{4}\dot{\delta\alpha_{y}} & = & \beta_{x}+\frac{A_{3}}{A_{2}}r_{\rho0}^{2}\zeta_{x},\\
\frac{\dot{\delta r_{y}}}{r_{\rho0}}+\frac{F_{4}}{4}\dot{\delta\alpha_{x}}+\frac{F_{3}}{4}\dot{\delta\alpha_{y}} & = & \beta_{y}+\frac{A_{3}}{A_{2}}r_{\rho0}^{2}\zeta_{y},\\
I_{10}\dot{\delta\alpha_{xy}} & = & 2A_{1}\beta_{xy},
\end{eqnarray}
and other five equations for variational parameter in the amplitude
\begin{eqnarray}
A_{1}r_{\rho0}\dot{\beta}_{x}+A_{2}r_{\rho0}^{3}\dot{\zeta}_{x}+2V_{\rho}\delta r_{x}+V_{\rho\rho}\delta r_{y}+V_{\rho\alpha}\delta\alpha_{x}+V_{\alpha\rho}\delta\alpha_{y} & \!\!\!\!=\!\!\!\! & 0,\\
A_{1}r_{\rho0}\dot{\beta}_{y}+A_{2}r_{\rho0}^{3}\dot{\zeta}_{y}+V_{\rho\rho}\delta r_{x}+2V_{\rho}\delta r_{y}+V_{\alpha\rho}\delta\alpha_{x}+V_{\rho\alpha}\delta\alpha_{y} & \!\!\!\!=\!\!\!\! & 0,\\
A_{1}r_{\rho0}^{2}\!\left(\!\dot{\beta}_{x}F_{1}\!+\!\dot{\beta}_{y}F_{2}\!\right)\!+\!\frac{1}{2}A_{2}r_{\rho0}^{4}\!\left(\!\dot{\zeta}_{x}F_{3}\!+\!\dot{\zeta}_{y}F_{4}\!\right)\!+\! V_{\rho\alpha}\delta r_{x}\!+\! V_{\alpha\rho}\delta r_{y}\!+\!2V_{\alpha}\delta\alpha_{x}\!+\! V_{\alpha\alpha}\delta\alpha_{y} & \!\!\!\!=\!\!\!\! & 0,\\
A_{1}r_{\rho0}^{2}\!\left(\!\dot{\beta}_{x}F_{2}\!+\!\dot{\beta}_{y}F_{1}\!\right)\!+\!\frac{1}{2}A_{2}r_{\rho0}^{4}\!\left(\!\dot{\zeta}_{x}F_{4}\!+\!\dot{\zeta}_{y}F_{3}\!\right)\!+\! V_{\alpha\rho}\delta r_{x}\!+\! V_{\rho\alpha}\delta r_{y}\!+\! V_{\alpha\alpha}\delta\alpha_{x}\!+\!2V_{\alpha}\delta\alpha_{y} & \!\!\!\!=\!\!\!\! & 0,\\
r_{\rho0}^{2}I_{10}\dot{\beta}_{xy}+V_{xy}\delta\alpha_{xy} & \!\!\!\!=\!\!\!\! & 0.
\end{eqnarray}
We can reduce these ten equations into 4 coupled equations plus one
uncoupled equation. The equation for $\delta\alpha_{xy}$ is uncoupled
from the others according to 
\begin{equation}
\ddot{\delta\alpha_{xy}}+\frac{2A_{1}V_{xy}}{I_{10}^{2}r_{\rho0}^{2}}\delta\alpha_{xy}=0,
\end{equation}
i.e., the motion represented by the deviation $\delta\alpha_{xy}$
is independent of the other collective modes. Those four equations
lead to the linearized matrix equation
\[
M\ddot{\delta}+V\delta=0,
\]
\begin{equation}
\begin{pmatrix}M_{\rho} & 0 & M_{\rho\alpha} & M_{\alpha\rho}\\
0 & M_{\rho} & M_{\alpha\rho} & M_{\rho\alpha}\\
M_{\rho\alpha} & M_{\alpha\rho} & M_{\alpha} & M_{\alpha\alpha}\\
M_{\alpha\rho} & M_{\rho\alpha} & M_{\alpha\alpha} & M_{\alpha}
\end{pmatrix}\begin{pmatrix}\ddot{\delta r_{x}}\\
\ddot{\delta r_{y}}\\
\ddot{\delta\alpha_{x}}\\
\ddot{\delta\alpha_{y}}
\end{pmatrix}+\begin{pmatrix}2V_{\rho} & V_{\rho\rho} & V_{\rho\alpha} & V_{\alpha\rho}\\
V_{\rho\rho} & 2V_{\rho} & V_{\alpha\rho} & V_{\rho\alpha}\\
V_{\rho\alpha} & V_{\alpha\rho} & 2V_{\alpha} & V_{\alpha\alpha}\\
V_{\alpha\rho} & V_{\rho\alpha} & V_{\alpha\alpha} & 2V_{\alpha}
\end{pmatrix}\begin{pmatrix}\delta r_{x}\\
\delta r_{y}\\
\delta\alpha_{x}\\
\delta\alpha_{y}
\end{pmatrix}=0,\label{eq:eigensystem}
\end{equation}
where the entries in the matrix $M$ are given by
\begin{eqnarray}
M_{\rho} & = & 2A_{1},\\
M_{\alpha} & = & \frac{A_{1}A_{2}^{2}r_{\rho0}^{2}}{2\left(A_{2}^{2}-A_{1}A_{3}\right)}\left[F_{1}F_{3}+F_{2}F_{4}-\frac{F_{3}^{2}}{4}-\frac{F_{4}^{2}}{4}-\frac{A_{1}A_{3}}{A_{2}^{2}}\left(F_{1}^{2}+F_{2}^{2}\right)\right],\\
M_{\alpha\alpha} & = & \frac{A_{1}A_{2}^{2}r_{\rho0}^{2}}{2\left(A_{2}^{2}-A_{1}A_{3}\right)}\left[F_{1}F_{4}+F_{2}F_{3}-\frac{F_{3}F_{4}}{2}-\frac{2A_{1}A_{3}}{A_{2}^{2}}F_{1}F_{2}\right],\\
M_{\rho\alpha} & = & A_{1}F_{1}r_{\rho0},\\
M_{\alpha\rho} & = & A_{1}F_{2}r_{\rho0}.
\end{eqnarray}
 Matrix $V$ results from the energy part of the Lagrangian, i.e.
from Eqs. (\ref{eq:Etrap}), (\ref{eq:Ekin}), and (\ref{eq:Eint}).
This determinant may be either positive or negative reflecting the
system stability. In the other hand, the determinant of $M$ cannot
be negative or zero, since it results from our choice for the wave-function
phase. The equation (\ref{eq:eigensystem}) seems the Newton's equation
therefore we can say that matrix $M$ has an effect of mass-like,
and matrix $V$ works as a potential \cite{peristalticmode}. Solving
the characteristic equation, 
\begin{equation}
\det\left(M^{-1}V-\varpi^{2}I\right)=0,\label{eq:ce}
\end{equation}
results in the frequencies of the collective modes of oscillation.
Eq.(\ref{eq:ce}) is a quartic equation of $\varpi^{2}$. This means
that we have four pairs of frequencies $\pm\varpi_{n}^{2}$ being
one pair for each oscillatory mode. Among these four modes, two of
them have a static vortex representing the collective modes for cloud:
they are the breathing mode $B_{c}$, and the quadrupole mode $Q_{c}$.
In other words, these modes are similar to collective oscillations
of the vortex-free state, where the difference is in a small shift
in their frequencies depending on the charge of the vortex, as it
is shown in Fig.\ref{fig:MC}. Therefore $B_{c}$ decreases the frequency
value while $Q_{c}$ has the opposite effect shifting to higher frequency
value. Note that for a vortex-free condensate $\ell=0$, Eq.(\ref{eq:ce})
is a quadratic equation in $\varpi^{2}$. That means the system presents
only two modes ($B_{c}$ with $\varpi=2$, and $Q_{c}$ with $\varpi=\sqrt{2}$)
in absence of vortex, whose frequencies are constant with respect
to the interaction parameter $\gamma$. There are still other two
modes which couple vortex dynamics with collective modes. They are
another breathing mode $B_{v}$ and another quadrupole mode $Q_{v}$.
In this breathing mode $B_{v}$, the vortex-core sizes oscillate out
of phase with cloud radii, while $\delta r_{x}$ ($\delta\alpha_{x}$)
and $\delta r_{y}$ ($\delta\alpha_{y}$) are in phase. In the quadrupole
mode $Q_{v}$, both these sizes $\delta\alpha_{i}$ and $\delta R_{i}$
are oscillating in phase while $\delta r_{x}$ ($\delta\alpha_{x}$)
and $\delta r_{y}$ ($\delta\alpha_{y}$) have a $\pi$-phase difference
between their oscillations. These modes are sketched in fig.\ref{fig:MV}.
The second quadrupole mode $Q_{v}$ has an imaginary frequency (fig.\ref{fig.2DcmQV}),
i.e. $Q_{v}$-mode is one possible channel to a multi-charged vortex
decay into unitary vortices. Therefore, the multi-charged vortex decay
can be explained by the appearance and growth of this unstable quadrupole
mode due to quantum or thermal fluctuations. These fluctuations work
inducing collective modes, which are coupled to the vortex dynamics
through their sound waves.

This model is completely consistent with CE Bogoliubov modes for $\ell=2$,
which are composed by only the CE mode associated to two-fold symmetry
being our quadrupole mode $Q_{v}$ \cite{stab03}. However when $\ell>2$
this calculation is incomplete since we considered only breathing
and quadrupole modes. Hence for a complete description it is necessary
to add others symmetries for each higher order of $\ell$, which is
not a trivial task. Because the Ansatz requires more degrees of freedom,
that means we should increase the number of variational parameters.

\begin{figure}
\centering

\subfloat[$\omega_{\rho}t=0$]{\includegraphics[scale=0.5]{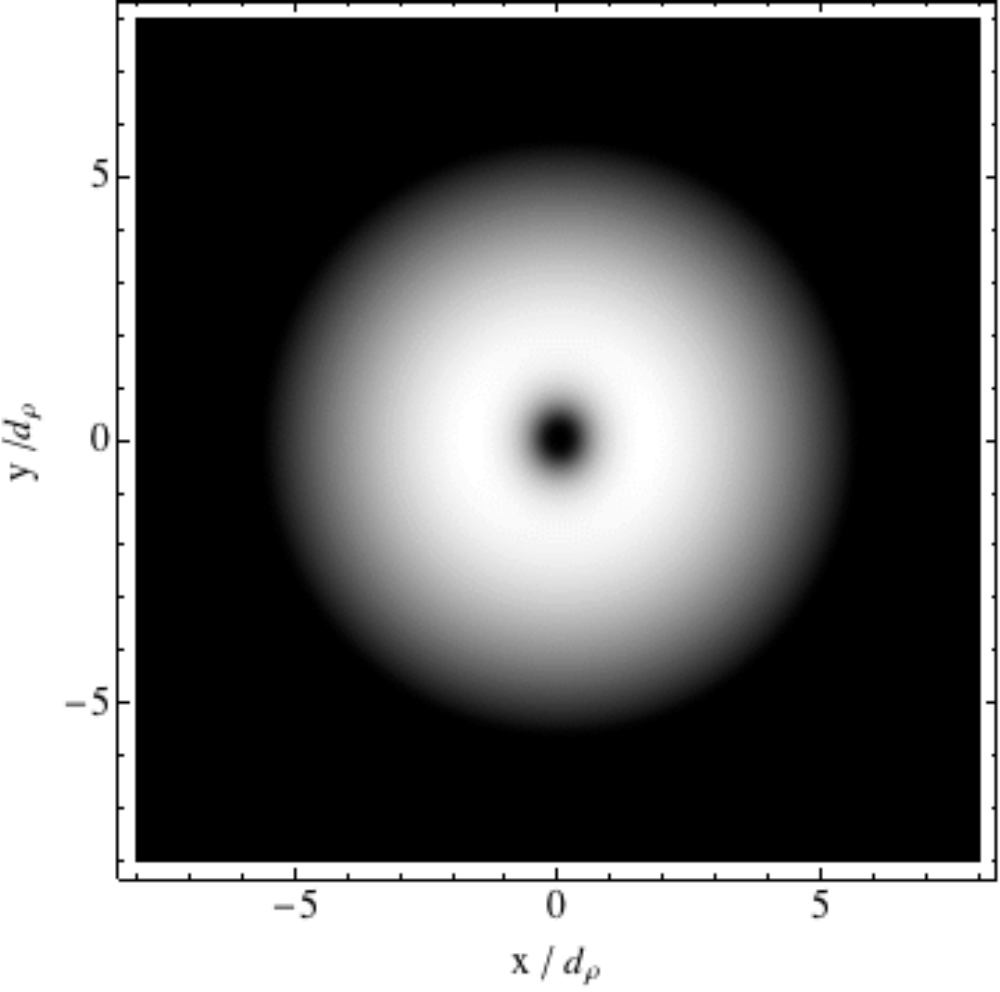}

}\subfloat[$\omega_{\rho}t=0$]{\includegraphics[scale=0.5]{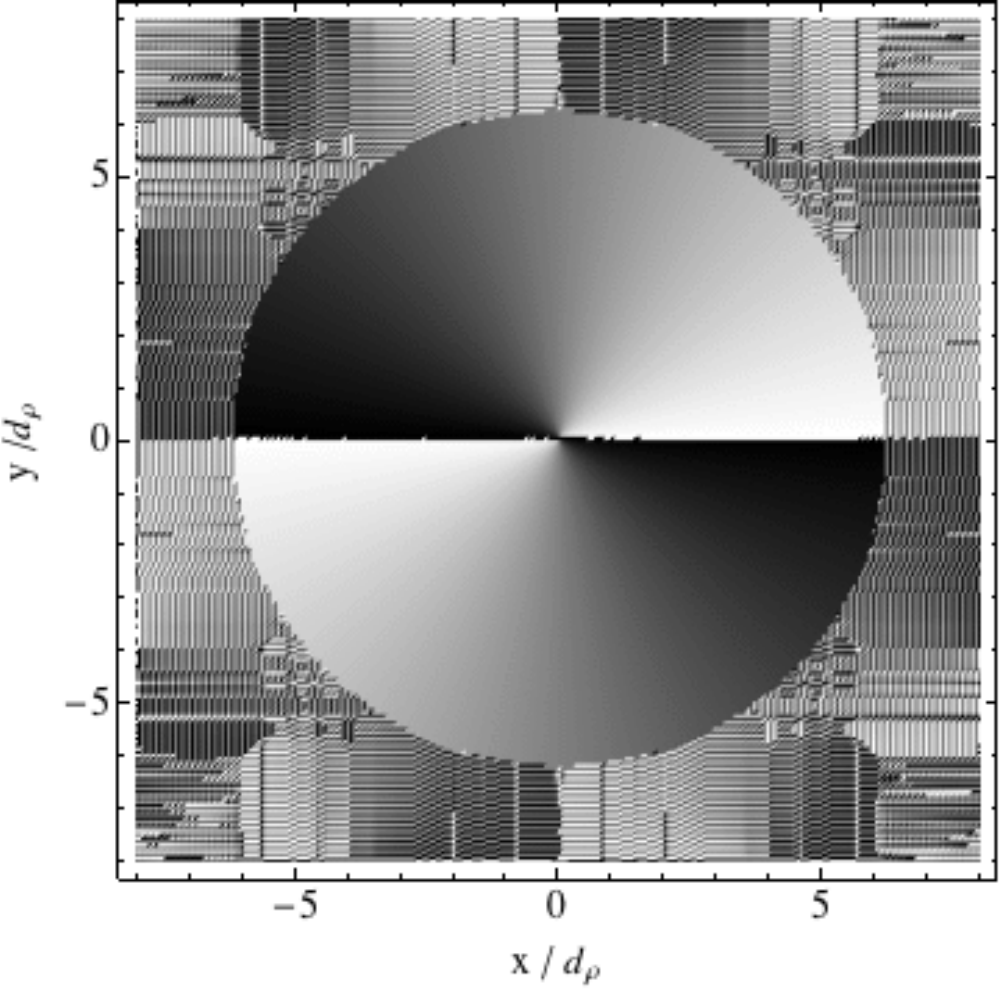}

}

\subfloat[$\omega_{\rho}t=20.2$]{\includegraphics[scale=0.5]{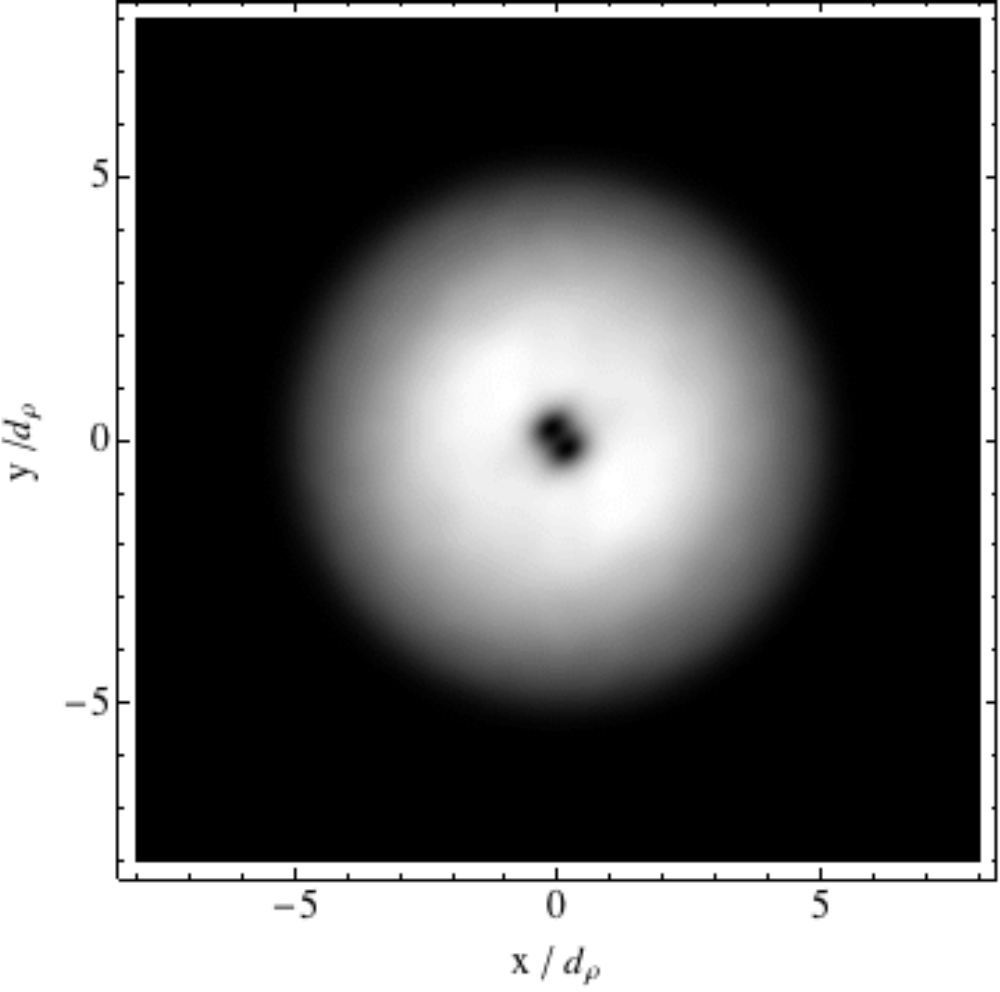}

}\subfloat[$\omega_{\rho}t=20.2$]{\includegraphics[scale=0.5]{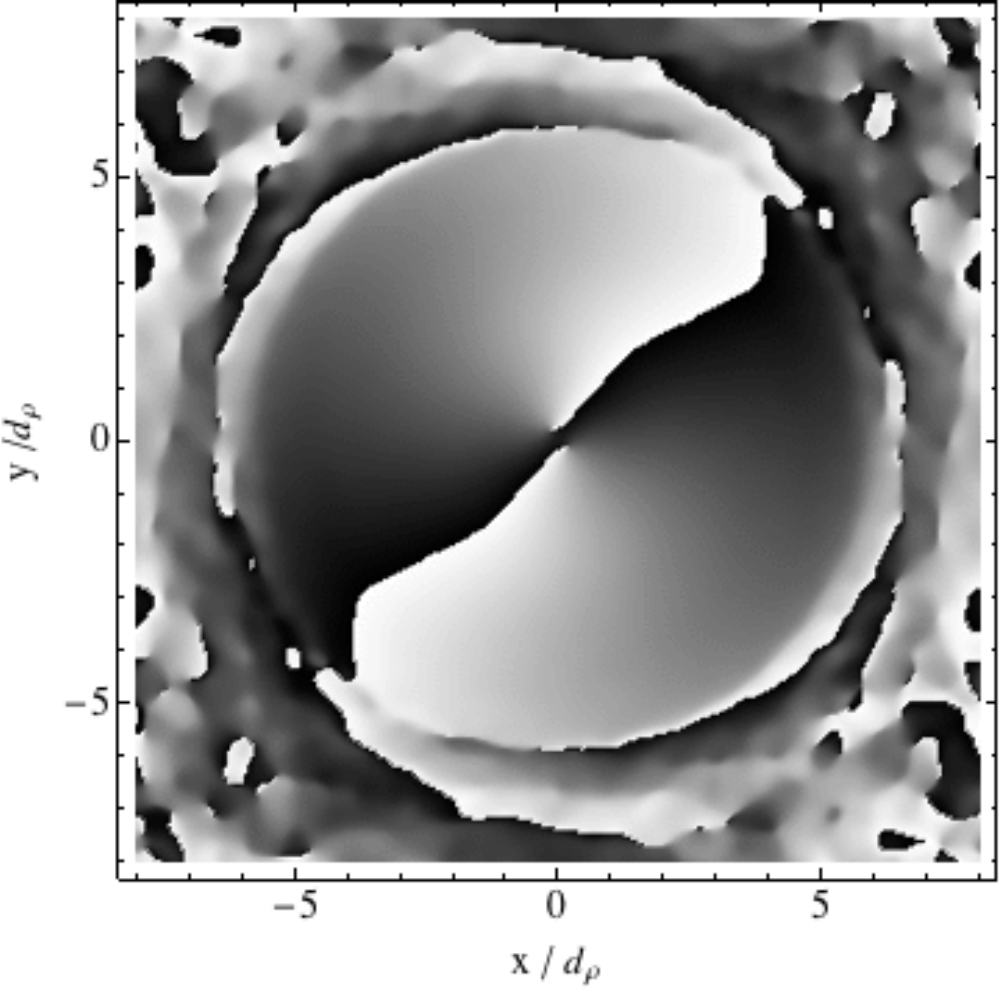}

}

\subfloat[$\omega_{\rho}t=100$]{\includegraphics[scale=0.5]{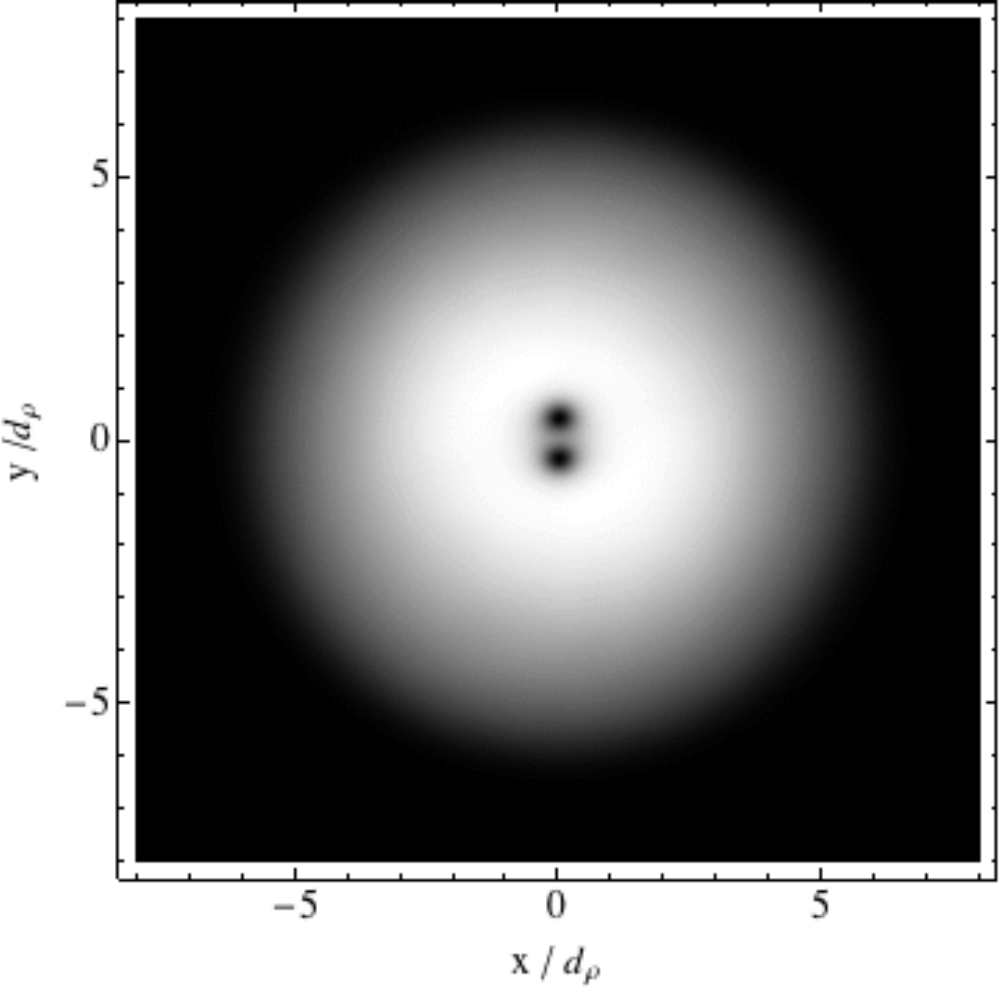}

}\subfloat[$\omega_{\rho}t=100$]{\includegraphics[scale=0.5]{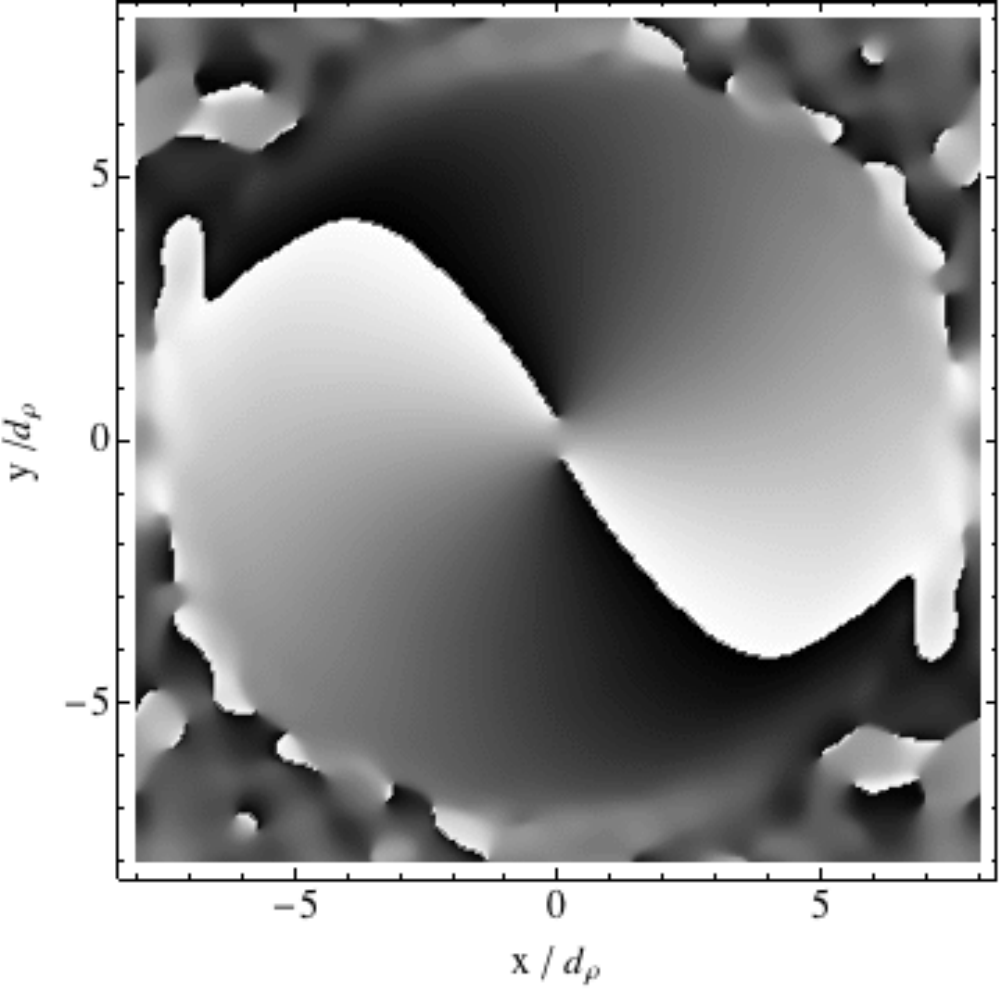}

}

\protect\caption{(Color online) Time evolution of the density (a,b,c) and phase (d,
e, f) of condensate with a doubly-charged vortex. We have used $\tilde{\mu}=20.198$,
$Na_{s}/d_{z}=100$, $\epsilon=0.001$, and a factor of $0.01$ multiplying
of the amplitude of deviations.}

\label{fig:2vD}
\end{figure}

\begin{figure}
\centering

\subfloat[$\omega_{\rho}t=0$]{\includegraphics[scale=0.5]{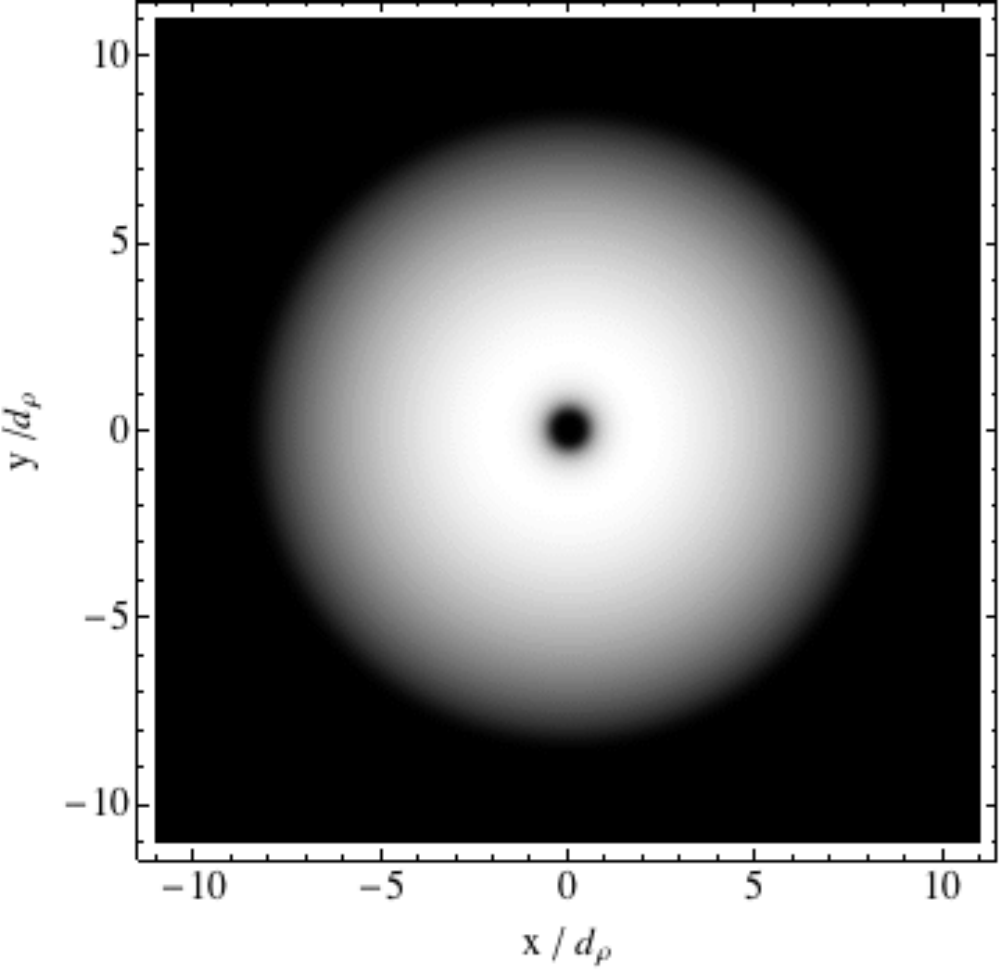}

}\subfloat[$\omega_{\rho}t=0$]{\includegraphics[scale=0.5]{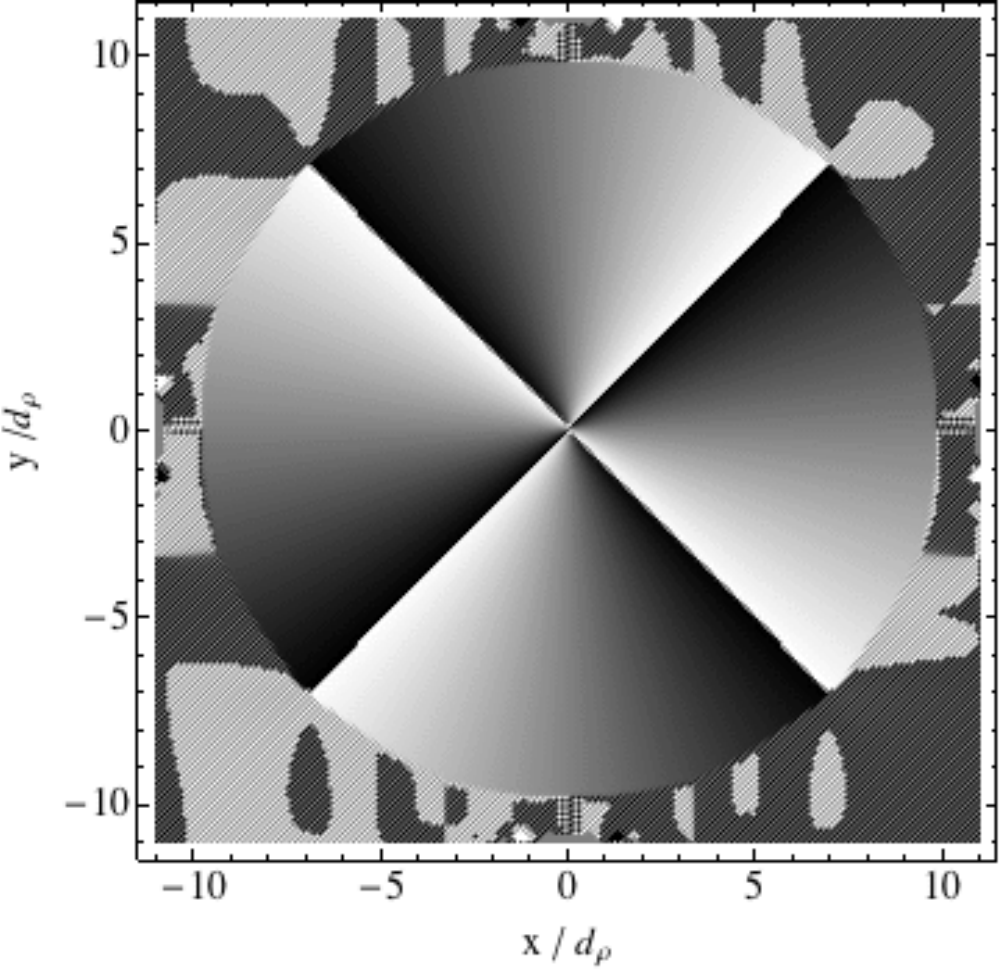}}

\subfloat[$\omega_{\rho}t=22$]{\includegraphics[scale=0.5]{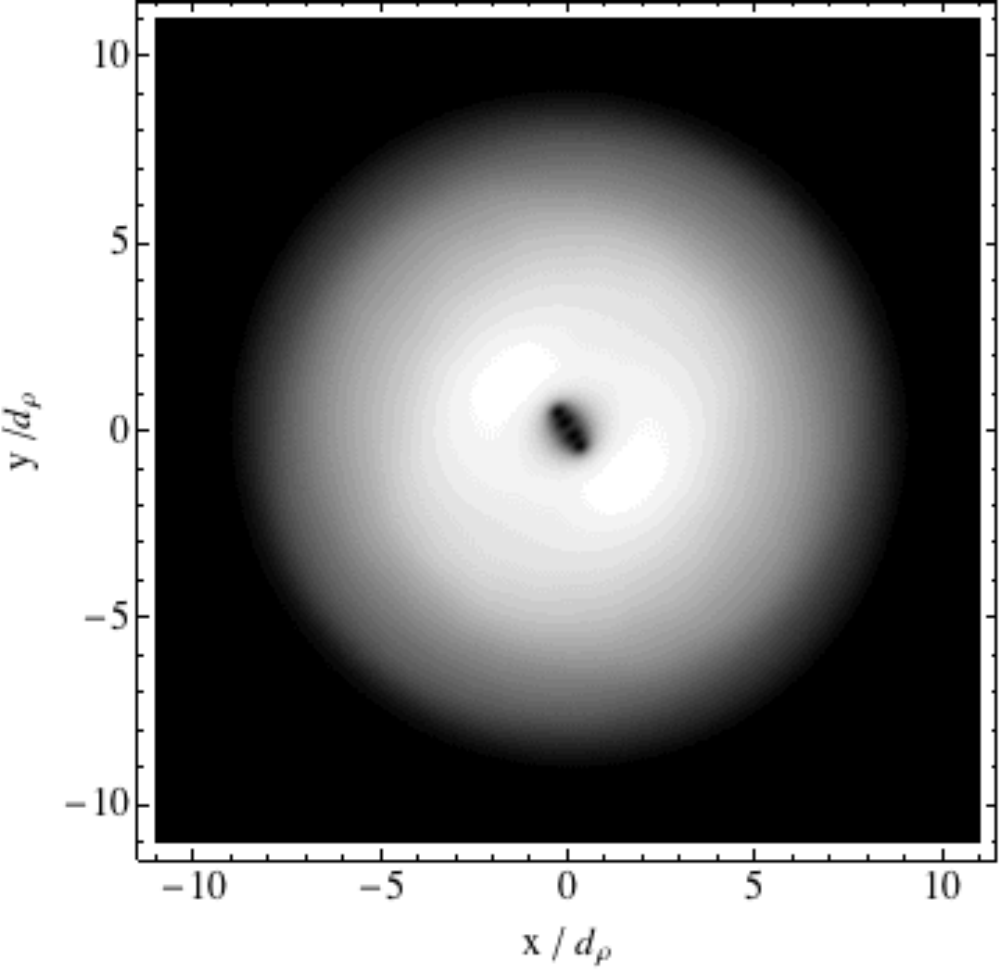}

}\subfloat[$\omega_{\rho}t=22$]{\includegraphics[scale=0.5]{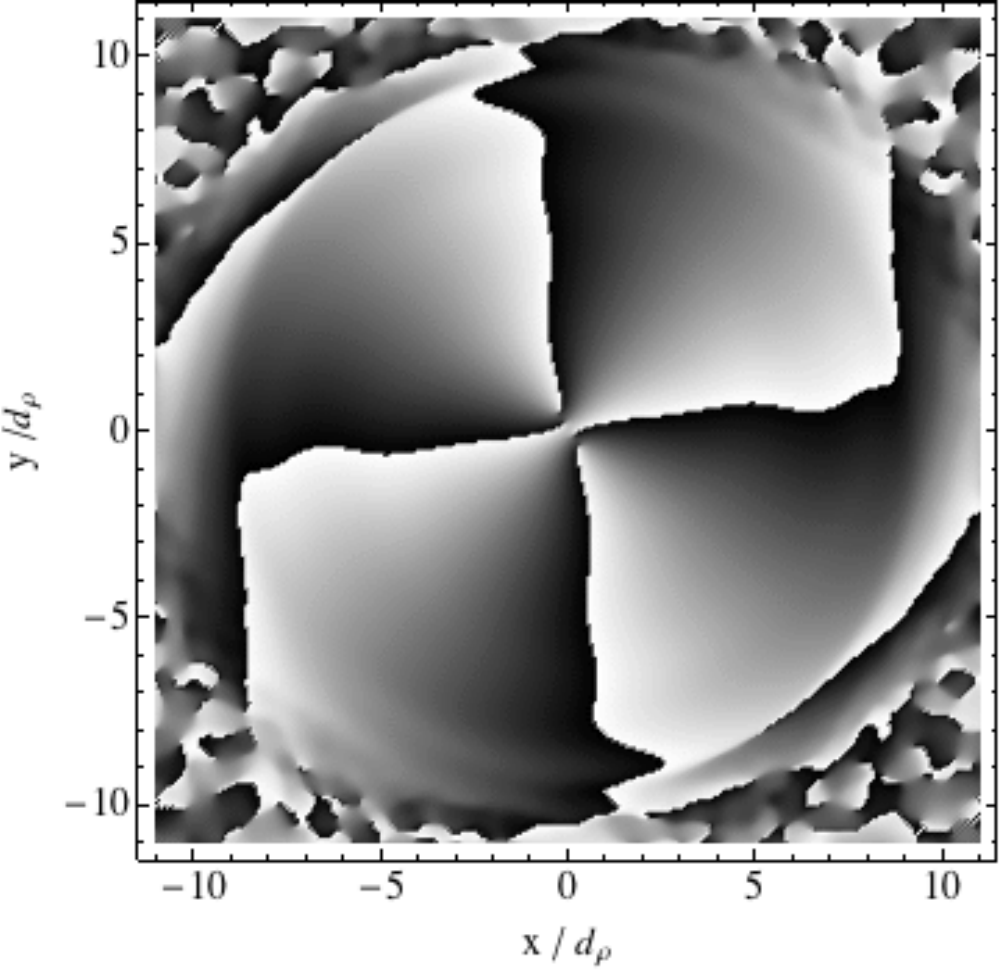}

}

\subfloat[$\omega_{\rho}t=100$]{\includegraphics[scale=0.5]{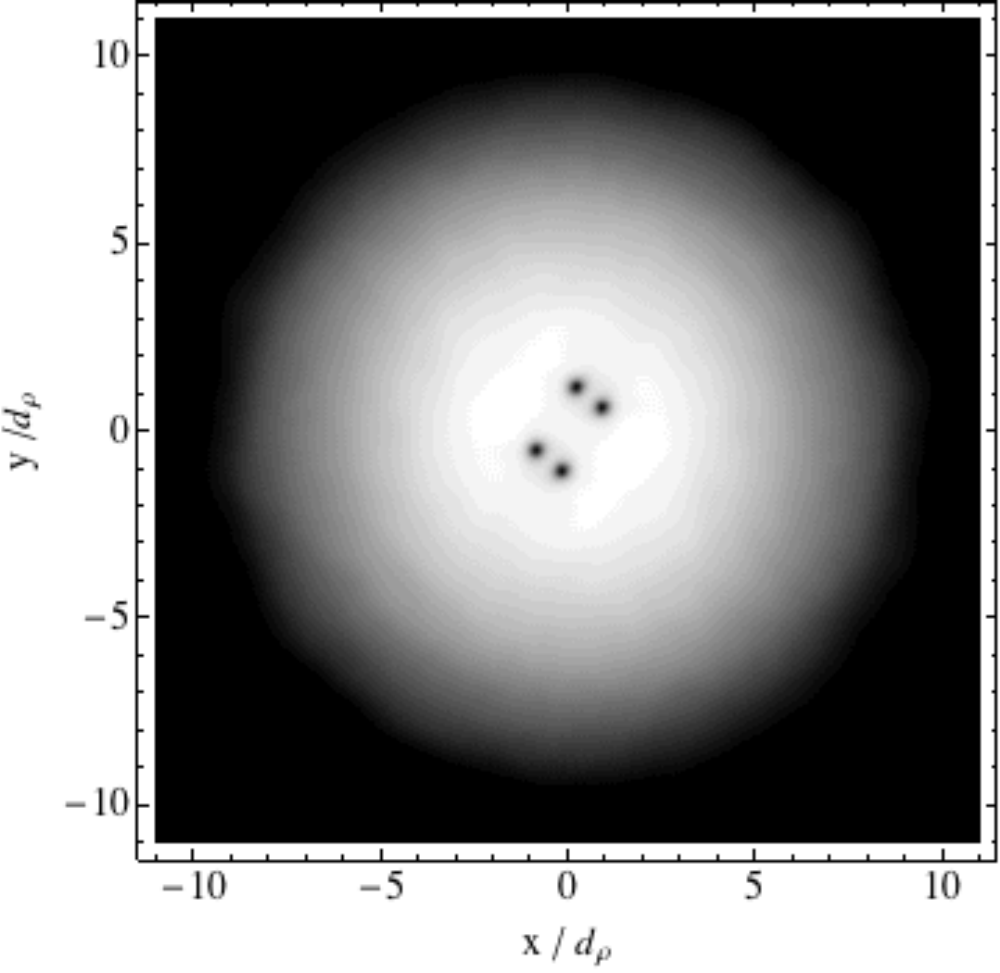}

}\subfloat[$\omega_{\rho}t=100$]{\includegraphics[scale=0.5]{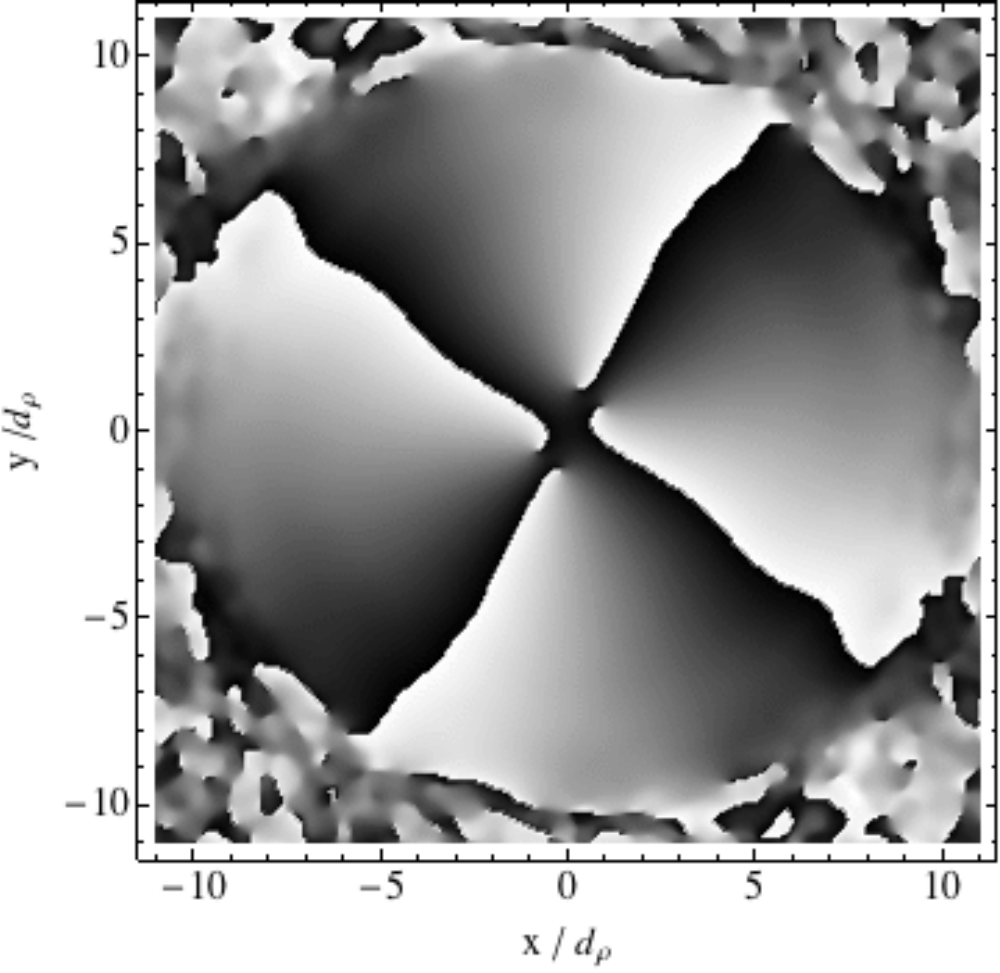}}

\protect\caption{(Color online) Time evolution of the density and phase of condensate
with a quadruply-charged vortex. We have used $\tilde{\mu}=45.9552$,
$Na_{s}/d_{z}=520$, $\epsilon=0.001$, and a factor of $0.001$ multiplying
the amplitude of deviations.}

\label{fig:4vD}
\end{figure}

In order to check our results we proceed the full numerical calculation
of the Gross-Pitaevskii equation (with the usual phenomenological
dissipation $\epsilon$ used since Ref.\cite{vortexLatticeFormation}).
The reason of this dissipative description is the prevention of non-physical
waves created by the grid edge. The initial state is calculated by
evolving a trial function in imaginary-time with the parameters given
by the equilibrium point from Eq. (\ref{eq:se}). We introduce the
eigenvector from Eq. (\ref{eq:eigensystem}) corresponding to the
unstable quadrupole mode ($Q_{v}$). This trial function is given
by
\begin{equation}
\Phi_{\ell}\propto\left\{ \frac{\left[x/\left(\xi_{0}+\delta\xi_{x}\right)\right]+i\left[y/\left(\xi_{0}+\delta\xi_{y}\right)\right]}{\sqrt{\left[x/\left(\xi_{0}+\delta\xi_{x}\right)\right]^{2}+\left[y/\left(\xi_{0}+\delta\xi_{y}\right)\right]^{2}+1}}\right\} ^{\ell}\sqrt{1-\left[\frac{x}{\left(R_{0}+\delta R_{x}\right)}\right]^{2}-\left[\frac{y}{\left(R_{0}+\delta R_{y}\right)}\right]^{2}}.\label{eq:numTrial}
\end{equation}
Furthermore we have done the evolution in real-time where we could
check the multi-charged vortex decaying to an initial state containing
only the deviations of $Q_{v}$-mode. In figure \ref{fig:2vD}, is
shown the evolution of the condensate in real-time for a doubly-charged
vortex, such that it starts to split around $\omega_{\rho}t=20.2$.
In figure \ref{fig:4vD}, we notice that the life-time of quadruply-charged
vortex is around $\omega_{\rho}t=22$. It is necessary to observe
that these life-times are different depending on the amplitude of
deviations and imaginary-time evolution. It is also possible to induce
the decaying by shaping an anisotropic trap, however our semi-analytic
approach is valid only for an isotropic trap.

It is interesting to observe the way in which multi-charged vortices
decay by $Q_{v}$-mode excitations, which makes the multi-charged
vortices split into a straight line of vortices with unitary angular
momentum. For instance, we see in figure \ref{fig:4vD} the quadruply-charged
vortex splitting into four vortices and forming a straight line, then
evolving based on its interaction with the velocity fields until the
final configuration.

\section{Stability diagram due to a static Gaussian potential}

\indent

\label{sec:staticLG}

Some articles on numerical simulations propose to stabilize an multi-charged
vortex by turning on a Gaussian laser beam at the middle of the vortex-core.
It means basically that we need to add an external potential with
Gaussian shape to the harmonic potential, i.e. 
\begin{eqnarray}
V_{\bot}\left(\mathbf{r}_{\bot}\right) & = & V_{trap}\left(\mathbf{r}_{\bot}\right)+V_{G}\left(\mathbf{r}_{\bot}\right)\nonumber \\
 & = & \frac{1}{2}m\omega_{\rho}^{2}\rho^{2}+\frac{1}{2}V_{0}e^{-\rho^{2}/\xi_{0}^{2}},
\end{eqnarray}
where the Gaussian width must be proportional to the vortex-core radius
($w=\sqrt{2}\xi_{0}$). An apparent objection to our approach could
lie on the fact that optical resolution limit of a laser beam is around
of some microns, while single-charged vortex core is usually smaller
than $0.5\text{\ensuremath{\mu}m}$. However, multi-charged vortices
may attain much larger sizes depending on its charge, the trap anisotropy,
number of atoms, and atomic species. For instance, a quadruply-charged
vortex in a $^{85}Rb$ condensate ($N=10^{5}$, $a_{s}=100a_{0}$,
$\omega_{\rho}=10\text{Hz}$, and $\omega_{z}=100\text{Hz}$) has
$5.9\mu m$. By applying a Gaussian beam with $w=10\mu m$ inside
of this vortex, its radius grows to $7.1\mu m$. Thus we use this
procedure in our semi-analytical method in order to draw a stability
diagram, and show that it is enough to stabilize the quadrupole mode
$Q_{v}$. So we have to calculate now the Lagrangian part corresponding
to the Gaussian potential,
\begin{equation}
L_{G}=\int V_{G}\left(\mathbf{r}_{\bot}\right)\left|\Phi\left(\mathbf{r}_{\bot}\right)\right|^{2}d\mathbf{r}_{\bot}.
\end{equation}
By expanding in Taylor series the integral of Gaussian potential $V_{LG}\left(\mathbf{r}_{\bot}\right)$
we have
\begin{eqnarray}
\int e^{-\rho^{2}/\xi_{0}^{2}}\left(\Phi^{*}\Phi\right)d\mathbf{r}_{\bot} & = & N_{0}R_{x}R_{y}\left[A_{7}+\frac{I_{26}}{R_{0}}\left(\delta R_{x}+\delta R_{y}\right)+\frac{I_{27}}{R_{0}^{2}}\left(\delta R_{x}^{2}+\delta R_{y}^{2}\right)+\frac{I_{28}}{R_{0}^{2}}\delta R_{x}\delta R_{y}\right.\nonumber \\
 &  & +I_{29}\left(\delta\alpha_{x}+\delta\alpha_{y}\right)+I_{30}\left(\delta\alpha_{x}^{2}+\delta\alpha_{y}^{2}\right)+I_{31}\delta\alpha_{x}\delta\alpha_{y}\nonumber \\
 &  & \left.\!\!\!\!\!\!+\frac{I_{32}}{R_{0}}\left(\delta R_{x}\delta\alpha_{x}+\delta R_{y}\delta\alpha_{y}\right)+\frac{I_{33}}{R_{0}}\left(\delta R_{x}\delta\alpha_{y}+\delta R_{y}\delta\alpha_{x}\right)+I_{34}\delta\alpha_{xy}^{2}\right],
\end{eqnarray}
where Lagrangian part becomes
\begin{eqnarray}
L_{G} & = & -\frac{\tilde{V}_{0}A_{7}}{2A_{0}}\left\{ 1+\frac{I_{26}}{A_{7}R_{0}}\left(\delta R_{x}+\delta R_{y}\right)+\frac{I_{27}}{A_{7}R_{0}^{2}}\left(\delta R_{x}^{2}+\delta R_{y}^{2}\right)\right.\nonumber \\
 &  & +\frac{I_{28}}{A_{7}R_{0}^{2}}\delta R_{x}\delta R_{y}+\left(\frac{I_{29}}{A_{7}}-\frac{I_{1}}{A_{0}}\right)\left(\delta\alpha_{x}+\delta\alpha_{y}\right)\nonumber \\
 &  & +\left[\frac{I_{30}}{A_{7}}-\frac{I_{2}}{A_{0}}-\frac{I_{1}}{A_{0}}\left(\frac{I_{29}}{A_{7}}-\frac{I_{1}}{A_{0}}\right)\right]\left(\delta\alpha_{x}^{2}+\delta\alpha_{y}^{2}\right)\nonumber \\
 &  & +\left[\frac{I_{31}}{A_{7}}-\frac{I_{3}}{A_{0}}-2\frac{I_{1}}{A_{0}}\left(\frac{I_{29}}{A_{7}}-\frac{I_{1}}{A_{0}}\right)\right]\delta\alpha_{x}\delta\alpha_{y}\nonumber \\
 &  & +\frac{1}{R_{0}}\left(\frac{I_{32}}{A_{7}}-\frac{I_{26}I_{1}}{A_{7}A_{0}}\right)\left(\delta R_{x}\delta\alpha_{x}+\delta R_{y}\delta\alpha_{y}\right)\nonumber \\
 &  & \left.+\frac{1}{R_{0}}\left(\frac{I_{33}}{A_{7}}-\frac{I_{26}I_{1}}{A_{7}A_{0}}\right)\left(\delta R_{x}\delta\alpha_{y}+\delta R_{y}\delta\alpha_{x}\right)+\left(\frac{I_{34}}{A_{7}}-\frac{I_{4}}{A_{0}}\right)\delta\alpha_{xy}^{2}\right\} .\label{eq:Lpin}
\end{eqnarray}
Notice that we have terms of first order in deviations in Eq. (\ref{eq:Lpin}),
it means that the stationary solution is modified when the condensate
is under the influence of a Gaussian potential. The first-order contribution
in (\ref{eq:Lpin}) becomes
\begin{equation}
L_{G}^{\left(1\right)}=-\frac{\tilde{V}_{0}A_{7}}{2A_{0}}\left\{ \overset{s_{\rho}}{\overbrace{\frac{I_{26}}{A_{7}R_{0}}}}\left(\delta R_{x}+\delta R_{y}\right)+\overset{s_{\alpha}}{\overbrace{\left(\frac{I_{29}}{A_{7}}-\frac{I_{1}}{A_{0}}\right)}}\left(\delta\alpha_{x}+\delta\alpha_{y}\right)\right\} ,
\end{equation}
while the second-order terms are 
\begin{eqnarray}
L_{G}^{\left(2\right)} & = & -\frac{\tilde{V}_{0}A_{7}}{2A_{0}}\left\{ \stackrel{p_{\rho}}{\overbrace{\frac{I_{27}}{A_{7}R_{0}^{2}}}}\left(\delta R_{x}^{2}+\delta R_{y}^{2}\right)+\stackrel{p_{\rho\rho}}{\overbrace{\frac{I_{28}}{A_{7}R_{0}^{2}}}}\delta R_{x}\delta R_{y}\right.\nonumber \\
 &  & +\stackrel{p_{\alpha}}{\overbrace{\left[\frac{I_{30}}{A_{7}}-\frac{I_{2}}{A_{0}}-\frac{I_{1}}{A_{0}}\left(\frac{I_{29}}{A_{7}}-\frac{I_{1}}{A_{0}}\right)\right]}}\left(\delta\alpha_{x}^{2}+\delta\alpha_{y}^{2}\right)\nonumber \\
 &  & +\stackrel{p_{\alpha\alpha}}{\overbrace{\left[\frac{I_{31}}{A_{7}}-\frac{I_{3}}{A_{0}}-2\frac{I_{1}}{A_{0}}\left(\frac{I_{29}}{A_{7}}-\frac{I_{1}}{A_{0}}\right)\right]}}\delta\alpha_{x}\delta\alpha_{y}\nonumber \\
 &  & +\stackrel{p_{\rho\alpha}}{\overbrace{\frac{1}{R_{0}}\left(\frac{I_{32}}{A_{7}}-\frac{I_{26}I_{1}}{A_{7}A_{0}}\right)}}\left(\delta R_{x}\delta\alpha_{x}+\delta R_{y}\delta\alpha_{y}\right)\nonumber \\
 &  & \left.+\stackrel{p_{\alpha\rho}}{\overbrace{\frac{1}{R_{0}}\left(\frac{I_{33}}{A_{7}}-\frac{I_{26}I_{1}}{A_{7}A_{0}}\right)}}\left(\delta R_{x}\delta\alpha_{y}+\delta R_{y}\delta\alpha_{x}\right)\right\} .\label{eq:Lpin2}
\end{eqnarray}
The equilibrium points are changed to
\begin{equation}
S_{\rho}+s_{\rho}=0,\text{ and }S_{\alpha}+s_{\alpha}=0.
\end{equation}
Each terms in (\ref{eq:Lpin2}) adds a contribution to a different
element in the matrix $V$ of the linearized Euler-Lagrange equation
(\ref{eq:eigensystem}) which then becomes
\begin{equation}
M\ddot{\delta}+\left(V+V_{G}\right)\delta=0,
\end{equation}
where
\begin{equation}
V_{G}=\frac{\tilde{V}_{0}A_{7}}{2A_{0}}\begin{pmatrix}2p_{\rho} & p_{\rho\rho} & p_{\rho\alpha} & p_{\alpha\rho}\\
p_{\rho\rho} & 2p_{\rho} & p_{\alpha\rho} & p_{\rho\alpha}\\
p_{\rho\alpha} & p_{\alpha\rho} & 2p_{\alpha} & p_{\alpha\alpha}\\
p_{\alpha\rho} & p_{\rho\alpha} & p_{\alpha\alpha} & 2p_{\alpha}
\end{pmatrix}.\label{eq:VLG}
\end{equation}

\begin{figure}
\centering
\subfloat[$\ell=2$]{\includegraphics[scale=0.7]{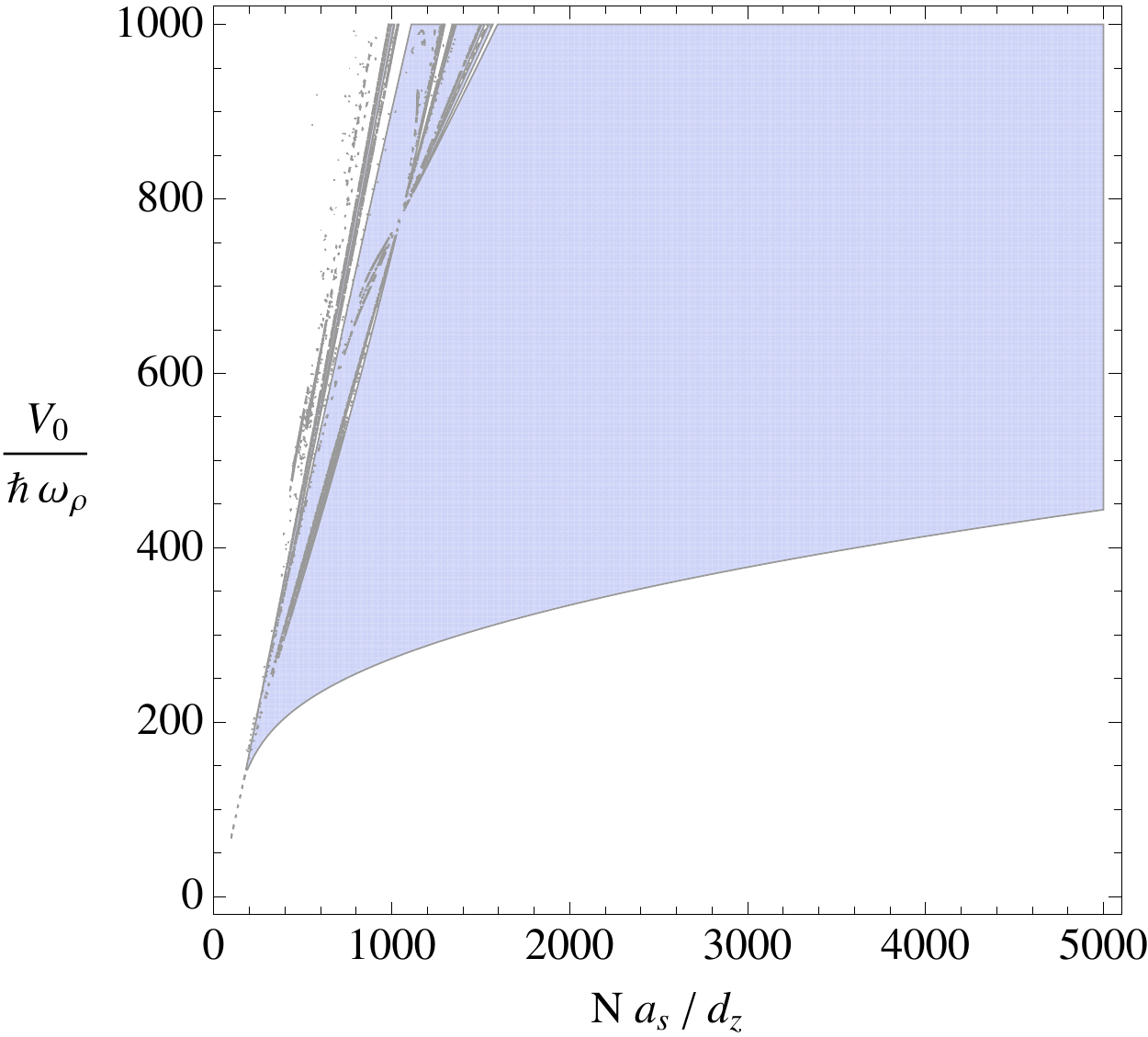}

}\subfloat[$\ell=4$]{\includegraphics[scale=0.7]{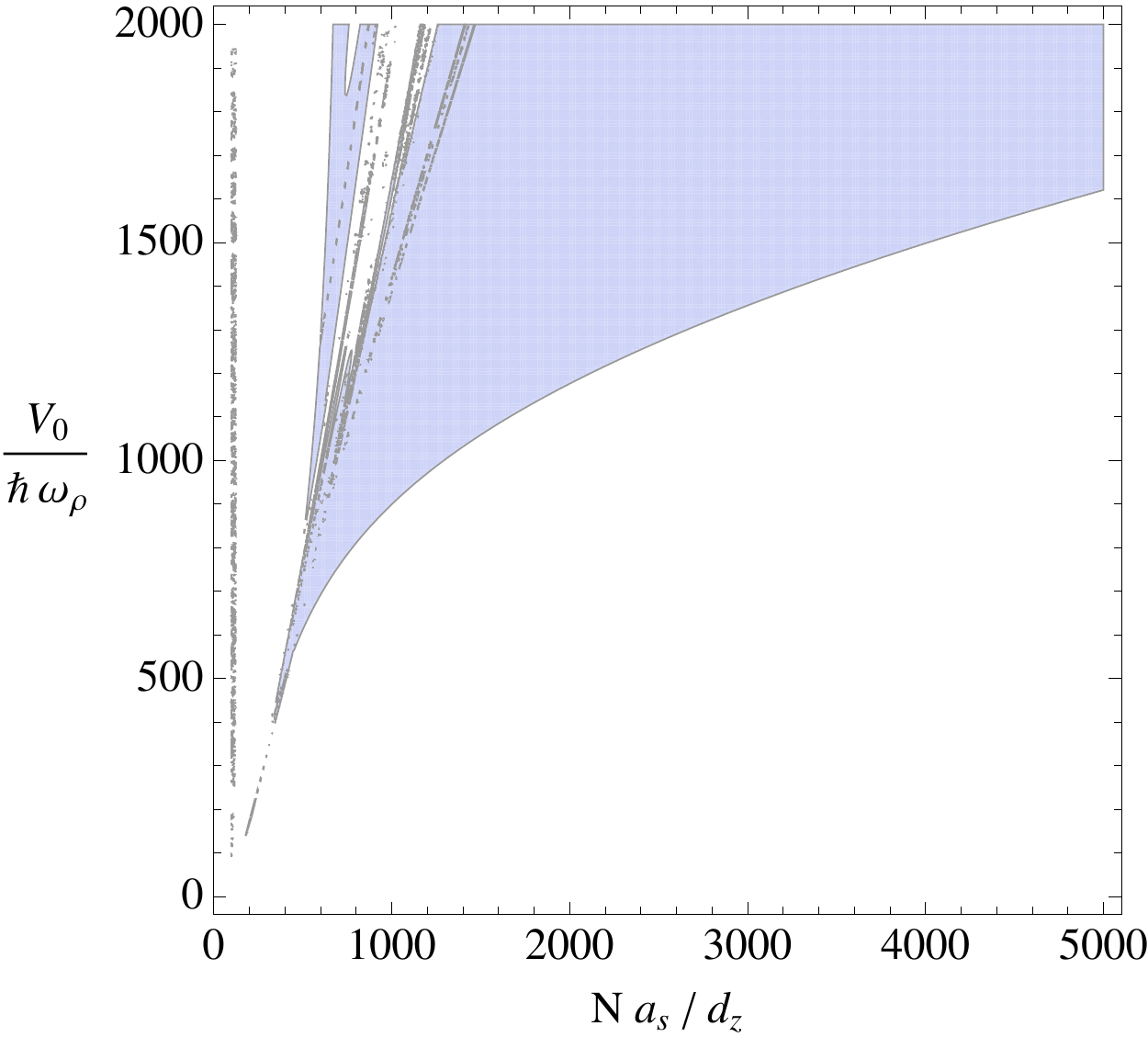}

}

\protect\caption{(Color online) Diagram of magnitude of pinning potential by atomic
interaction for vortex with $\ell=2$ and $\ell=4$. Hatched region
represents stable eigensystem meaning the vortex-core become stable.}

\label{fig:PinDiagram}
\end{figure}

Since the stability of the eigensystem depends only on the $Q_{v}$-frequency,
we can build a stability diagram of $V_{0}/\hbar\omega_{\rho}$ versus
$Na_{s}/d_{z}$. In fig.\ref{fig:PinDiagram}, this diagram is shown
considering two cases, $\ell=2$ and $\ell=4$. As the angular momentum
$\ell$ gets larger the stable region decreases. Hence the pinning
potential can prevent the vortices from splitting for some values
of $V_{0}/\hbar\omega_{\rho}$ depending on $Na_{s}/d_{\rho}$.

\begin{figure}
\centering

\subfloat[$\omega_{\rho}t=1$]{\includegraphics[scale=0.5]{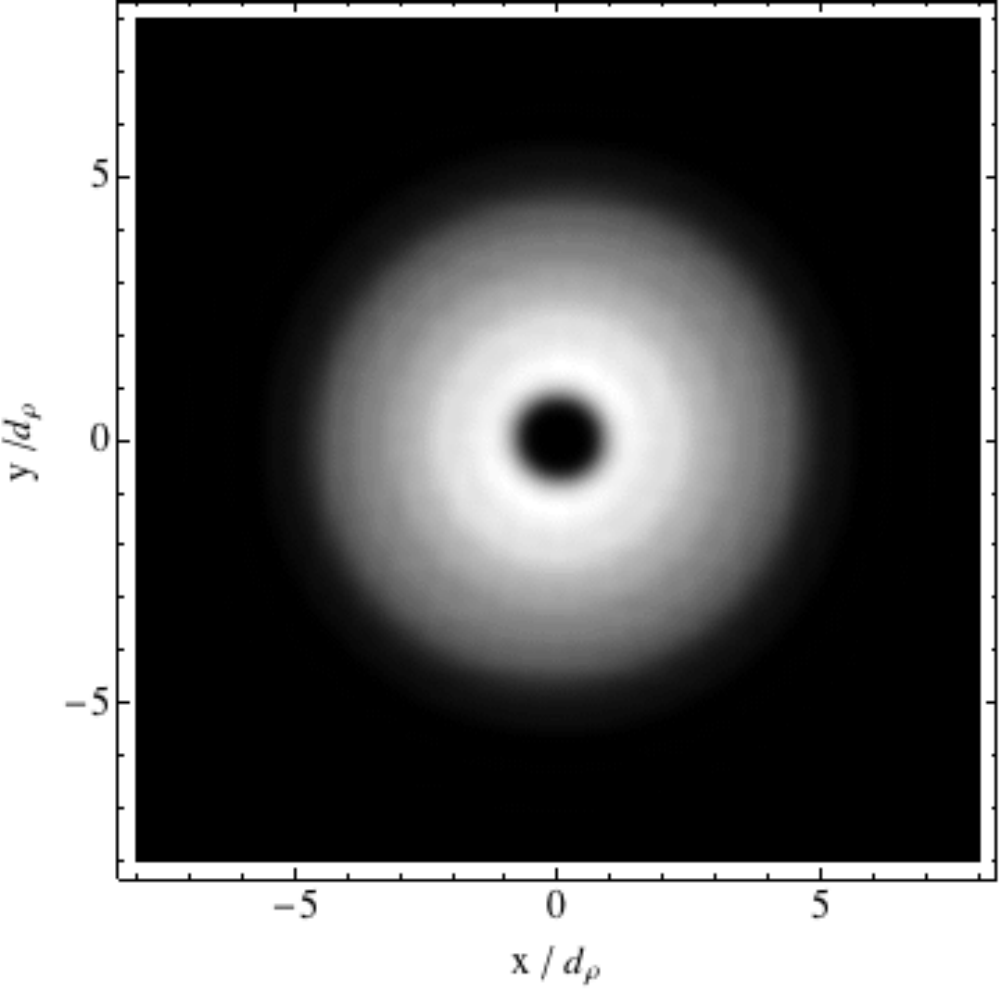}

}\subfloat[$\omega_{\rho}t=1$]{\includegraphics[scale=0.5]{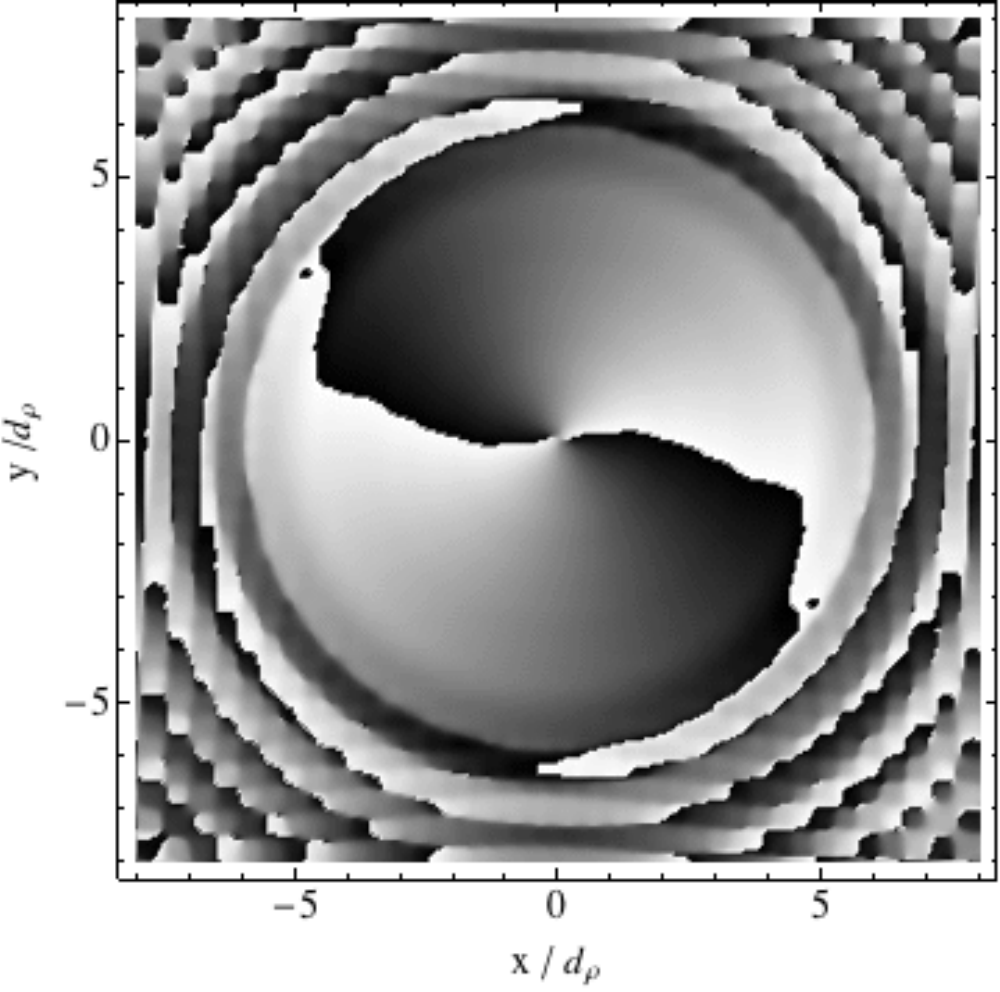}

}

\subfloat[$\omega_{\rho}t=21$]{\includegraphics[scale=0.5]{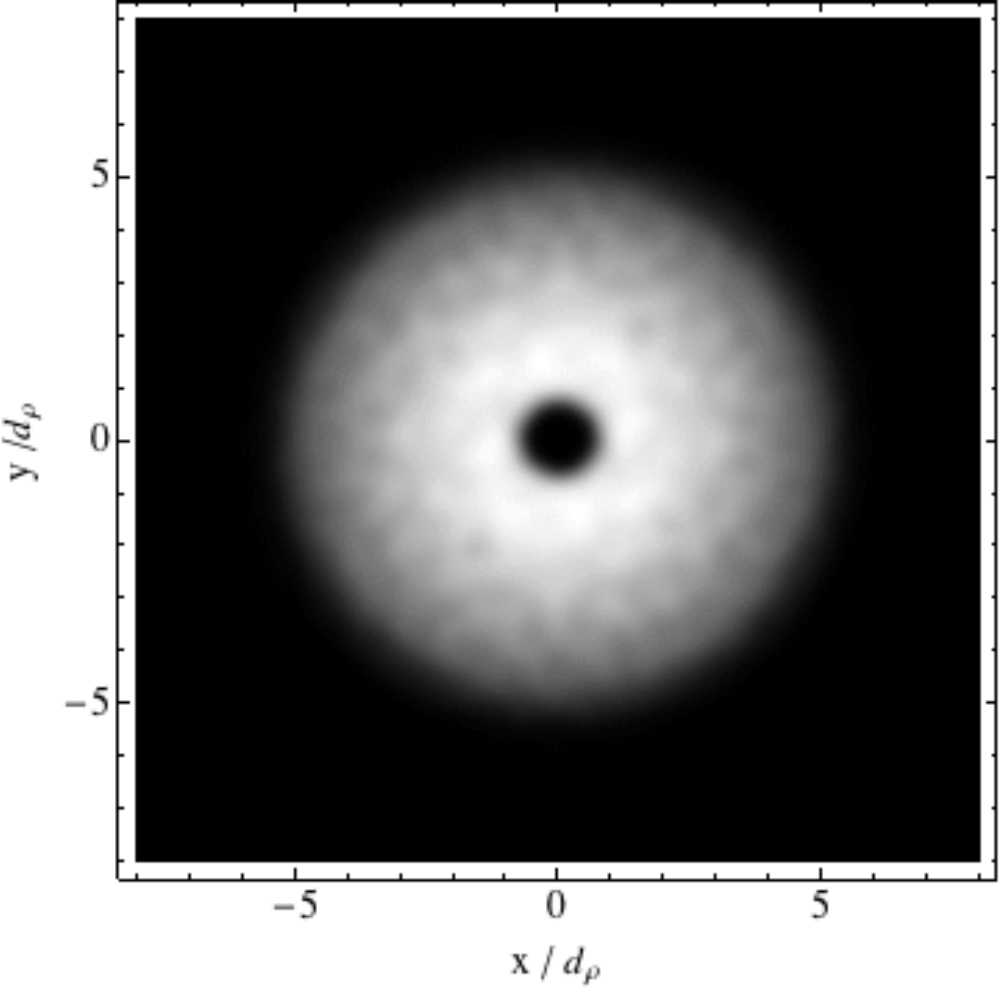}

}\subfloat[$\omega_{\rho}t=21$]{\includegraphics[scale=0.5]{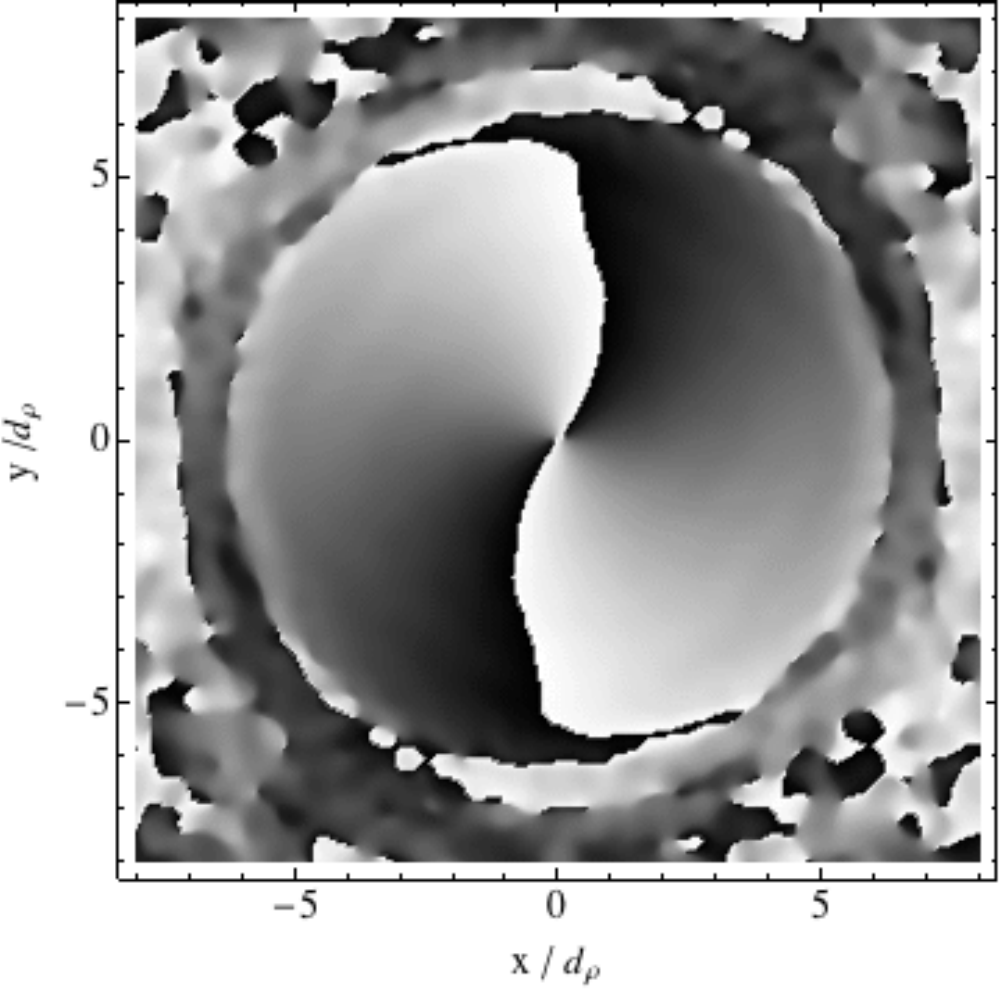}

}

\subfloat[$\omega_{\rho}t=100$]{\includegraphics[scale=0.5]{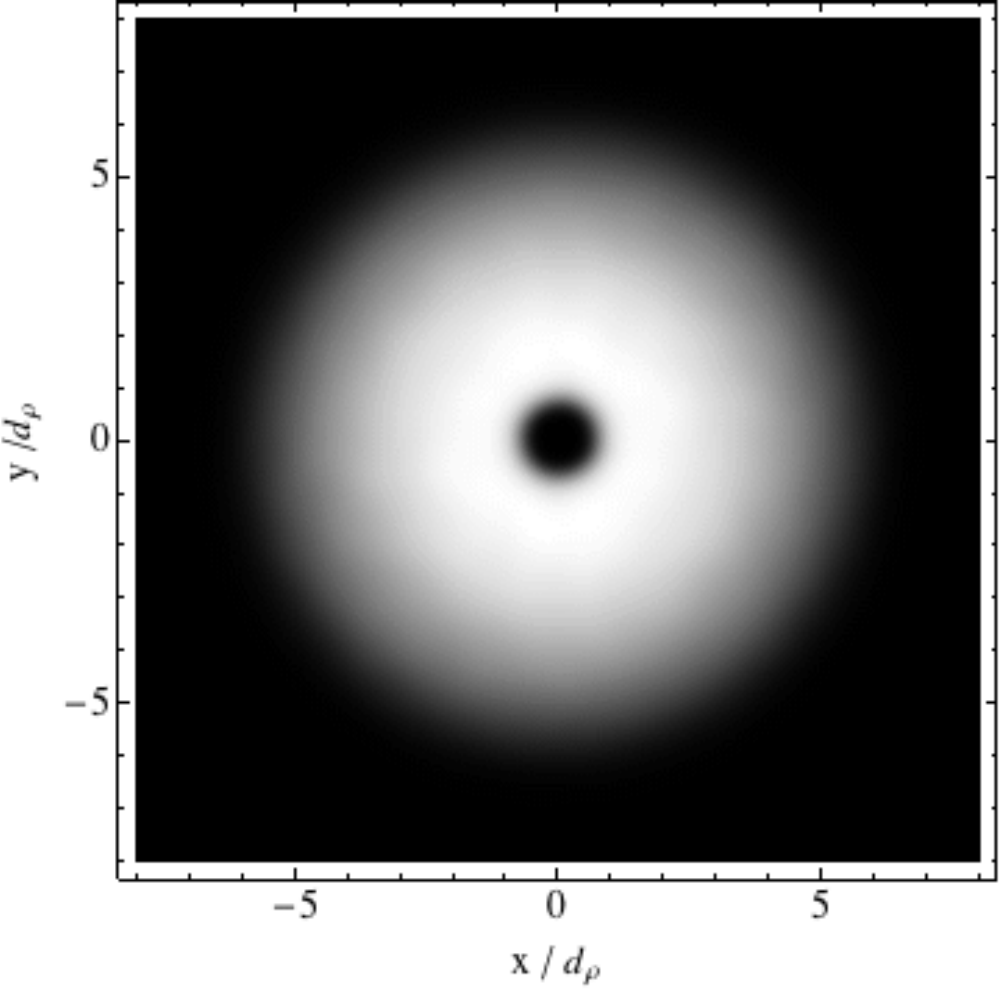}

}\subfloat[$\omega_{\rho}t=100$]{\includegraphics[scale=0.5]{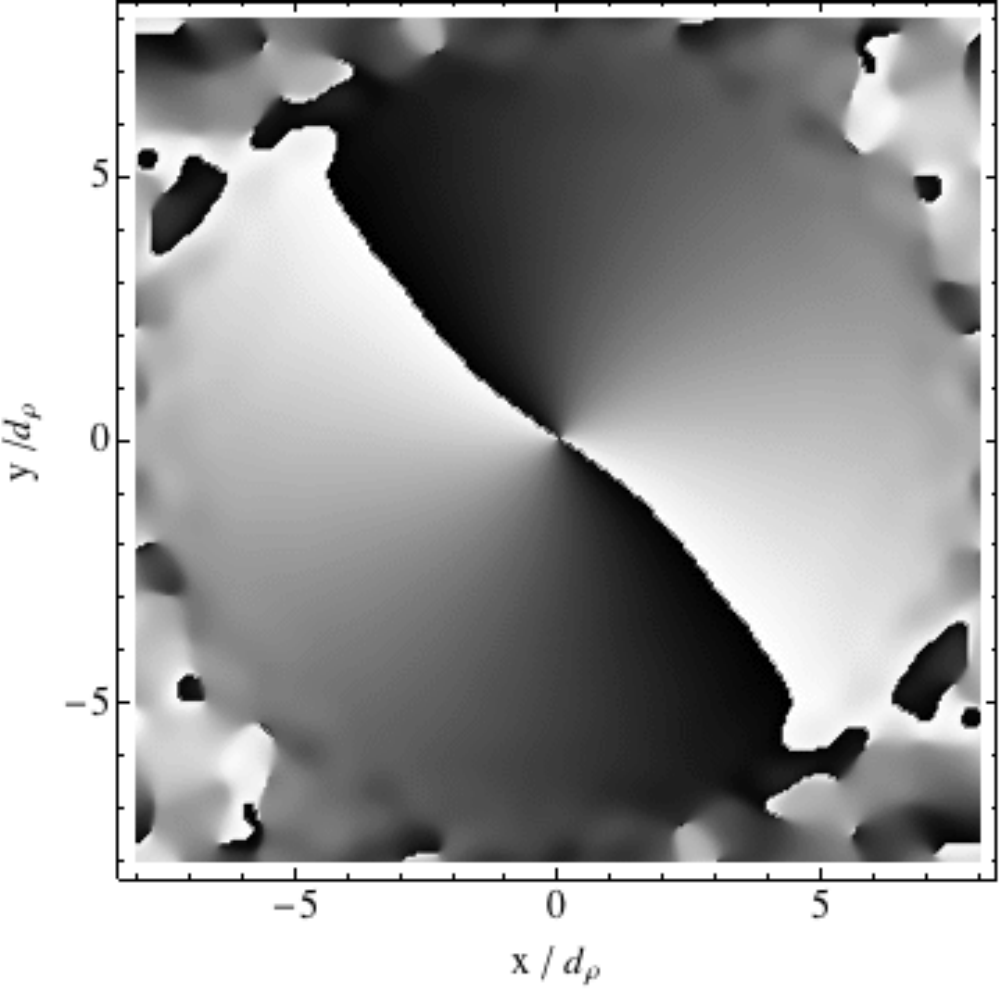}

}

\protect\caption{(Color online) Time evolution of the density (a, c, e) and phase (b,
d, f) of condensate with a doubly-charged vortex. We used $\mu/\hbar\omega_{\rho}=20.198$,
$Na_{s}/d_{z}=100$, $V_{0}/\hbar\omega_{\rho}=150$, $\epsilon=0.001$,
and a factor of $0.01$ multiplying the amplitude of deviations.}

\label{fig:2vDpin}
\end{figure}

\begin{figure}
\centering

\subfloat[$\omega_{\rho}t=1$]{\includegraphics[scale=0.5]{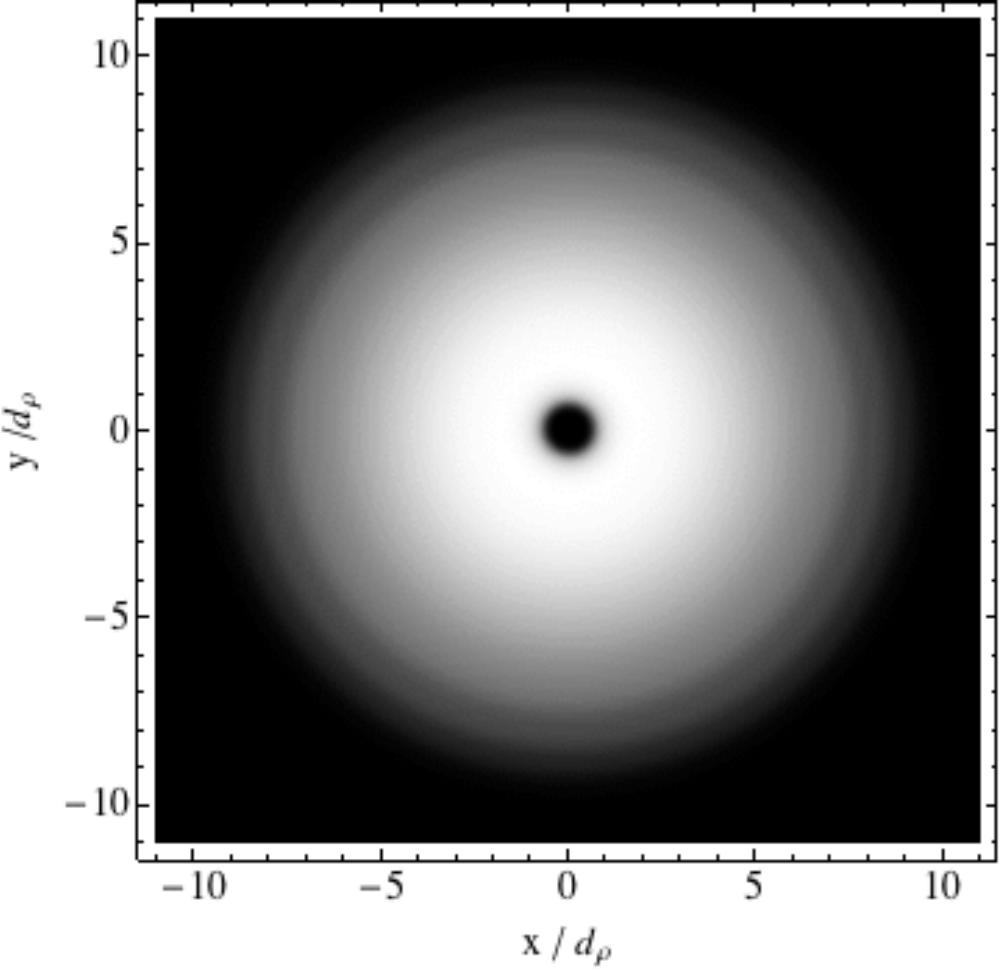}

}\subfloat[$\omega_{\rho}t=1$]{\includegraphics[scale=0.5]{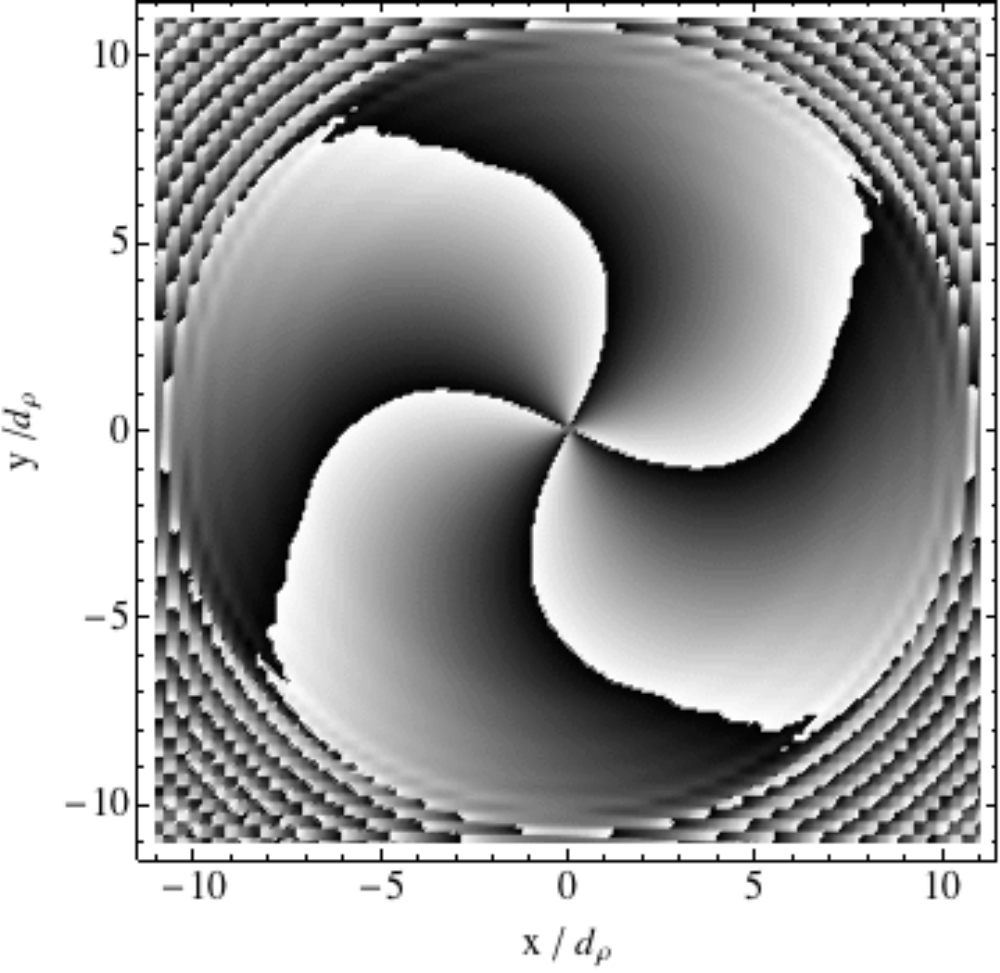}

}

\subfloat[$\omega_{\rho}t=22$]{\includegraphics[scale=0.5]{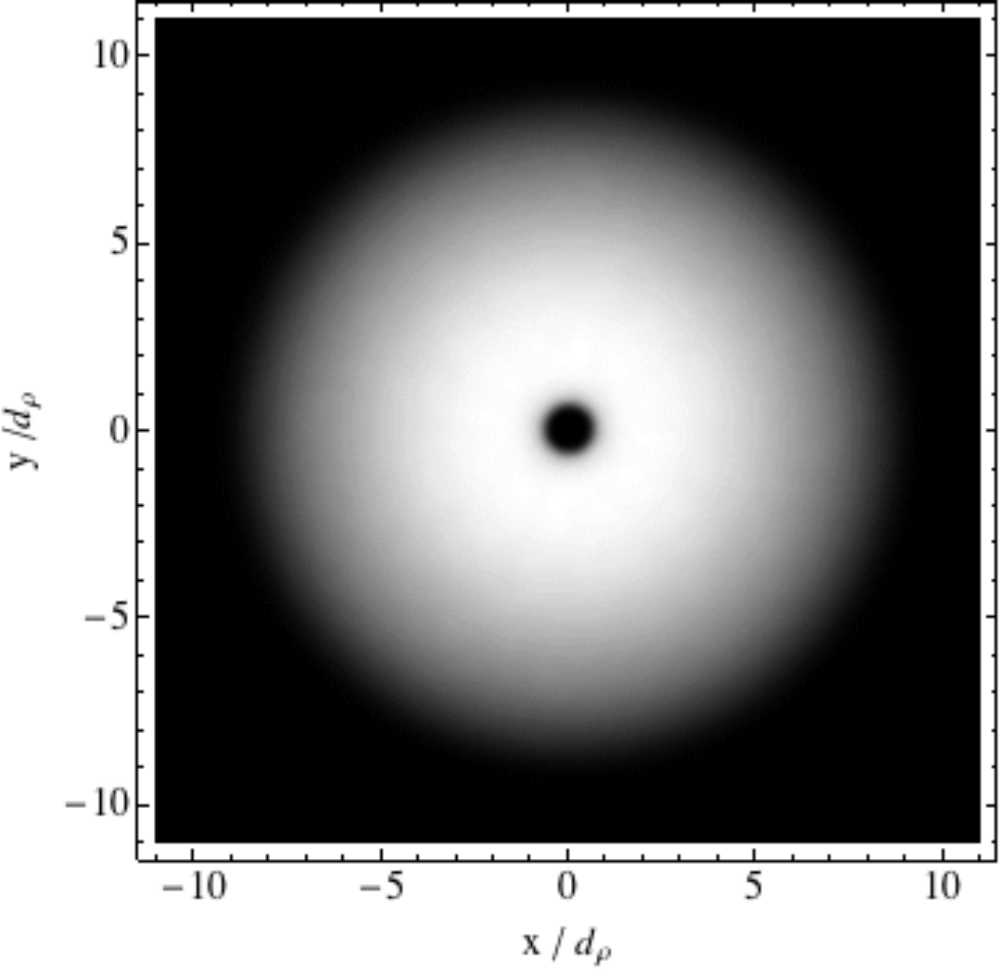}

}\subfloat[$\omega_{\rho}t=22$]{\includegraphics[scale=0.5]{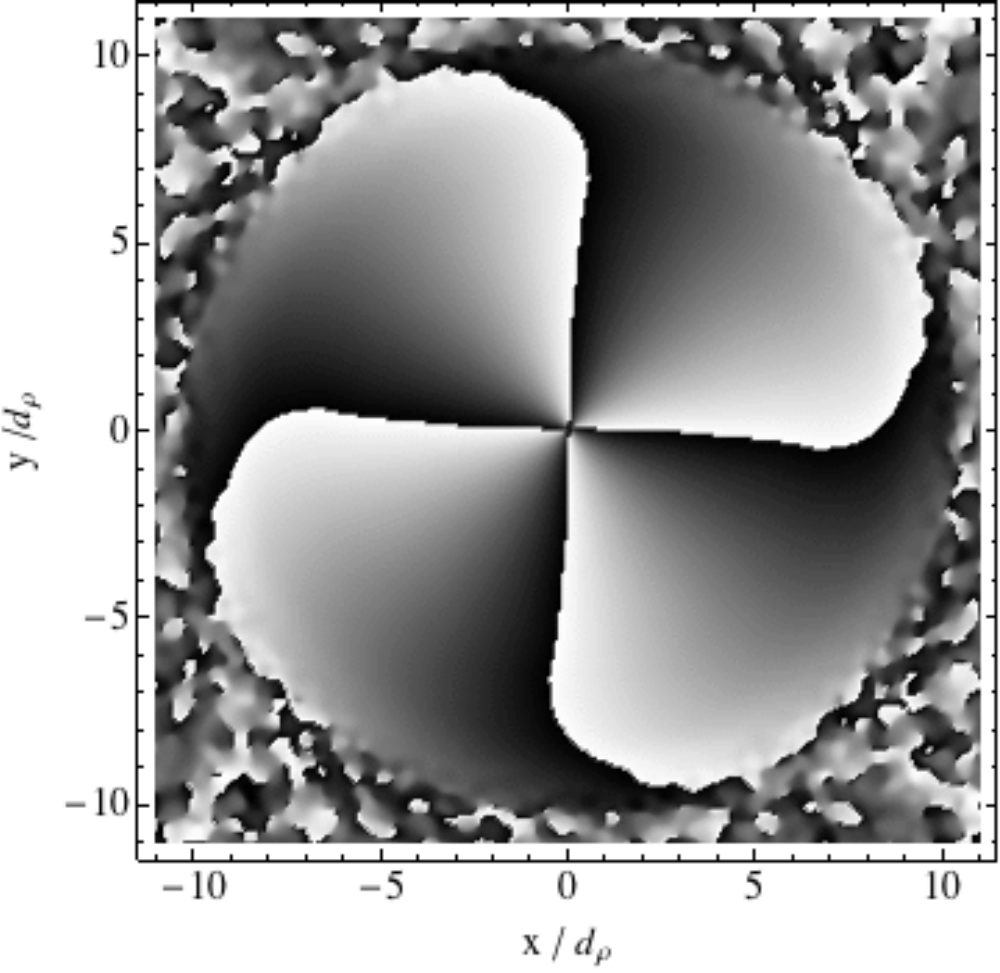}

}

\subfloat[$\omega_{\rho}t=100$]{\includegraphics[scale=0.5]{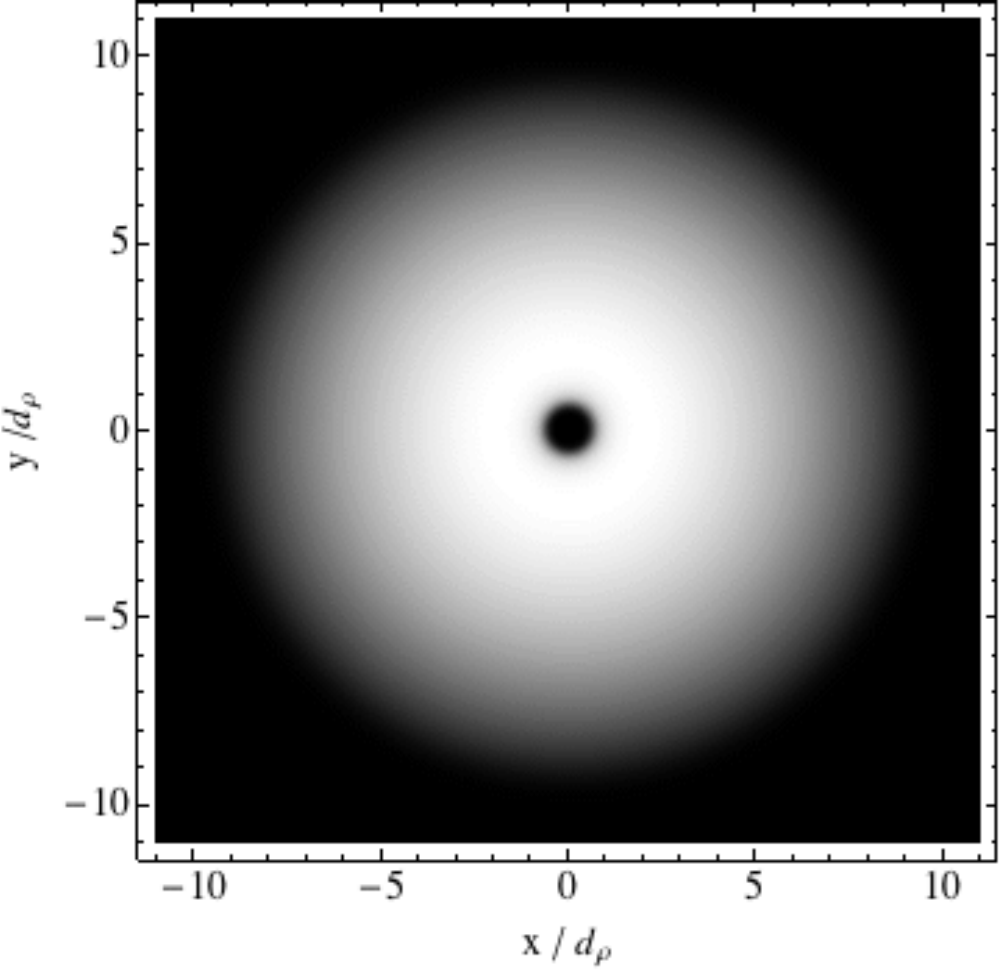}

}\subfloat[$\omega_{\rho}t=100$]{\includegraphics[scale=0.5]{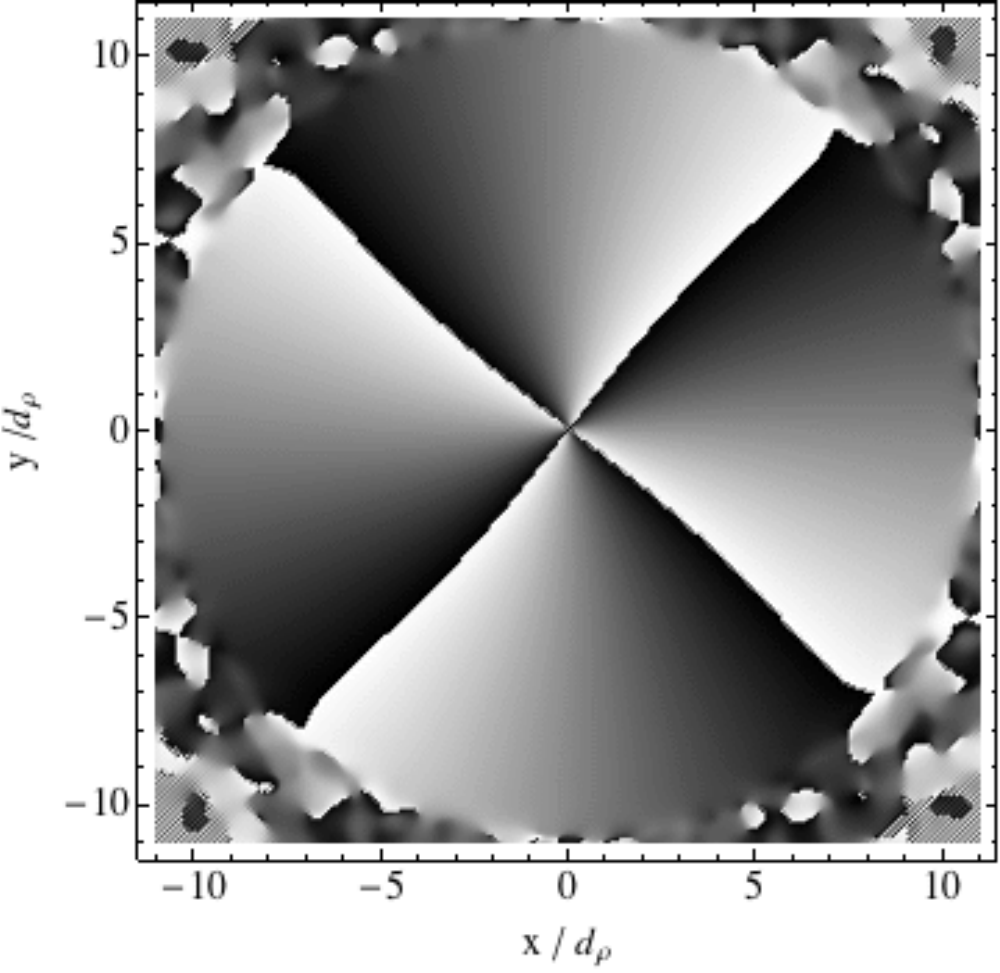}

}

\protect\caption{(Color online) Time evolution of the density (a, c, e) and phase (b,
d, f) of condensate with a quadruply-charged vortex. We used $\mu/\hbar\omega_{\rho}=45.9552$,
$Na_{s}/d_{z}=520$, $V_{0}/\hbar\omega_{\rho}=500$, $\epsilon=0.001$,
and a factor of $0.001$ multiplying the amplitude of deviations.}

\label{fig:4vDpin}
\end{figure}

In order to validate these stability diagrams, we make a numerical
simulation of the Gross-Pitaevskii equation. When the Gaussian potential
is turned on, we have seen that it provokes some phonon-waves on the
condensate surface and increases a little the vortex-core size besides
preventing the vortex decay. Figure \ref{fig:2vDpin} shows phonon-waves
rising and vanishing due to dissipation. The same phenomena may be
seen in figure \ref{fig:4vDpin}.

The vortex decay happens when the sound waves couple the quadrupole
mode from the edge of the condensate with the vortex-core, which breaks
the polar symmetry of vortex. Therefore, the pinning potential acts
as a wall reflecting these sound waves, and preventing the vortex
symmetry break.

\section{Diagram of stability due to a dynamic Gaussian potential}

\indent

\label{sec:dynamicLG}

\begin{figure}
\centering

\includegraphics{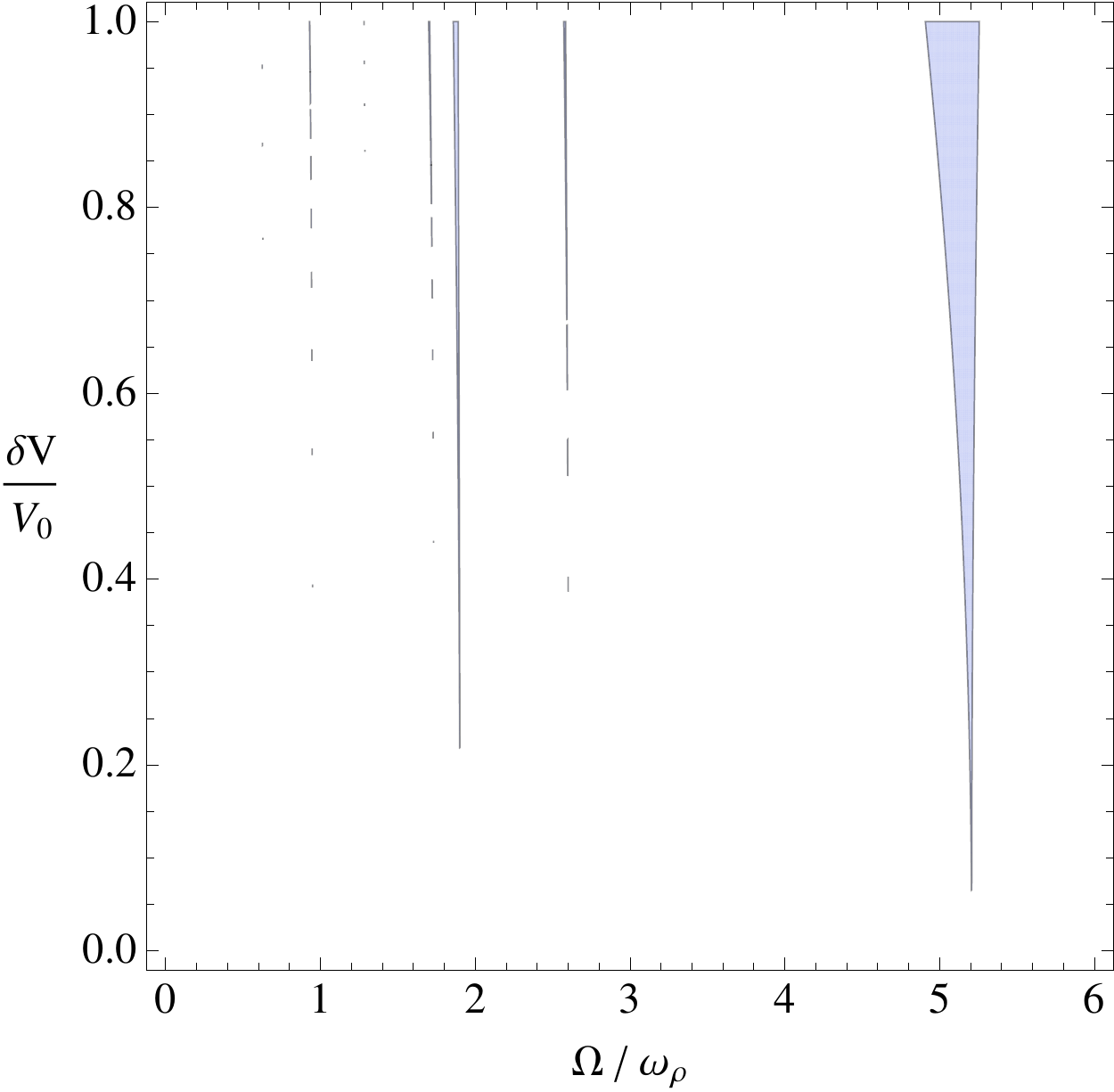}\protect\caption{(Color online) Diagram of amplitude versus frequency where hatched
stable regions are found for a condensate containing a triply-charged
vortex subjected to \textbf{height modulation}. We used $Na_{s}/d_{z}=125$.
Convergence is obtained already with two iterations.}

\label{fig:pD}
\end{figure}

In section \ref{sec:staticLG}, we have seen that it is possible to
make a multi-charged vortex stable using a static Gaussian potential.
In addition, we calculated a diagram of height versus interaction
strength which shows the stable region. Here we propose to stabilize
a multi-charged vortex with a sinusoidal modulation of height of the
Gaussian potential with an amplitude given by $\delta V$,
\begin{equation}
V_{0}\left(t\right)=V_{0}-\delta V\cos\left(\Omega t\right),
\end{equation}
at the specific region of interaction strength where the static potential
is not capable of stabilizing the vortex, i.e. $0<Na_{s}/d_{\rho}\leq160$.
The equation for this case is given by
\begin{equation}
M\ddot{\delta}+\left\{ V+V_{G}\left[1-\frac{\delta V}{V_{0}}\cos\left(\tilde{\Omega}\tilde{t}\right)\right]\right\} \delta=0,
\end{equation}
where matrices $M$ and $V$ can be found at Eq.(\ref{eq:eigensystem}),
and $V_{LG}$ is given by Eq.(\ref{eq:VLG}). By scaling the time
as follows 
\begin{equation}
\frac{\Omega}{\omega_{\rho}}\tilde{t}\rightarrow2\tau,
\end{equation}
we obtain the Mathieu equation
\begin{equation}
\frac{\tilde{\Omega}^{2}}{4}\ddot{\delta}+\left\{ A-2\frac{\delta V}{V_{0}}Q\cos\left(2\tau\right)\right\} \delta=0,\label{eq:Mathieu}
\end{equation}
where $A=M^{-1}\left(V+V_{LG}\right)$ and $Q=\left(1/2\right)M^{-1}V_{LG}$
are constants depending on the initial conditions. This equation becomes
solvable by using Floquet theory \cite{mathieu1,mathieu2,floquet,Floquet1,Floquet2}.
The basic idea of this theory is that if a linear differential equation
has periodic coefficients, the solutions will be a linear periodic
combination of functions times exponentially increasing (or decreasing)
functions. Thus linear independent solutions of the Mathieu equation
for any pair of $A$ and $B$ can be expressed as
\begin{equation}
\delta\left(\tau\right)=e^{\pm\eta\tau}P\left(\pm\tau\right),
\end{equation}
where $\eta$ is called the characteristic exponent which is a constant
depending on both $A$ and $Q$, and $P\left(\tau\right)$ is $\pi$-periodic
in $\tau$ that which can be written as an infinity series
\begin{equation}
\delta\left(\tau\right)=e^{\eta\tau}\sum_{n=-\infty}^{\infty}b_{2n}e^{2ni\tau},\label{eq:FAnsatz}
\end{equation}
with $b_{2n}$ being a Fourier component. Doing the substitution of
(\ref{eq:FAnsatz}) into (\ref{eq:Mathieu}), we have
\begin{equation}
\left[A+\frac{\tilde{\Omega}^{2}}{4}\left(\eta+2ni\right)^{2}I\right]b_{2n}-Q\left(b_{2n+2}+b_{2n-2}\right)=0.\label{eq:M1}
\end{equation}
At this point it is wise to define ladder operators $L_{2n}^{\pm}b_{2n}=b_{2n\pm2}$
which yields
\begin{equation}
L_{2n}^{\pm}=\left\{ A+\frac{\tilde{\Omega}^{2}}{4}\left[\eta+2i\left(n\pm1\right)\right]^{2}I-QL_{2n\pm2}^{\pm}\right\} ^{-1}Q.\label{eq:Ladder}
\end{equation}
By using (\ref{eq:Ladder}) to write (\ref{eq:M1}) in terms of $b_{0}$
only, we obtain an iteration algorithm wherein we replace the ladder
operator over and over inside itself which then becomes
\begin{eqnarray}
\left(A+\frac{\tilde{\Omega}^{2}}{4}\eta^{2}I-Q\left\{ \left[A+\frac{\tilde{\Omega}^{2}}{4}\left(\eta+2i\right)^{2}I-\cdots\right]^{-1}\right.\right.\nonumber \\
\left.\left.+\left[A+\frac{\tilde{\Omega}^{2}}{4}\left(\eta-2i\right)^{2}I-\cdots\right]^{-1}\right\} Q\right)b_{0} & = & 0.\label{eq:FloquetDet}
\end{eqnarray}
Since we are not interested in trivial solutions for $b_{0}$, the
determinant of (\ref{eq:FloquetDet}) must vanish. Thus the stability
diagram for a modulation of the Gaussian potential with frequency
$\Omega$ and amplitude $V_{0}$ is presented in fig. \ref{fig:pD},
where its resonant behavior does not depend on the initial conditions
\cite{PR1}.

The edges between stable and unstable domains (also called as Floquet
fringes) were calculated by making $\eta=0$. Since the equilibrium
configuration rarely has solution for $V_{0}/\hbar\omega_{\rho}\geq Na_{s}/d_{z}$,
we only build the stability diagram for $V_{0}/\hbar\omega_{\rho}<Na_{s}/d_{z}$.
The iterative algorithm converges very fast, and does not require
more than two iterations.

The stable regions, also called resonance region, can lead the system
to lose coherence if the excitation time is long enough (hundreds
of milliseconds according to number of atoms) which leads to destruction
of the condensate state.

\begin{figure}
\centering
\subfloat[$\omega_{\rho}t=14$]{\includegraphics[scale=0.5]{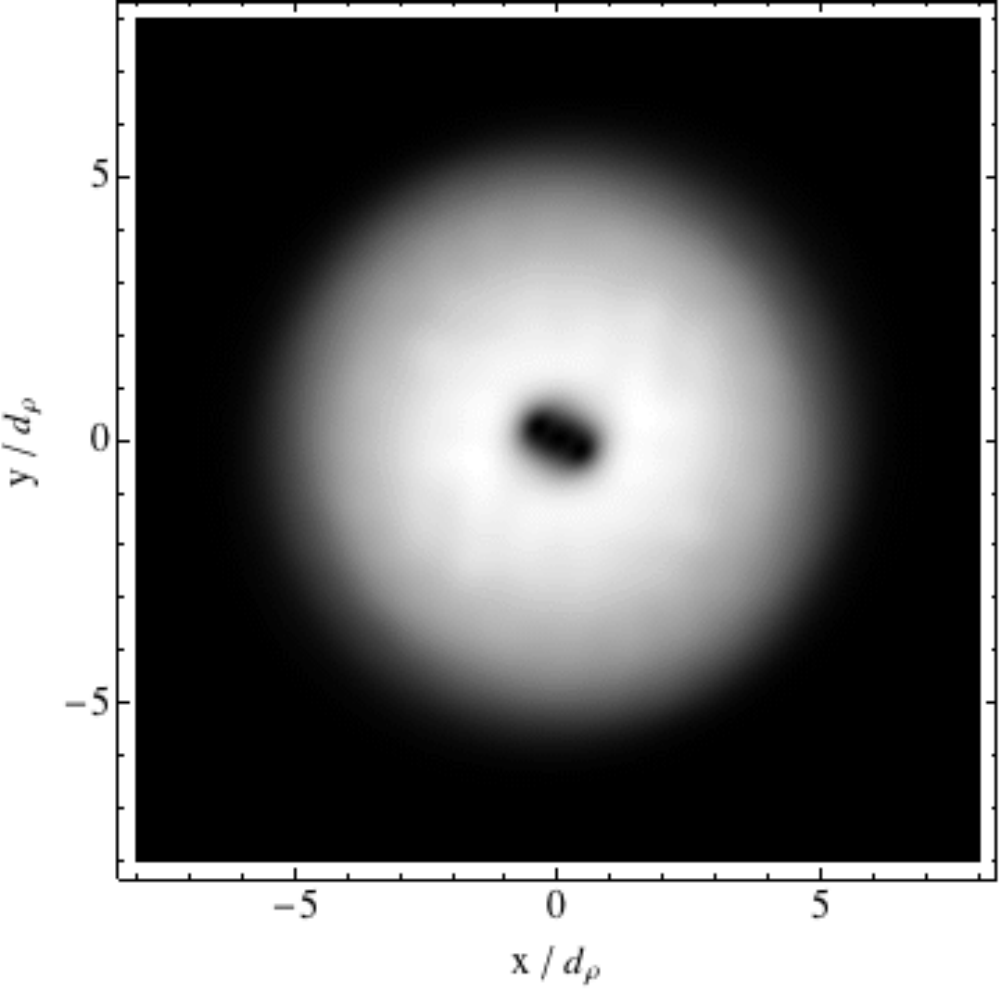}

}\subfloat[$\omega_{\rho}t=14$]{\includegraphics[scale=0.5]{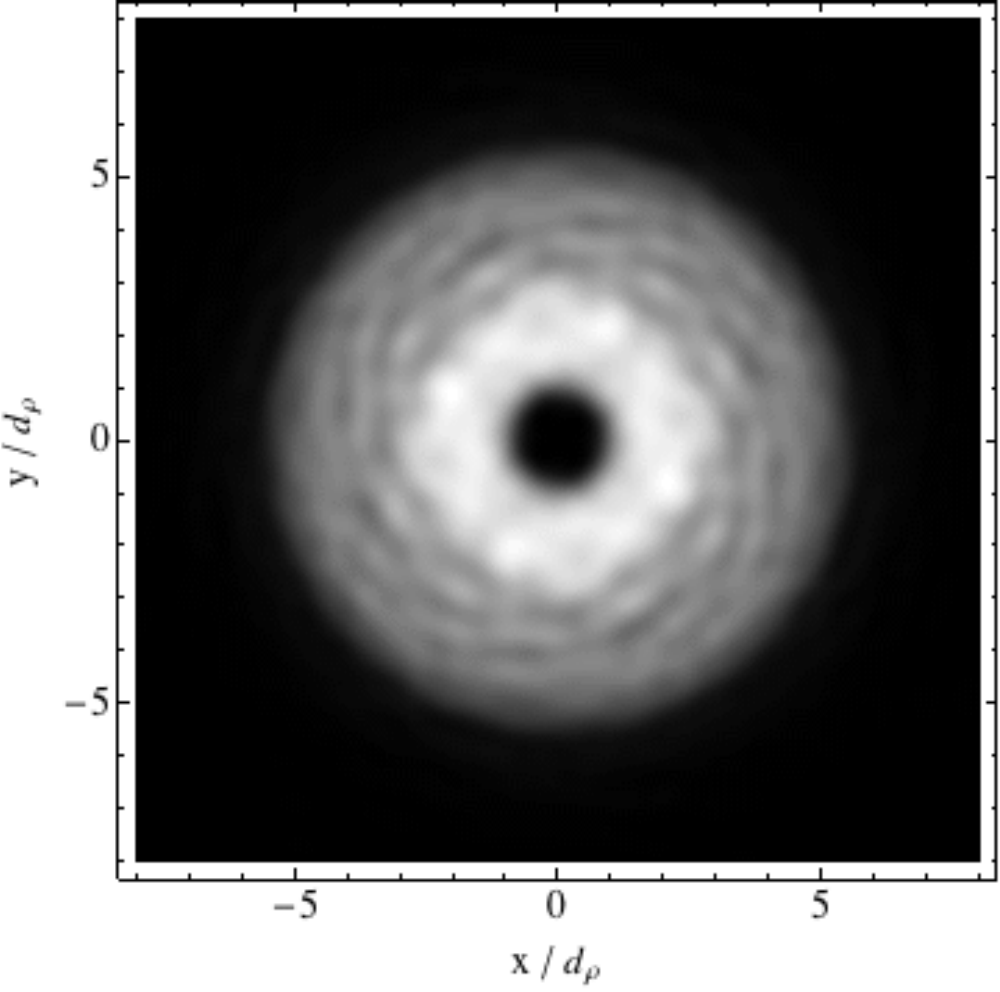}

}

\subfloat[$\omega_{\rho}t=19$]{\includegraphics[scale=0.5]{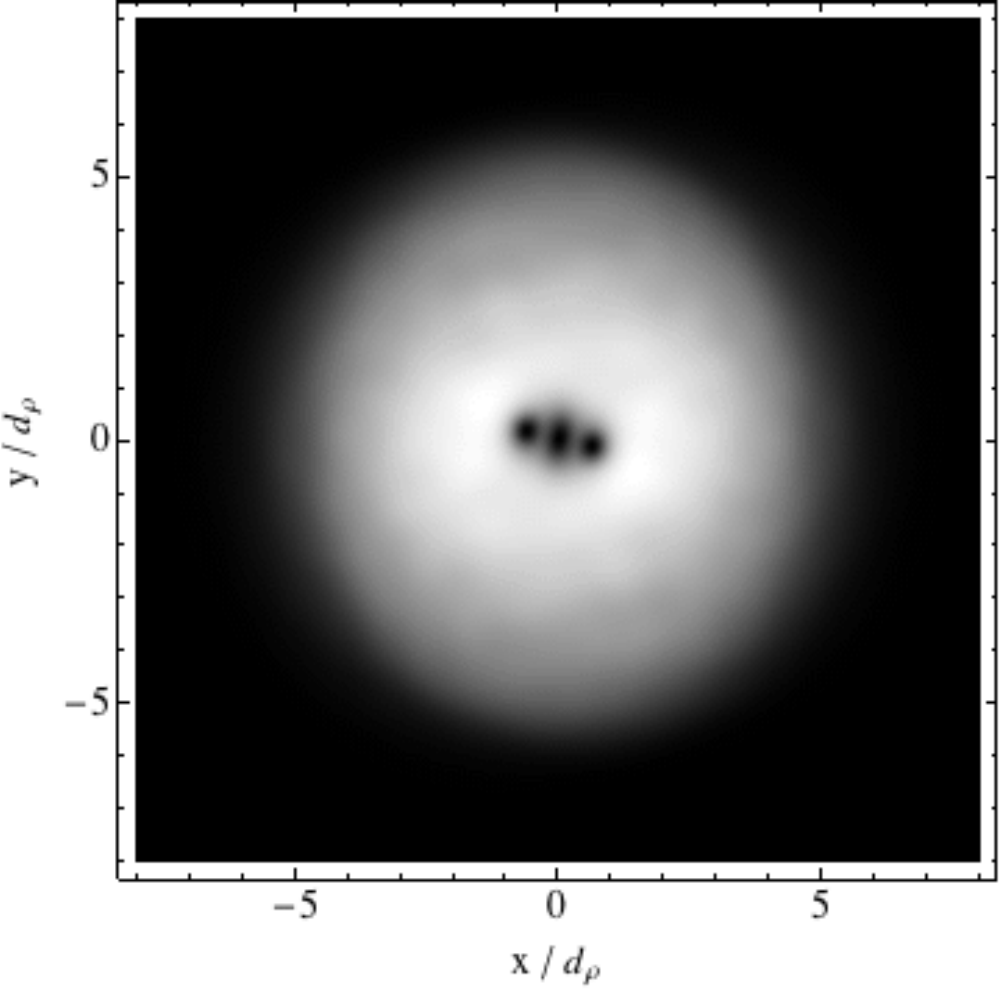}

}\subfloat[$\omega_{\rho}t=19$]{\includegraphics[scale=0.5]{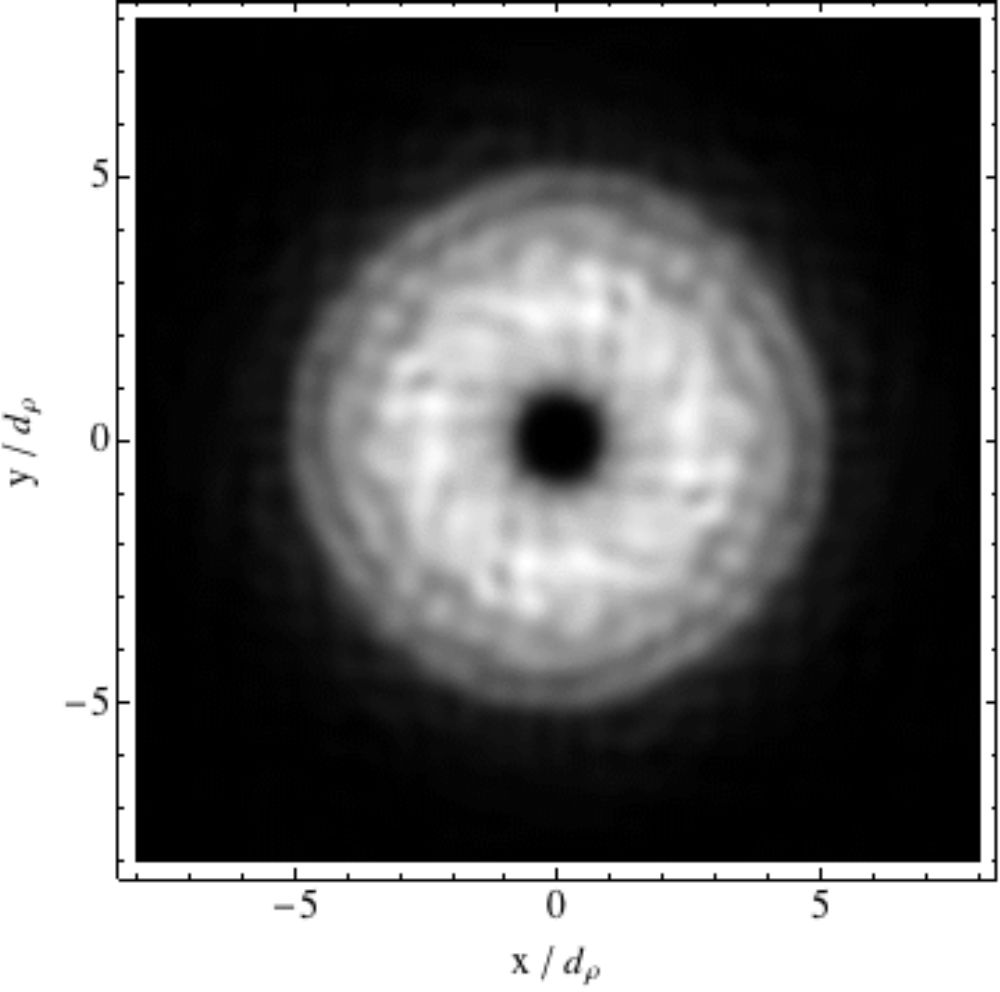}

}

\subfloat[$\omega_{\rho}t=100$]{\includegraphics[scale=0.5]{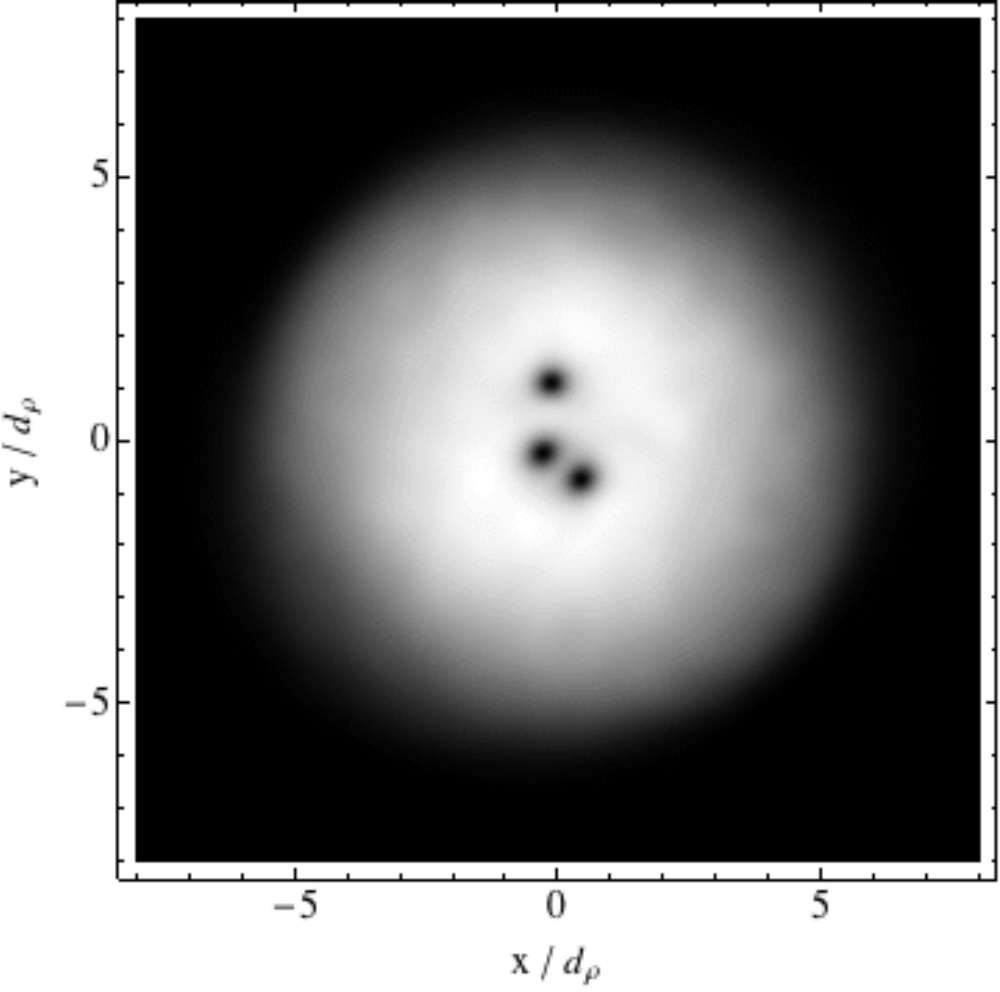}

}\subfloat[$\omega_{\rho}t=100$]{\includegraphics[scale=0.5]{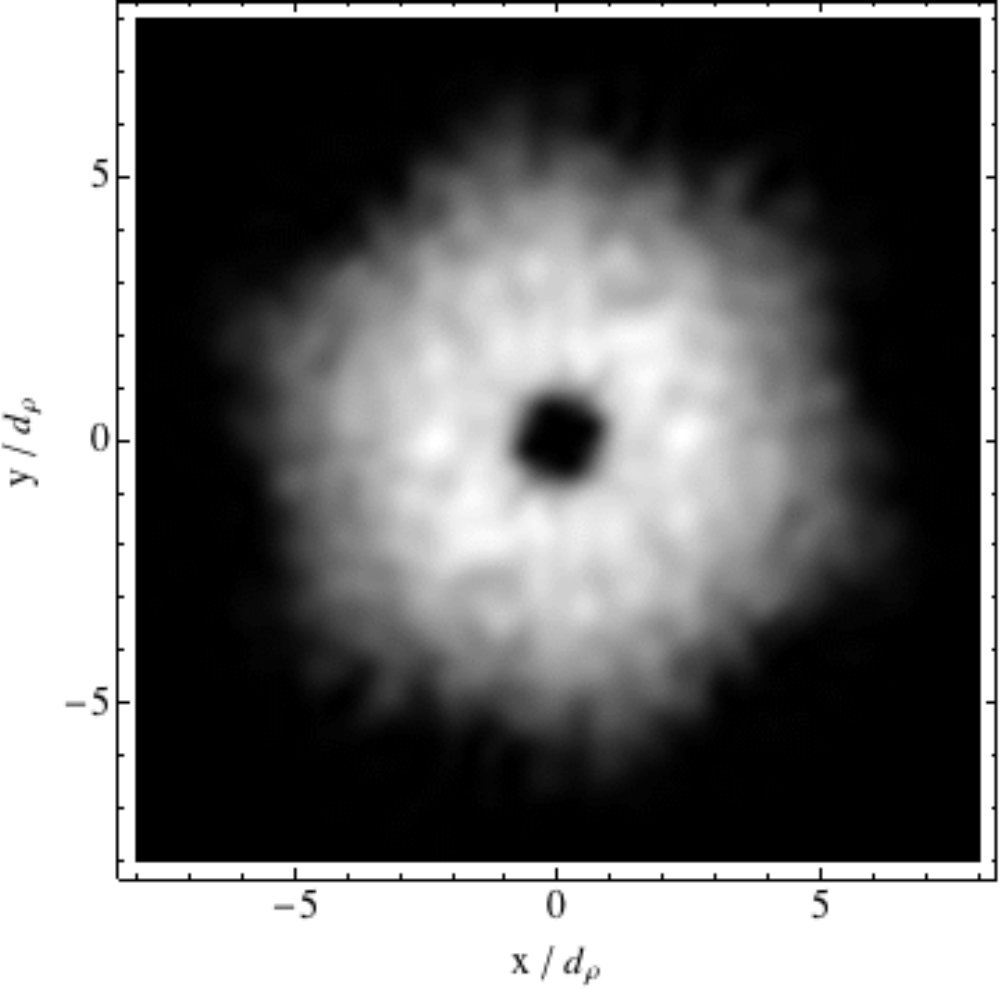}

}

\protect\caption{(Color online) Time evolution of condensate density with a triply-charged
vortex for both free Gaussian potential (a, c, e) and dynamical potential
(b, d, f). We used $\mu/\hbar\omega_{\rho}=20$, $Na_{s}/d_{z}=125$,
$V_{0}/\hbar\omega_{\rho}=50$, $\delta V/V_{0}=0.5$, $\Omega/\omega_{\rho}=5.2$,
$\epsilon=0.001$, and a factor of 0.01 multiplying the amplitude
of deviations.}

\label{fig.mPin}
\end{figure}

The dynamical mechanism works exciting the resonant mode by the oscillatory
potential placed at the center of the condensate that suppresses completely
the $Q_{v}$-mode, when the correct frequency and amplitude are considered.
Since this mode no longer exists, the vortex becomes stable (Fig.\ref{fig.mPin}).
It is what happens for the case where the static potential cannot
stabilize the vortex by itself. On the other hand, in the case of
static potential is enough to prevent the vortex decay, the modulation
of the height plays an opposite role inducing the vortex decay in
resonance regions.

\section{Conclusions}

In this paper we have studied the stability of collective modes as
well as its dynamical stability for a quasi-2D Bose-Einstein condensate
with a multi-charged vortex. The presence of a $\ell$-charged vortex
causes a shift in the frequencies of the cloud collective modes, however
such changes are not substantial. The vortex rotational mode is an
independent degree of freedom and does not affect vortex stability.
The vortex dynamics couples with collective excitations, and it can
be the cause for the $\ell$-charged vortex decay. Its decay has as
responsible the quadrupole oscillation $Q_{v}$, which is one channel
that leads the $\ell$-charged vortex to decay into $\ell$ singly
vortices. This quadrupole is the main channel to doubly-charged vortex
decay into two singly vortices. By applying a static Gaussian potential
we can prevent the decay of a vortex for specific potential amplitudes,
whereas for some regions in the parameter space can be stabilized
by a time periodic modulation of the laser potential.
\begin{acknowledgments}
We acknowledge financial support from the National Council for the
Improvement of Higher Education (CAPES) and from the State of S�o
Paulo Foundation for Research Support (FAPESP).
\end{acknowledgments}

\pagebreak{}

\appendix

\section{Functions $A_{i}\left(\ell,\alpha_{0}\right)$ and $I_{i}\left(\ell,\alpha_{0}\right)$}

\label{sec:A}

Similar functions to $A_{i}\left(\ell,\alpha_{0}\right)$ for a 3D
case have been calculated in Ref.\cite{rpteles2}. Since it is a Thomas-Fermi
wave-function the procedure to evaluate each integral is the same,
where we start changing the scale of both $x$ and $y$ coordinates
according to $x\rightarrow R_{x}x$ and $y\rightarrow R_{y}y$. By
doing this $\xi_{i}$ becomes $\alpha_{i}=\xi_{i}/R_{i}$, i.e. the
integral becomes dimensionless. Now it is convenient to change the
coordinates from cartesians to polar ($x=\rho\cos\phi$ and $y=\rho\sin\phi$)
where the integration domains are $0\leq\rho\leq1$ and $0\leq\phi\leq2\pi$,
in this way we have
\begin{equation}
A_{0}\left(\ell,\alpha_{0}\right)=\frac{\pi}{\alpha_{0}^{2\ell}}\left[\frac{_{2}F_{1}\left(\ell,\ell+1;\ell+2;-\alpha_{0}^{-2}\right)}{\ell+1}-\frac{_{2}F_{1}\left(\ell,\ell+2;\ell+3;-\alpha_{0}^{-2}\right)}{\ell+2}\right],
\end{equation}
\begin{equation}
A_{1}\left(\ell,\alpha_{0}\right)=\frac{1}{2}\left(-1\right)^{\ell}\pi\alpha_{0}^{4}\left[B\left(-\alpha_{0}^{-2};\ell+2,\ell-1\right)+\alpha_{0}^{2}B\left(-\alpha_{0}^{-2};\ell+3,\ell-1\right)\right],
\end{equation}
\begin{equation}
A_{2}\left(\ell,\alpha_{0}\right)=\frac{3}{8}\pi\alpha_{0}^{-2\ell}\left[\frac{_{2}F_{1}\left(\ell,\ell+3;\ell+4;-\alpha_{0}^{-2}\right)}{\ell+3}-\frac{_{2}F_{1}\left(\ell,\ell+4;\ell+5;-\alpha_{0}^{-2}\right)}{\ell+4}\right],
\end{equation}
\begin{equation}
A_{3}\left(\ell,\alpha_{0}\right)=\frac{5}{16}\left(-1\right)^{\ell}\pi\alpha_{0}^{8}\left[B\left(-\alpha_{0}^{-2};\ell+4,\ell-1\right)+\alpha_{0}^{2}B\left(-\alpha_{0}^{-2};\ell+5,\ell-1\right)\right],
\end{equation}
\begin{equation}
A_{4}\left(\ell,\alpha_{0}\right)=\frac{\pi\left(1+\alpha_{0}^{2}\right)^{-\ell}}{2\ell\left(1+\ell\right)},
\end{equation}
\begin{equation}
A_{5}\left(\ell,\alpha_{0}\right)=\frac{\pi}{\alpha_{0}^{2\ell}}\left[\frac{_{2}F_{1}\left(\ell,\ell;\ell+1;-\alpha_{0}^{-2}\right)}{\ell}-\frac{_{2}F_{1}\left(\ell,\ell+1;\ell+2;-\alpha_{0}^{-2}\right)}{\ell+1}\right],
\end{equation}
\begin{equation}
A_{6}\left(\ell,\alpha_{0}\right)=\frac{\pi}{\alpha_{0}^{4\ell}}\left[\frac{_{2}F_{1}\left(2\ell,2\ell+1;2\ell+2;-\alpha_{0}^{-2}\right)}{2\ell+1}-\frac{_{2}F_{1}\left(2\ell,2\ell+2;2\ell+3;-\alpha_{0}^{-2}\right)}{2\ell+2}\right],
\end{equation}
\begin{equation}
I_{1}\left(\ell,\alpha_{0}\right)=\pi\ell\alpha_{0}\left\{ \left(1+\alpha_{0}^{2}\right)^{-\ell}+\left(-1\right)^{\ell}\left[1+\left(1+\ell\right)\alpha_{0}^{2}\right]B\left(-\alpha_{0}^{-2};\ell+1,-\ell\right)\right\} ,
\end{equation}
\begin{equation}
I_{2}\left(\ell,\alpha_{0}\right)=\frac{3}{4}\pi\ell\left[\left(1+\alpha_{0}^{2}\right)^{-\ell-1}-\left(\frac{\ell+1}{\alpha_{0}^{2\ell+2}}\right)\frac{_{2}F_{1}\left(\ell+2,\ell+2;\ell+3;-\alpha_{0}^{-2}\right)}{\ell+2}\right],
\end{equation}
\begin{eqnarray}
I_{3}\left(\ell,\alpha_{0}\right) & = & \frac{\pi\ell}{2\alpha_{0}^{2\ell+2}}\left[\left(\frac{1+3\ell}{\ell+1}\right)\left(1+\alpha_{0}^{-2}\right)^{-\ell-1}\right.\nonumber \\
 &  & \left.-\left(\frac{2}{\alpha_{0}^{2}}+3\ell+3\right)\frac{_{2}F_{1}\left(\ell+2,\ell+2;\ell+3;-\alpha_{0}^{-2}\right)}{\ell+2}\right],
\end{eqnarray}
\begin{eqnarray}
I_{4}\left(\ell,\alpha_{0}\right) & = & \frac{\pi\ell}{4\alpha_{0}^{2\ell-4}}\left[\left(\frac{3\ell+1}{\ell+1}\right)\left(1+\alpha_{0}^{-2}\right)^{-\ell-1}\right.\nonumber \\
 &  & \left.-\left(\frac{2}{\alpha_{0}^{2}}+3\ell+3\right)\frac{_{2}F_{1}\left(\ell+2,\ell+2;\ell+3;-\alpha_{0}^{-2}\right)}{\ell+2}\right],
\end{eqnarray}
\begin{equation}
I_{5}\left(\ell,\alpha_{0}\right)=-\frac{3}{4}\left(-1\right)^{\ell}\pi\alpha_{0}^{3}\left[B\left(-\alpha_{0}^{-2};\ell+2,-\ell\right)+\alpha_{0}^{2}B\left(-\alpha_{0}^{-2};\ell+3,-\ell\right)\right]
\end{equation}
\begin{eqnarray}
I_{6}\left(\ell,\alpha_{0}\right) & = & \frac{\pi\ell}{16\alpha_{0}^{2\ell}}\left\{ \left[\frac{11\left(2\ell+1\right)+12\alpha_{0}^{-2}}{\left(\ell+1\right)^{-1}}\right]\frac{_{2}F_{1}\left(\ell+2,\ell+2;\ell+3;-\alpha_{0}^{-2}\right)}{\ell+2}\right.\nonumber \\
 &  & \left.-\frac{\alpha_{0}^{-2}+11\left(\ell+1\right)}{\left(1+\alpha_{0}^{-2}\right)^{\ell+1}}\right\} ,
\end{eqnarray}
\begin{equation}
I_{7}\left(\ell,\alpha_{0}\right)=\frac{\pi\ell}{16\alpha_{0}^{2\ell}}\left[\frac{_{2}F_{1}\left(\ell+2,\ell+2;\ell+3;-\alpha_{0}^{-2}\right)}{\left(\ell+1\right)^{-1}}-\frac{\ell+1-\alpha_{0}^{-2}}{\left(1+\alpha_{0}^{-2}\right)^{\ell+1}}\right],
\end{equation}
\begin{eqnarray}
I_{8}\left(\ell,\alpha_{0}\right) & = & \frac{\pi\ell}{4\alpha_{0}^{2\ell}}\left\{ \left[\frac{2\ell+4+3\alpha_{0}^{-2}}{\left(\ell+1\right)^{-1}}\right]\frac{_{2}F_{1}\left(\ell+2,\ell+2;\ell+3;-\alpha_{0}^{-2}\right)}{\ell+2}\right.\nonumber \\
 &  & \left.-\frac{\alpha_{0}^{-2}+2\ell+2}{\left(1+\alpha_{0}^{-2}\right)^{\ell+1}}\right\} ,
\end{eqnarray}
\begin{eqnarray}
I_{9}\left(\ell,\alpha_{0}\right) & = & \frac{\pi\ell}{8\alpha_{0}^{2\ell-6}}\left[\left(\ell+1\right)\left(\frac{3}{\alpha_{0}^{2}}+4+2\ell\right)\frac{_{2}F_{1}\left(\ell+2,\ell+2;\ell+3;-\alpha_{0}^{-2}\right)}{\ell+2}\right.\nonumber \\
 &  & \left.-\frac{\alpha_{0}^{-2}+2\ell+2}{\left(1+\alpha_{0}^{-2}\right)^{\ell+1}}\right],
\end{eqnarray}
\begin{equation}
I_{10}\left(\ell,\alpha_{0}\right)=\!\frac{\pi\ell}{4\alpha_{0}^{2\ell-2}}\!\!\left[\!\frac{_{2}F_{1}\left(\ell+1,\ell+2;\ell+3;-\alpha_{0}^{-2}\right)}{\ell+2}\!-\!\frac{_{2}F_{1}\left(\ell+1,\ell+3;\ell+4;-\alpha_{0}^{-2}\right)}{\ell+3}\!\right]\!\!,
\end{equation}
\begin{equation}
I_{11}\left(\ell,\alpha_{0}\right)=\frac{5}{8}\left(-1\right)^{\ell}\pi\ell\alpha_{0}^{5}\left[B\left(-\alpha_{0}^{-2};\ell+3,-\ell\right)+\alpha_{0}^{2}B\left(-\alpha_{0}^{-2};\ell+4,-\ell\right)\right],
\end{equation}
\begin{equation}
I_{12}\left(\ell,\alpha_{0}\right)=\frac{\pi}{4}\frac{2\alpha_{0}^{-1}+\left(3\ell+2\right)\alpha_{0}}{\ell\left(\ell+1\right)\left(1+\alpha_{0}^{2}\right)^{\ell+1}},
\end{equation}
\begin{equation}
I_{13}\left(\ell,\alpha_{0}\right)=\frac{\pi}{4}\frac{2\alpha_{0}^{-1}-\left(\ell-2\right)\alpha_{0}}{\ell\left(\ell+1\right)\left(1+\alpha_{0}^{2}\right)^{\ell+1}},
\end{equation}
\begin{equation}
I_{14}\left(\ell,\alpha_{0}\right)=\frac{\pi}{8}\frac{3\ell+8+4\alpha_{0}^{-2}+\left[4+\ell\left(5\ell+8\right)\right]\alpha_{0}^{2}}{\ell\left(\ell+1\right)\left(1+\alpha_{0}^{2}\right)^{\ell+2}},
\end{equation}
\begin{equation}
I_{15}\left(\ell,\alpha_{0}\right)=\frac{\pi}{8}\frac{\left(\ell-2\right)\alpha_{0}^{2}-3}{\left(\ell+1\right)\left(1+\alpha_{0}^{2}\right)^{\ell+2}},
\end{equation}
\begin{equation}
I_{16}\left(\ell,\alpha_{0}\right)=\frac{\pi}{4}\frac{\left(\ell^{2}-\ell+2\right)\alpha_{0}^{2}-2\alpha_{0}^{-2}-2\ell-4}{\ell\left(\ell+1\right)\left(1+\alpha_{0}^{2}\right)^{\ell+2}},
\end{equation}
\begin{equation}
I_{17}\left(\ell,\alpha_{0}\right)=\frac{\pi}{8}\frac{2\alpha_{0}^{4}-\left(2\ell-2\right)\alpha_{0}^{6}+\left(\ell^{2}-\ell+2\right)\alpha_{0}^{8}}{\ell\left(\ell+1\right)\left(1+\alpha_{0}^{-2}\right)^{\ell+2}},
\end{equation}
\begin{equation}
I_{18}\left(\ell,\alpha_{0}\right)=\frac{3\pi}{2\alpha_{0}}\left(1+\alpha_{0}^{2}\right)^{-\ell}\left[1-\ell\frac{_{2}F_{1}\left(1,1;\ell+2;-\alpha_{0}^{-2}\right)}{\ell+1}\right],
\end{equation}
\begin{equation}
I_{19}\left(\ell,\alpha_{0}\right)=\frac{3\pi}{4\alpha_{0}^{2}}\left(1+\alpha_{0}^{2}\right)^{-\ell},
\end{equation}
\begin{equation}
I_{20}\left(\ell,\alpha_{0}\right)=\frac{\pi}{2\alpha_{0}^{2}}\left(1+\alpha_{0}^{2}\right)^{-\ell-1}\left[\left(\frac{\ell-1}{\ell+1}\right)\alpha_{0}^{2}-1+2\ell\frac{_{2}F_{1}\left(1,1;\ell+3;-\alpha_{0}^{-2}\right)}{\ell+2}\right],
\end{equation}
\begin{equation}
I_{21}\left(\ell,\alpha_{0}\right)=\frac{\pi\alpha_{0}^{4}}{4}\left(1+\alpha_{0}^{2}\right)^{-\ell-1}\left[\left(\frac{\ell-1}{\ell+1}\right)\alpha_{0}^{2}-1+2\ell\frac{_{2}F_{1}\left(1,1;\ell+3;-\alpha_{0}^{-2}\right)}{\ell+2}\right],
\end{equation}
\begin{eqnarray}
I_{22}\left(\ell,\alpha_{0}\right) & = & \frac{\pi\ell}{\alpha_{0}^{4\ell-1}}\left[\frac{_{2}F_{1}\left(2\ell+2,2\ell+1;2\ell+3;-\alpha_{0}^{-2}\right)}{\ell+1}\right.\nonumber \\
 &  & \left.-2\frac{_{2}F_{1}\left(2\ell+1,2\ell+1;2\ell+2;-\alpha_{0}^{-2}\right)}{2\ell+1}\right],
\end{eqnarray}
\begin{equation}
I_{23}\left(\ell,\alpha_{0}\right)=\frac{3}{4}\pi\ell\left[\frac{2}{\left(1+\alpha_{0}^{2}\right)^{2\ell+1}}-\left(\frac{2\ell+1}{\alpha_{0}^{4\ell+2}}\right)\frac{_{2}F_{1}\left(2\ell+2,2\ell+2;2\ell+3;-\alpha_{0}^{-2}\right)}{\ell+1}\right],
\end{equation}
\begin{eqnarray}
I_{24}\left(\ell,\alpha_{0}\right) & = & \frac{\pi\ell}{\alpha_{0}^{4\ell}}\left[\left(\frac{2\ell-1}{2\ell+1}\right)\alpha_{0}^{-2}\left(1+\alpha_{0}^{-2}\right)^{-2\ell-1}\right.\nonumber \\
 &  & +\frac{4}{\alpha_{0}^{4}}\left(\frac{\ell+1}{2\ell+2}\right)\frac{_{2}F_{1}\left(2\ell+2,2\ell+3;2\ell+4;-\alpha_{0}^{-2}\right)}{2\ell+3}\nonumber \\
 &  & \left.-\left(\frac{2\ell-1}{\alpha_{0}^{2}}+\frac{2}{\alpha_{0}^{4}}\right)\frac{_{2}F_{1}\left(2\ell+2,2\ell+2;2\ell+3;-\alpha_{0}^{-2}\right)}{2\ell+2}\right],
\end{eqnarray}
\begin{eqnarray}
I_{25}\left(\ell,\alpha_{0}\right) & = & \frac{\pi\ell}{2\alpha_{0}^{4\ell}}\left\{ \left(\frac{2\ell-1}{2\ell+1}\right)\alpha_{0}^{4}\left(1+\alpha_{0}^{-2}\right)^{-2\ell-1}\right.\nonumber \\
 &  & -\left[2+\left(2\ell-1\right)\alpha_{0}^{4}\right]\frac{_{2}F_{1}\left(2\ell+2,2\ell+2;2\ell+3;-\alpha_{0}^{-2}\right)}{2\ell+2}\nonumber \\
 &  & \left.+4\alpha_{0}^{2}\left(\frac{\ell+1}{2\ell+2}\right)\frac{_{2}F_{1}\left(2\ell+2,2\ell+3;2\ell+4;-\alpha_{0}^{-2}\right)}{2\ell+3}\right\} .
\end{eqnarray}
Where $_{p}F_{q}\left(a_{1},\ldots,a_{p};b_{1},\ldots,b_{q};x\right)$
are the hypergeometric functions, and $B\left(x;a,b\right)$ are beta
functions. The functions derived from Gaussian potential have not
an easy general form, then we write them in integral form:
\begin{equation}
A_{7}\left(\ell,\alpha_{0}\right)=2\pi\int_{0}^{1}e^{-\rho^{2}/\alpha_{0}^{2}}\left(\frac{\rho^{2}}{\rho^{2}+\alpha_{0}^{2}}\right)^{\ell}\left(1-\rho^{2}\right)\rho d\rho,
\end{equation}
\begin{equation}
I_{26}\left(\ell,\alpha_{0}\right)=-\frac{2\pi}{\alpha_{0}^{2}}\int_{0}^{1}e^{-\rho^{2}/\alpha_{0}^{2}}\left(\frac{\rho^{2}}{\rho^{2}+\alpha_{0}^{2}}\right)^{\ell}\left(1-\rho^{2}\right)\rho^{3}d\rho,
\end{equation}
\begin{equation}
I_{27}\left(\ell,\alpha_{0}\right)=\frac{\pi}{\alpha_{0}^{4}}\int_{0}^{1}e^{-\rho^{2}/\alpha_{0}^{2}}\left(\frac{\rho^{2}}{\rho^{2}+\alpha_{0}^{2}}\right)^{\ell}\left(\frac{3}{2}\rho^{2}-\alpha_{0}^{2}\right)\left(1-\rho^{2}\right)\rho^{3}d\rho,
\end{equation}
\begin{equation}
I_{28}\left(\ell,\alpha_{0}\right)=\frac{\pi}{\alpha_{0}^{4}}\int_{0}^{1}e^{-\rho^{2}/\alpha_{0}^{2}}\left(\frac{\rho^{2}}{\rho^{2}+\alpha_{0}^{2}}\right)^{\ell}\left(1-\rho^{2}\right)\rho^{5}d\rho,
\end{equation}
\begin{equation}
I_{29}\left(\ell.\alpha_{0}\right)=-2\pi\ell\alpha_{0}\int_{0}^{1}e^{-\rho^{2}/\alpha_{0}^{2}}\left(\frac{\rho^{2}}{\rho^{2}+\alpha_{0}^{2}}\right)^{\ell+1}\left(1-\rho^{2}\right)d\rho,
\end{equation}
\begin{equation}
I_{30}\left(\ell,\alpha_{0}\right)=\frac{3}{2}\pi\ell\left(\ell+1\right)\alpha_{0}^{2}\int_{0}^{1}e^{-\rho^{2}/\alpha_{0}^{2}}\left(\frac{\rho^{2}}{\rho^{2}+\alpha_{0}^{2}}\right)^{\ell}\frac{\left(1-\rho^{2}\right)}{\left(\rho^{2}+\alpha_{0}^{2}\right)^{2}}\rho d\rho,
\end{equation}
\begin{equation}
I_{31}\left(\ell,\alpha_{0}\right)=\pi\ell\int_{0}^{1}e^{-\rho^{2}/\alpha_{0}^{2}}\left(\frac{\rho^{2}}{\rho^{2}+\alpha_{0}^{2}}\right)^{\ell}\left[\frac{\left(\ell-1\right)\alpha_{0}^{2}-2\rho^{2}}{\left(\rho^{2}+\alpha_{0}^{2}\right)^{2}}\right]\left(1-\rho^{2}\right)\rho d\rho,
\end{equation}
\begin{equation}
I_{32}\left(\ell,\alpha_{0}\right)=\frac{3\pi\ell}{\alpha_{0}}\int_{0}^{1}e^{-\rho^{2}/\alpha_{0}^{2}}\left(\frac{\rho^{2}}{\rho^{2}+\alpha_{0}^{2}}\right)^{\ell+1}\left(1-\rho^{2}\right)\rho d\rho,
\end{equation}
\begin{equation}
I_{33}\left(\ell,\alpha_{0}\right)=\frac{\pi\ell}{\alpha_{0}}\int_{0}^{1}e^{-\rho^{2}/\alpha_{0}^{2}}\left(\frac{\rho^{2}}{\rho^{2}+\alpha_{0}^{2}}\right)^{\ell+1}\left(1-\rho^{2}\right)\rho d\rho,
\end{equation}
\begin{equation}
I_{34}\left(\ell,\alpha_{0}\right)=\frac{1}{2}\pi\ell\alpha_{0}^{6}\int_{0}^{1}e^{-\rho^{2}/\alpha_{0}^{2}}\left(\frac{\rho^{2}}{\rho^{2}+\alpha_{0}^{2}}\right)^{\ell}\left[\frac{\left(\ell-1\right)\alpha_{0}^{2}-2\rho^{2}}{\left(\alpha_{0}+\rho^{2}\right)^{2}}\right]\left(1-\rho^{2}\right)\rho d\rho.
\end{equation}
 \pagebreak{}

\bibliographystyle{plain}
\bibliography{manuscript}

\end{document}